\documentclass[onecolumn,showpacs,preprintnumbers,amsmath,amssymb]{elsart}
 \input amssym.tex
\newcommand{\hs}{\hspace*{0.5cm}}

\newcommand{\be}{\begin{equation}}
\newcommand{\ee}{\end{equation}}
\newcommand{\bea}{\begin{eqnarray}}
\newcommand{\eea}{\end{eqnarray}}
\newcommand{\nn}{\nonumber}
\newcommand{\crn}{\nonumber \\}

\newcommand{\al}{\alpha}

\newcommand{\va}{\varphi}

\newcommand{\fr}{\frac}
\newcommand{\bc}{\begin{center}}
\newcommand{\ec}{\end{center}}

\newcommand {\ba}{\begin{array}}
\newcommand {\ea}{\end{array}}
\newcommand{\ben}{\begin{enumerate}}
\newcommand{\een}{\end{enumerate}}
\newcommand{\ftsz}{\footnotesize}
\usepackage{axodraw,epsfig, graphicx}
\usepackage{bm}
\usepackage{dcolumn}
\begin{document}
\begin{frontmatter}
\title{Lepton flavor violating processes $\tau
\rightarrow\mu\gamma$, $\tau\rightarrow 3\mu$ and $Z\rightarrow
\mu\tau$ in the supersymmetric economical 3-3-1 model}
\author[]{L.T. Hue\thanksref{H1}}
\author[]{D.T. Huong\thanksref{H}}
\author[]{H.N. Long\thanksref{L}}
\address{Institute of Physics, Vietnam Academy of Science and
Technology,
 10 Dao Tan, Ba Dinh, Hanoi, Vietnam}
\thanks[H1]{Email: lthue@iop.vast.ac.vn}
\thanks[H]{Email: dthuong@iop.vast.ac.vn}
\thanks[L]{Email: hnlong@iop.vast.ac.vn}

\begin{abstract}
In this work, we study the charged lepton flavor  violating (cLFV)
decays
 $\tau\rightarrow \mu\gamma$, $\tau\rightarrow 3\mu$ and
 $Z\rightarrow\mu\tau$   in the framework of the
 supersymmetric economical 3-3-1 model.
 Analytic formulas for branching ratios (BR)
 of these decays are presented.  We assume that there exist lepton flavor violation (LFV) sources in both
 right-  and left-handed slepton sectors. This leads  to the  strong enhancement of
 cLFV decay rates. We also show that the effects of  the LFV source  to
 the cLFV  decay rates in the left-handed slepton sector are greater than
 those in  the right-handed slepton sector.
 By numerical investigation, we  show that  the
 model  under consideration contains the relative light mass spectrum of
  sleptons  which   satisfies  the  current
 experimental  bounds
  on   LFV  processes in the limit of small $\tan \gamma$.
The interplay between monopole and dipole operators also was studied.
\end{abstract}
\end{frontmatter}

\section{\label{intro}Introduction.}
It is known that lepton flavor (LF) numbers are strictly conversed
in Standard Model (SM). However, the observation from of neutrino
experiments \cite{Fukuda} strongly suggests that there is lepton
flavor mixing in lepton sector. It means that (cLFV)
 processes may also occur at some
level. Many current experiments especially pay attention to search
for LFV processes in the charged lepton sector such as tau  decays
\cite{belle1,babar}, $Z\rightarrow \mu\tau$ decays
\cite{delphi,beringer}, etc. because
 LFV observed in experiment
is evidence of new physics beyond the SM.  Along with experiments,
many models beyond SM that contain
 LFV processes have been constructed. One class of the simple extended SM
  models with LFV is the class of
models with non-zero neutrino masses. This kind of models contains
a new type of Yukawa couplings of right-handed neutrinos which are
sources of cLFV. But  cLFV processes in these models have been
proved to be very suppressed \cite{cheng,robert}. However, in
supersymmetric models (SUSY), the situation can be changed.
Besides LFV origin affected by the new Yukawa couplings involving
right-handed neutrinos, SUSY models also contain
 other sources of cLFV.   Particularly, the large mixings
of slepton
 mass parameters in the soft term greatly
enhance the rates for cLFV processes. Because of the appearance of
new cLFV sources in SUSY models, the cLFV decay rates can reach to
experimental bounds
 even if the mixing
angles in the slepton sector are small. The cLFV processes caused
by this kinds of LFV were studied very early \cite{lee}. Other
interesting properties supporting the study cLFV phenomenology in
SUSY models were discussed in many recent works, for example
\cite{esteves}. Many investigations about the cLFV in SUSY models
such as \cite{Anna2,hisano,xiao} indicated that the values of
BR$(\tau\rightarrow 3\mu)$ and BR$(\tau\rightarrow\mu\gamma)$ may
exceed the bounds of present experimental results. So it is
necessary to find regions of parameter space  satisfying the
experimental results \cite{babar,delphi,beringer}: \be
\mathrm{BR}(\tau^-\rightarrow \mu^-\gamma)< 4.4\times 10^{-8},
\label{etaumugamma1}\ee \be \mathrm{BR}(\tau^-\rightarrow
\mu^-\mu^+\mu^-)< 2.1\times 10^{-8}, \label{etau3mu1}\ee \be
\mathrm{BR}(Z\rightarrow \mu^+\tau^-)< 1.2\times 10^{-5}.
\label{ztaumu1}\ee
 Supersymmetric unified theories predict large rates for
 cLFV processes except the decay rate
 of $Z\rightarrow\mu\tau$.
 The branching ratio of this decay   in  models  beyond SM
 is predicted to $10^{-9}-10^{-8}$ level \cite{Jose}.
 It is very suppressed compared to the experimental bound (\ref{ztaumu1}).
In fact, the GigaZ
option will be approaching to the sensitivity of $10^{-8}$ level
in the $Z\rightarrow\mu\tau$ decay mode \cite{Aguilar}.
If observed in a future experiment, it
will be evidence of  physics in SUSY models.
 One of the SUSY models in which cLFV processes are
thoroughly investigated is the Minimal Supersymmetric Standard
Model (MSSM) \cite{Anna2,Babu}. In  the mentioned  work, the
authors  have shown that if there exists LFV in the left-handed
sector, in order to get the experimental bound on the LFV decay
rates of muon and tauon decays, the mass parameters of sleptons
should be shifted toward TeV scale, especially in the case of
large $\tan\beta$. Because the model contains many LFV effective
operators, the interplay of different effective operators (dipole,
monopole coming from neutral boson exchanges and Higgs exchanges)
creates many interesting consequences. The detail  is discussed in
works \cite{Anna2,Babu}.

 The current experimental results \cite{Fukuda} show
that neutrinos are massive,
which contradict what assumed in the SM. Other words speaking, the SM must
be extended. Among
 extensions, the models based on the  $\mathrm{SU}(3)_\mathrm{C}
\otimes \mathrm{SU}(3)_\mathrm{L}\otimes
\mathrm{U}(1)_\mathrm{X} $ (3-3-1) gauge
group \cite{331m,331r} have the following interesting
features:
\begin{itemize}
\item To be anomaly free, the number of triplets should be equal
to number of anti-triplets. This leads to that number of
generations is multiple of the color number which is three.
Combining with condition of the QCD asymptotic freedom
 the requiring number of quark
generations should be less than five. Thus, in the 3-3-1 models,
number of generations
 is three.
\item The models give an explanation of electric charge quantization  \cite{lthdt},
 dark matter and CP violation.
\item  One of the three quark families
has to transform
under $\mathrm{SU}(3)_L$ differently from the other two. This
leads to an explanation why the top quark is uncharacteristically
heavy.
\end{itemize}

The weakness of the 3-3-1 models is the complication in the Higgs
sector: in the minimal 3-3-1 model \cite{331m}, content of Higgs
sector  includes three  triplets and one sextet, while in the
3-3-1 model with right-handed neutrino \cite{331r}, there are
three triplets. To solve the mentioned weakness, the attempt is
reduction in the scalar sector. The first result is the economical
3-3-1 model \cite{e331s,e331} - the version with right-handed
neutrino and only two Higgs triplets. Scalar sector in the minimal
3-3-1 model has been reduced to minimum with two Higgs triplets,
and such version is called the reduced minimal 3-3-1 model
\cite{rm331}. It is emphasized that the problem on neutrino masses
is not totally solved in the last version. The situation seems
better in a supersymmetric version of the reduced minimal 3-3-1
model \cite{srm331}. An supersymmetric version of the economical
3-3-1 model has been established in \cite{Dong1} and called the
Supersymmetric Economical 3-3-1 model (SUSYE331).
 The SUSYE331 has the same advantages  as its non-supersymmetric version.
 However, to generate masses for fermions, the
 non-supersymmetric models
  with minimal Higgs content need
 effective   nonrenormalizable  interactions, while the SUSY versions do not
(the interested readers can find in Refs. \cite{Dong1,Dong3}).

 In  \cite{Giang}, the LFV decays of neutral Higgs bosons in
the SUSYE331 were considered.
In this work, we are interested in the cLFV processes in
the SUSYE331,  namely: cLFV decays of the tauon and  $Z$
bosons.

We remind that in the SUSYE331,  there are more leptons, Higgses
and  gauge bosons as well as their supersymmetric partners than
those of the MSSM. This implies that the model contains many
sources of LFV in the slepton sector. This suggests that the
branching ratios of LFV Higgs and Z decays are greatly enhanced
and may be detectable in near future. In contrast, bounds from
experiments (such as shown in (\ref{etaumugamma1}) and
(\ref{etau3mu1})) will strictly constrain the values of parameters
causing the cLFV. So it is really important to investigate the
parameter space where LVF decay rates can be satisfied the present
bound of experiments. In many previous works \cite{Anna2,Babu},
the authors showed that in the MSSM the interested region of
parameter space is set at TeV scale with the large value of $\tan
\beta$. Because of appearance of new LFV sources in  the SUSYE331,
we can find the interested region of parameter space even in the
limit of small
 slepton  mass parameters. In this work, we are
interested in the processes: $\tau^-\rightarrow \mu^-\gamma$,
$\tau^-\rightarrow \mu^-\mu^+\mu^-$ and $Z\rightarrow\mu\tau$.
 In particular, we incorporate the mixing of
 slepton  mass, of charginos, Higgsino, gauginos as well as the interactions
between gauge boson,
 slepton, lepton and the Yukawa interactions. This leads
to many types of enhanced  diagrams.  We will consider how large
contribution from each type of  diagrams can be. We also will
extend previous work in \cite{Anna2}  to our considered model such
as: constructing  the analytical formulas of effective operators
from the  diagrams,  from which we obtain the formulas of the
branching ratios of decays BR($\tau\rightarrow \mu\gamma$),
BR($\tau\rightarrow \mu\mu\mu$) and BR($Z\rightarrow\mu\tau$).
After that, we investigate some numerical
 results of these branching
ratios in the limit of  small $\tan\gamma$ and  the slepton
parameters are set below $1$ TeV which may be detected by current
colliders.  As in the previous works on SUSYE331, $\tan\gamma$ is
defined as the ratio of two vacuum expectations of two neutral
Higgs components  $\rho$ and $\rho'$, namely
$\tan\gamma=\langle\rho^0\rangle/\langle\rho^{\prime0}\rangle$.
Here, the quantity $\tan\gamma$ in the SUSYE331 plays a similar
role as the $\tan\beta$ in  the MSSM \cite{Giang}. In the MSSM
model, the mass of the lightest Higgs depends on $\tan\beta$ and
the mass of the standard gauge bosons.
 If we combine the theoretical result for the upper bound on
the lightest Higgs-boson mass
with the direct experimental search, it leads to exclude the
limit of small $\tan\beta$
 for the case where
soft parameters are set to TeV scale. However, this kind of
constraints on the $\tan\gamma$ does not happen in the SUSYE331
model. On the other hand, the large values of $\tan\gamma$ do not
support the region of the soft parameter space below 1 TeV. Hence,
in this  work, we concentrate  on numerical studying in the limit
of small $\tan \gamma$.
  In this paper we concentrate on only two aims. First,
we establish analytic formulas to calculating some cLFV processes
in the framework of SUSYE331. Second, we prove that there  exist
regions satisfying the current bounds of cLFV experiments.
Especially, on the basis of numerical
studying, we find some interested regions of parameter space that
satisfy the experimental bound on the cLFV decay rates.
We also discuss on the interplay between mono and dipole
operators which may lead to some interesting
consequences.

This paper is arranged as follows:
In section \ref{particles} we
review the particles content. The effective operators as well as
branching ratios of the mentioned cLFV decay processes are
presented in section \ref{effective}. Section \ref{num}  is
devoted for discussion
 on results of numerical investigation on parameter space and the last
section is the conclusion. All the analytic formulas of
 effective couplings appearing in effective
operators and interacting terms needed for our calculation are
shown in Appendices \ref{InLAno}-\ref{apbtaumu}.

\section{\label{particles} Particles content}
In this part, let us quickly review the
particle content in the
SUSYE331 model, which were given
 in the  previous papers \cite{Dong1,Dong3,Giang}.
The fermion superfields
 are given by
 \be
\widehat{L}_{a L}=\left(\widehat{\nu}_{a}, \widehat{l}_{a},
\widehat{\nu}^c_{a}\right)^T_{L} \sim (1,3,-1/3),\hs
  \widehat {l}^{c}_{a L} \sim (1,1,1), \hs a=1,2,3, \label{l2}
\end{equation}
 \be \widehat Q_{1L}= \left(\widehat { u}_1,\
                        \widehat {d}_1,\
                        \widehat {u}^\prime
 \right)^T_L \sim (3,3,1/3),\nn \ee
\be \widehat {u}^{c}_{1L},\ \widehat { u}^{ \prime c}_{L} \sim
(3^*,1,-2/3),\widehat {d}^{c}_{1L} \sim (3^*,1,1/3 ), \label{l5}
\ee
\begin{equation}
\begin{array}{ccc}
 \widehat{Q}_{\alpha L} = \left(\widehat
 {d}_{\alpha}, - \widehat{u}_{\alpha},
 \widehat{d^\prime}_{\alpha}\right)^T_{L}
  \sim (3,3^*,0), \hs \al=2,3, \label{l3}
\end{array}
\end{equation}
\begin{equation}
\widehat{u}^{c}_{\alpha L} \sim \left(3^*,1,-2/3 \right),\hs
\widehat{d}^{c}_{\alpha L},\ \widehat{d}^{\prime c}_{\alpha L}
\sim \left(3^*,1,1/3 \right),\label{l4}
\end{equation}
and the Higgs superfields are written as \be \widehat{\chi}= \left
( \widehat{\chi}^0_1, \widehat{\chi}^-, \widehat{\chi}^0_2
\right)^T\sim (1,3,-1/3), \hs \widehat{\rho}= \left
(\widehat{\rho}^+_1, \widehat{\rho}^0, \widehat{\rho}^+_2\right)^T
\sim  (1,3,2/3), \label{l8} \ee
 \be \widehat{\chi}^\prime= \left
(\widehat{\chi}^{\prime o}_1, \widehat{\chi}^{\prime
+},\widehat{\chi}^{\prime o}_2 \right)^T\sim ( 1,3^*,1/3), \,
\widehat{\rho}^\prime = \left (\widehat{\rho}^{\prime -}_1,
  \widehat{\rho}^{\prime o},  \widehat{\rho}^{\prime -}_2
\right)^T\sim (1,3^*,-2/3). \label{l10} \ee

It is noted that
$\widehat{\psi}^c_L=(\widehat{\psi}_R)^c\equiv
\widehat{\psi}_R^{\dagger}$ and  $u^\prime$, $d^\prime $ are
exotic quarks. The values in
each parenthesis show the quantum numbers of the
$(SU(3)_c, SU(3)_L, U(1)_X)$ group, respectively. The $
\mathrm{SU}(3)_L \otimes \mathrm{U}(1)_X$ gauge group is broken
as follows:
 \be \mathrm{SU}(3)_L \otimes
\mathrm{U}(1)_X \stackrel{w,w'}{\longrightarrow}\ \mathrm{SU}(2)_L
\otimes \mathrm{U}(1)_Y\stackrel{v,v',u,u'}{\longrightarrow}
\mathrm{U}(1)_{Q},\label{stages}\ee
with VEVs
given by
\be
 \sqrt{2} \langle\chi\rangle^T = \left(u, 0, w\right),  \hs \hs \sqrt{2}
 \langle\chi^\prime\rangle^T = \left(u^\prime,  0,
 w^\prime\right),\ee \be
\sqrt{2}  \langle\rho\rangle^T = \left( 0, v, 0 \right), \hs \hs
\sqrt{2} \langle\rho^\prime\rangle^T = \left( 0, v^\prime,  0
\right).\ee

The SUSYE331 model contains seventeen vector superfields such
as $\widehat{V}_c^a$, $\widehat{V^a}$ and
$\widehat{V}^\prime$. The vector superfields contain the usual gauge bosons
 given in \cite{Dong1,Dong3}. The total supersymmetric
 Lagrangian was given in \cite{Dong3}. In this work
 only terms relevant to our work   are
collected in appendix \ref{InLAno}.

 \section{ \label{effective}Effective operators and branching ratios}
In this section, we extend the previous work \cite{Anna2} in the
model under consideration. In the SUSYE331, there are six physical
vector bosons: the photon, two charged ($W^{\pm},~Y^{\pm}$), two
Hermitian neutral ($Z,Z'$)  and a Non-hermitian neutral $X$. The
effective couplings of $X\mu\tau$ interaction is very small so we
ignore them in the calculation.  Let us first write down the cLFV
effective operators for muon-tau  and photon or $Z$ bosons,
$Z^\prime$ bosons or other leptons.

\subsection{$\tau\mu\gamma$ effective operators}
First we write down the LFV
operators for $\tau, \mu, \gamma$. These
operators are divided into two terms
 \footnote{The operator containing
$D^{\gamma}$ is different from \cite{Anna2} a factor $i$. This is
because of definition of $\sigma^{\mu\nu}$ and
$\bar{\sigma}^{\mu\nu}$.   The definitions in
\cite{SpMartin1} are used.}:

\be e \left[C^{\gamma}_L \bar{\mu} \bar{\sigma}^{\mu} \tau+
C^{\gamma}_R \mu^c \sigma^{\mu} \bar{\tau}^c +
\mathrm{h.c.}\right]\square A_{\mu},\label{gamutau1} \ee
 \be e ~m_{\tau} \left[D^{\gamma}_L \bar{\mu} \bar{\sigma}^{\mu\nu}\bar{\tau}^c+
D^{\gamma}_R \mu^c \sigma^{\mu\nu} \tau +
\mathrm{h.c.}\right]F_{\mu\nu}. \label{gamutau2}\ee

The processes  associated with external photon line
depend only on the $D^{\gamma}_{L,R}$ while other processes
with virtual photon depend on both $C^{\gamma}$ and $D^{\gamma}$.
The Feynman diagrams
contributing to $C^{\gamma}$ and $D^{\gamma}$ are given
in appendix \ref{mutaugamma}. We would like to
emphasize that the number of diagrams in the considered model
is more than that of the MSSM because
 the SUSYE331 model contains
  new Higgs and
gauge bosons.   In order to obtain
 analytical formulas for $C^{\gamma}$ and $D^{\gamma}$,
we have used
 Feynman rules and some approximate expansion.
 For more details, the interested  reader can see in
 appendix \ref{mutaugamma}. As in the MSSM, the
$C^{\gamma}_{L(R)}$ does not depend  on $\tan\gamma$ and the
factor $D^{\gamma}_{L(R)}$ can be divided into three sub-terms
including sub-term  $D^{\gamma(a)}_{L(R)}$ which is independent on
$\tan\gamma$ and sub-terms $D^{\gamma(c,b)}_{L(R)}$ which are not,
more specifically
  \be D^{\gamma}_{L(R)}=D^{\gamma(a)}_{L(R)}
  +D^{\gamma(b)}_{L(R)}+D^{\gamma(c)}_{L(R)}
  \label{dgalr0}.\ee
We would like to mention that
only $D^{\gamma(c)}_{L(R)}$ contains left-right slepton mixing
parameters which, in this paper,  are denoted by $(A_{\tau},
A^{L}_{\mu\tau},A^{R}_{\mu\tau})$.

\subsection{\label{zeff1}$Z\tau\mu$ and $Z'\tau\mu$ effective
operator} First, let us consider the effective operator of the
$Z\tau\mu$. This kind of effective operators can be written in the
standard form  given in \cite{Anna2} as follows: \be g_Z~ m^2_Z
\left[A^{Z}_L \bar{\mu} \bar{\sigma}^{\mu} \tau+ A^{Z}_R
\bar{\mu}^c \sigma^{\mu} \bar{\tau}^c +
\mathrm{h.c.}\right]Z_{\mu},\label{zmutau0} \ee
\be g_Z \left[C^{Z}_L \bar{\mu} \bar{\sigma}^{\mu} \tau+ C^{Z}_R
\mu^c \sigma^{\mu} \bar{\tau}^c + \mathrm{h.c.}\right] \square
Z_{\mu},\label{zmutau1} \ee
 \be g_Z ~m_{\tau} \left[D^{Z}_L \bar{\mu} \bar{\sigma}^{\mu\nu}\tau^c+
 D^{Z}_R \mu^c \sigma^{\mu\nu} \tau + \mathrm{h.c.}\right]
Z_{\mu\nu}. \label{zmutau2}\ee
  where the mass of the  $Z$ boson in the SUSYE331 is determined
  as \cite{Dong1}:
   \bc\bea m^2_Z &\simeq & \frac{g^2\left(
   v^2+v^{\prime2}\right)}{4c^2_W}, \crn \hs  g_Z &\simeq& \frac{g}{c_W} ~~~~
  \mathrm{and}~~~~ g_{Z'} \simeq\frac{g}{\kappa_1c_W}.
  \label{mz1}\eea\ec
  ($g_Z$ and $g_{Z'}$  are defined from  covariant derivatives, as
 explained in  Appendices \ref{InLAno}-\ref{apbtaumu}.) The operators related to the factors
 $A^{Z}_{L,R}$  are chirality conserving (monopole). The leading
 contribution to the monopole operators comes from effective
 couplings of muon and tauon with two neutral Higgs bosons. In the
  model under consideration, we have four neutral Higgs bosons
 $\rho^0,~\rho^{\prime0},~\chi^{0}_1$ and $ \chi^{\prime0}_1$
 which can couple to the $Z$ boson.
However, investigation in \cite{e331} has noted the relation
$u,u'\ll v,v'\ll w,w'$. In this limit, we can neglect the coupling
of the $Z$ with $\chi$ and
 $\chi^{\prime}$. It means that  we obtain only the leading interactions of
 $\mu, \tau$ with $\rho, \rho^\prime$.
These leading terms lead to a consequence that the monopole
operators in
 (\ref{zmutau0}) can be extracted out
a factor $m^2_Z$. We would like to remind that the class of
diagrams containing $\mu, \tau$, $\rho^0$ and $\rho^{\prime0}$
 presented in Appendix \ref{alra}
 give contribution to $A^{Z}_{L,R}$
 and this factor can be written as sum
 of three parts:
  \be
 A^{Z}_{L(R)}=A^{Z(a)}_{L(R)}+A^{Z(b)}_{L(R)}+A^{Z(c)}_{L(R)},\label{azlr0}\ee
 where  analytical formulas of each term in (\ref{azlr0}) as well as $C^{Z}_{L(R)}$
 and $D^{Z}_{L(R)}$ are given in Appendix \ref{apzmutau}. The
 operator related to $C^{Z}_{L(R)}$ also are chirality conserving
(monopole)
 while the operators related to $D^{Z}_{L(R)}$ are chirality flipping
  (dipole).  More details about origin of these operators, the
  interested reader can see in \cite{Anna2}.

 A big difference compared to the MSSM  is that
 the  model under consideration
contains more $\mu \tau $ effective operators
 such as $\mu \tau Z^\prime $. Because of the couplings of $Z^\prime$ with
 all of neutral Higgs--$\chi^0_2, ~\chi^{\prime0}_2, ~\rho^0$ and
 $\rho^{\prime0}$--the
  monopole operators of $Z^\prime \mu \tau$
relate not only to the factor $m^2_{Z}$ in but also the
factor $m^2_{Z^\prime}$,
as indicated in two formulas (\ref{zpmutau0}) and
(\ref{zpmutaua20}).
It is noted that $A^{1Z'(a)}_{L(R)}$
comes from the leading interactions of $\mu \tau$ with
two neutral Higgs $\rho$ and $ \rho^\prime$ while the
$A^{1Z'(a)}_{L(R)}$
comes from the leading interactions of $\mu \tau$ with
two neutral Higgs $\chi_2^0$ and $ \chi^{\prime 0}_2$.
 For all of others effective
 operators mentioned in this work-$C^{Z,(Z')}_{L,(R)}$ and
 $D^{Z,(Z')}_{L,(R)}$-relate  with only two neutral Higgs $\rho$ and $\rho'$.
Above comments are enough for us to write the standard form of
$\mu \tau Z^\prime$ operators as follows:
\be g_{Z'}~ m^2_{Z} \left[A^{(1Z')}_L \bar{\mu} \bar{\sigma}^{\mu}
\tau+ A^{(1Z')}_R \bar{\mu}^c \sigma^{\mu} \bar{\tau}^c +
\mathrm{h.c.}\right]Z'_{\mu},\label{zpmutau0} \ee
\be g_{Z'}~ m^2_{Z'} \left[A^{(2Z')}_L \bar{\mu}
\bar{\sigma}^{\mu} \tau+ A^{(2Z')}_R \bar{\mu}^c \sigma^{\mu}
\bar{\tau}^c + \mathrm{h.c.}\right]Z'_{\mu},\label{zpmutaua20} \ee
 \be g_{Z'} \left[C^{Z'}_L \bar{\mu} \bar{\sigma}^{\mu}
\tau+ C^{Z'}_R \mu^c \sigma^{\mu} \bar{\tau}^c +
\mathrm{h.c.}\right] \square Z'_{\mu},\label{zpmutau1} \ee
 \be g_{Z'} ~m_{\tau} \left[ D^{Z'}_L \bar{\mu} \bar{\sigma}^{\mu\nu}\tau^c+
D^{Z'}_R \mu^c \sigma^{\mu\nu} \tau + \mathrm{h.c.}\right]
Z'_{\mu\nu}. \label{zpmutau2}\ee
Note that in the on-shell condition we have $\square (Z)
\rightarrow -m^2_{Z}$ and  $\square (Z') \rightarrow -m^2_{Z'}$,
where:
$$ m^2_{Z'}= \frac{g^2 c^2_W W^2}{4 c^2_W-1},\hs
W^2=w^2+w^{\prime2}.$$
The $A^{1Z'}_{L(R)}$ and $A^{2Z'}_{L(R)}$ are also written in the
form  as those for $A^{Z}_{L(R)}$. Forms of the $D^{Z}_{L(R)}$ and
$D^{Z'}_{L(R)}$ are the same of $D^{\gamma}_{L(R)}$ in
(\ref{dgalr0}).  All analytical formulas of effective operators in
this section are given in Appendix \ref{apzpmutau}.

 In addition, the SUSYE331 model
  has two others neutral gauge bosons, $Z'$ and
 $X$ compared to the  MSSM.  This appearance may give significant contribution
 to the $\tau\rightarrow 3\mu$ decay. The mentioned contribution may be
 similar to that of the $Z$ boson in some region of parameter
 space. This issue will be discussed in  more details in the next section.

\subsection{$\tau\mu \mu\mu$ effective operator}

Let us write down four-fermions $\tau\mu \mu\mu$ effective
operators. The $\tau\mu \mu\mu$ effective operators can be
constructed from the effective operators of $\mu, \tau$ with
photon or $Z, Z^\prime$ bosons as well as Higgs bosons. Besides
the contributions  coming from photon, $Z$ boson and Higgs
exchanges, there are other contributions  to the $\tau\mu \mu\mu$
effective operators which come from box-diagrams  shown in Fig.
(\ref{blrf}). In this part, we write down only the standard form
of $\tau\mu \mu\mu$ effective operator coming from box-diagrams as
\bea    &&\left[ (\bar{\mu} \bar{\sigma}^{\mu} \tau) \left(
B_L^{\mu_L} \,\bar{\mu} \bar{\sigma_\mu}\mu + B_L^{\mu_R} \, \mu^c
\sigma_{\mu} \bar{\mu}^c \right) \right.\crn&+&\left. ( \mu^c
\sigma^{\mu} \bar{\tau}^c) \left( B_R^{\mu_L} \,\bar{\mu}
\bar{\sigma_\mu} \mu + B_R^{\mu_R} \, \mu^c \sigma_\mu \bar{\mu}^c \right)
+ {\rm h.c.} \right] \label{efftau3mu1}. \eea
 Analytical forms of the coefficients $B_{L(R)}^{\mu_{L(R)}}$ are presented in
 Appendix \ref{apbtaumu}.

\subsection{Branching ratios}
General method to construct the branching ratios for the cLFV
decay from effective operator was written in \cite{Anna2}.
Especially, based on the basis of the effective operators, we
write effective Lagrangian of the muon and tauon with photon,
gauge bosons $Z, Z^\prime$ as well as lepton. From this effective
Lagrangian, we obtain the branching ratio for each of processes by
using the Feynman rules. In this section, we study the branching
ratios for the considered cLFV decays of the tau and the $Z$
bosons.

\subsubsection{$\tau\rightarrow\mu\gamma$}
Let us consider the cLFV decay  mode $\tau\rightarrow\mu\gamma$. It
is not hard to obtain the decay rate of $\tau\rightarrow
\mu\gamma$ from effective operators given in
Eqs. (\ref{gamutau1},\ref{gamutau2}). The
detailed calculation can be seen in
\cite{hisano}.  Comparing the
 branching of LFV decay $\tau\rightarrow \mu\gamma$
with that of the
$\tau\rightarrow \mu\bar{\nu}_{\mu}\nu_{\tau}$ decay given
in \cite{beringer,SpMartin1},
we can obtain the relation
between the two branching ratios of the two above processes.
The result is entirely consistent with the results given in \cite{Anna2}, namely

 \bea \mathrm{B R}(\tau\rightarrow
\mu^-\gamma)=\frac{48\pi^2\alpha}{G^2_F}\left[|D^{\gamma}_L|^2+
|D^{\gamma}_R|^2\right] \mathrm{BR}(\tau\rightarrow
\mu^-\bar{\nu}_\mu\nu_{\tau}), \label{brtaumugamma1}\eea
where $\alpha=\frac{e^2}{4\pi}$, $\frac{G_F}{\sqrt{2}}= \frac{g^2}{8 m^2_W}$
 and $\mathrm{Br}(\tau\rightarrow
\mu^-\bar{\nu}_\mu\nu_{\tau})\simeq 17.41\%$.

\subsubsection{$Z\rightarrow \mu\tau$}
In this subsection, we consider the decay mode of $Z \rightarrow
\mu \tau$. The decay rate of $Z\rightarrow \ell^+\ell^-$ in case
of the SUSY can also be determined from the general formula
established in \cite{SpMartin1}. In particular, the decay rate of
$Z \rightarrow \mu \tau$ in the  SUSYE331 model can be written as
follows:
\bea \mathrm{\Gamma} (Z\rightarrow \ell^+\ell^-)=
\frac{g^2_Zm_Z}{24\pi}
\left(1-\frac{4m_{\ell}^2}{m_Z^2}\right)^{1/2}
\left[(a_{\ell}^2+b_{\ell}^2)\left(1-\frac{4m_{\ell}^2}{m_Z^2}\right)+6
a_{\ell}b_{\ell}\frac{m_{\ell}^2}{m_Z^2} \right],
\crn\label{wzmutau1}\eea where
$$a_{\ell}\equiv
T^{3}_\ell-Q_{\ell}s^2_W =\frac{-1+2s^2_W}{2}\hs \mathrm{and}\hs
b_{\ell}=-Q_{\ell}s^2_W= s^2_W.$$ For the decay rate of
$Z\rightarrow \mu\tau$, because the form of effective operators of
$Z \mu \tau$ in the considered model is the same as that of the
MSSM, the form of decay rate of $Z \rightarrow \mu \tau $ is the
same as that established in \cite{Anna2}. Especially the result is
given as follows:
\bea \mathrm{\Gamma} (Z\rightarrow \mu^+\tau^-)=
\frac{g^2_Zm^5_Z}{24\pi}\left[\left|F^Z_L\right|^2
+\left|F^Z_R\right|^2+\frac{1}{2}
\left|\frac{m_{\tau}}{m_Z}D^Z_L\right|^2
+\frac{1}{2}\left|\frac{m_{\tau}}{m_Z}D^Z_R\right|^2\right],
\label{widthzmutau331}\eea where
$$ F^Z_{L,R}= A^Z_{L,R}-C^Z_{L,R}.$$

Comparing  the two results taken from Eqs.(\ref{wzmutau1})and
(\ref{widthzmutau331}), we obtain the relationship between the two
branching ratios corresponding to two processes $Z\rightarrow ll$
and
$Z\rightarrow \mu\tau$.
Our calculation is
consistent with results given in \cite{Anna2} such as:
\bea &&\mathrm{BR}(Z\rightarrow \mu\tau)\crn &&= c
m^4_Z\left[\left|F^Z_L\right|^2+\left|F^Z_R\right|^2+\frac{1}{2}
\left|\frac{m_{\tau}}{m_Z}D^Z_L\right|^2
+\frac{1}{2}\left|\frac{m_{\tau}}{m_Z}D^Z_R\right|^2\right]
\mathrm{B R}(Z\rightarrow \ell^+\ell^-)\label{bzmutau1},\eea
 where $c\equiv\left(a^2_{\ell}+ b^2_{\ell}\right)^{-1}
  =(1/4 -s^2_W+2s^4_W)^{-1}\simeq 7.9$ and $\mathrm{B R}(Z\rightarrow
 \ell^+\ell^-)\simeq 3.4\%$.

 \subsubsection{$Z'\rightarrow \mu\tau$}

Similar to the case of the $Z$ boson, the decay rate of
$Z'\rightarrow \ell^+\ell^-$ can be determined in the formula
below: \bea \mathrm{\Gamma} (Z'\rightarrow \ell^+\ell^-)&=&
\frac{g^2_{Z'}m_{Z'}}{24\pi}
\left(1-\frac{4m_{\ell}^2}{m_{Z'}^2}\right)^{1/2}
\left[(a_{\ell}^{\prime2}+b_{\ell}^{\prime2})
\left(1-\frac{4m_{\ell}^2}{m_{Z'}^2}\right)+6
a'_{\ell}b'_{\ell}\frac{m_{\ell}^2}{m_{Z'}^2}
\right]\crn&\simeq&\frac{g^2_{Z'}m_{Z'}}{24\pi}
(a^2_{\ell}+b^2_{\ell}), \label{wzmutau2}\eea where
$a_{\ell}^{\prime2}+b_{\ell}^{\prime2}=a_{\ell}^{2}+b_{\ell}^{2}=1/c$
and
 $$ a'_{\ell}=\left( 4c^2_W-1\right)\left(T^3_{\ell}
 -Q_{\ell}\right)+3c^2_WX_{\ell}=-a_{\ell},$$
  $$  b'_{\ell}= -\left( 4c^2_W-1\right)Q_{\ell_R}+3c^2_WX_{\ell_R}=-b_{\ell}.$$
 The decay rate $Z^\prime \rightarrow \mu \tau$ is similar to that
 of $Z \rightarrow \mu \tau$. It leads to the relation
 between
 two branchings as follows:
\bea \mathrm{BR}(Z'\rightarrow \mu\tau) &=& c~
m^4_{Z'}\left[\left|F^{Z'}_L\right|^2
+\left|F^{Z'}_R\right|^2+\frac{1}{2}
\left|\frac{m_{\tau}}{m_{Z'}}D^{Z'}_L\right|^2
+\frac{1}{2}\left|\frac{m_{\tau}}{m_{Z'}}D^{Z'}_R\right|^2\right]
\crn&\times&\mathrm{B R}(Z'\rightarrow
\ell^+\ell^-),\label{bzpmutau1}\eea
where
\be F^{Z'}_{\mathrm{L,R}}=
\frac{m_Z^2}{m^2_{Z'}}A^{Z'(1)}_{\mathrm{L,R}}+A^{Z'
(2)}_{\mathrm{L,R}}-C^{Z^\prime}_{\mathrm{L,R}}.\label{fzp1}\ee
\subsubsection{$\tau\rightarrow \mu\mu\mu$\label{huong1}}
  In the SUSYE331, the effective Lagrangian described
the $\tau\rightarrow \mu\mu\mu$ decay can be deduced from the
effective operators given in Eq.(\ref{efftau3mu1}) combining with
those induced by the effective operators of $\mu \tau$ with $Z$ or
$Z'$ or photon as well as Higgs boson.
 The general study was presented in \cite{Anna2}.
The contributions from the box-diagrams and vector  boson
exchanges to the branching of  the considered decay rate   are
 sub-leading ones  even  when $\tan \gamma$ is
large or small. However the contribution from the Higgs-bosons
exchange to decay rate is large for large $\tan \gamma$  (the
interested reader can see in \cite{Giang}). Hence, in this work we
will split each type of contributions to the considered branching
ratio. First, let us consider the case of absence of Higgs
exchange:  the effective Lagrangian can be deduced from  the
effective operators given in Eqs.(\ref{gamutau2}),
(\ref{zmutau1}), (\ref{zmutau2}),
 (\ref{zpmutau2}) and (\ref{efftau3mu1}).
The explicit formula of effective Lagrangian is
\bea \mathcal{L}^{\mathrm{eff}}_{\tau\mu\mu\mu} &=&
\left[\left(\bar{\mu}\bar{\sigma}^{\mu}\tau\right)
\left(F^{\mu_L}_L\bar{\mu}\bar{\sigma}_{\mu}\mu
+F^{\mu_R}_L\mu^c\sigma_{\mu}\bar{\mu^c}
\right)\right.\crn&+&\left.
\left(\mu^c\sigma^{\mu}\bar{\tau^c}\right)
\left(F^{\mu_L}_R\bar{\mu}\bar{\sigma}_{\mu}\mu
+F^{\mu_R}_R\mu^c\sigma_{\mu}\bar{\mu^c} \right)\right] \crn&+&
2e^2\left(D^{\gamma}_L\bar{\mu}\bar{\sigma}^{\mu\nu}\bar{\tau^c}
+D^{\gamma}_R\mu^c\sigma^{\mu\nu}\tau\right)\frac{m_{\tau}\partial_{\nu}}{\square}
\crn&\times&\left(\bar{\mu}\bar{\sigma}_{\mu}\mu
+\mu^c\sigma_{\mu}\bar{\mu^c}
\right)+\mathrm{h.c.},\label{effLagtau3mu1}\eea
 where
 \be A^{Z'}_{L(R)}= \frac{m_Z^2}{m^2_{Z'}}A^{Z'(1)}_{\mathrm{L,R}}+A^{Z'
(2)}_{\mathrm{L,R}} \label{azpp},\ee
 \be F^{\mu_L}_{L(R)}=
B^{\mu_L}_{L(R)}+\frac{1}{2} g_Z^2c_{2W} A^Z_{L(R)}- \frac{1}{2}
g_{Z'}^2c_{2W} A^{Z'}_{L(R)}-e^2C^{\gamma}_{L(R)}, \label{fmul}\ee
\be F^{\mu_R}_{L(R)}= B^{\mu_R}_{L(R)}+ g_Z^2s^2_W A^Z_{L(R)}-
g_{Z'}^2 s^2_WA^{Z'}_{L(R)}-e^2C^{\gamma}_{L(R)}. \label{fmur}\ee

Here we also assume that, as in the case of MSSM,  we ignore the
contributions of $C^{Z}_{L,R},~ C^{Z'}_{L,R},~ D^{Z}_{L,R}$ and
$D^{Z'}_{L,R}$ to the above effective Lagrangian.
 This leads to the branching ratios of decay $\tau\rightarrow
 \mu\mu\mu $ is
 \bea  \mathrm{B R}(\tau^-\rightarrow \mu^-\mu^+\mu^-)&=& \frac{1}{8G_F^2}
 \left[2\left|F^{\mu_L}_L\right|^2+\left|F^{\mu_R}_L\right|^2
 + \left|F^{\mu_L}_R\right|^2
 +2\left|F^{\mu_R}_R\right|^2\right.
 \crn&+& \left.4e^2\mathrm{Re}\left(D^{\gamma}_L\left(2\bar{F}^{\mu_L}_L+
 \bar{F}^{\mu_R}_L\right)+D^{\gamma}_R\left(\bar{F}^{\mu_L}_R+
 2\bar{F}^{\mu_R}_R\right)\right)\right.\crn&+&\left.
 8e^2\left(\left|D^{\gamma}_L\right|^2+\left|D^{\gamma}_R\right|^2\right)
 \left(\log\frac{m_{\tau}^2}{m_{\mu}^2}-\frac{11}{4}\right)
 \right]\crn&\times& \mathrm{B R}(\tau^-\rightarrow
 \mu^-\bar{\nu}_{\mu}\nu_{\tau}).
 \label{brtau3mu1}\eea
\subsection{ $H\mu\tau$ contribution to $\tau\rightarrow \mu\mu\mu$}
  Contribution of  Higgs exchange in the SUSYE331 model was
 investigated in \cite{Giang},
 where the corresponding effective Lagrangian for
 this,  is  given by:
\bea \mathcal{L}^{eff}_{\tau\mu\mu\mu}&=& -2 \sqrt{2}G_F m_\mu
m_\tau \tan\gamma \mathcal{C} (\mu^c\mu+ \bar{\mu}
\bar{\mu}^c)\crn &\times& (\Delta^{\rho}_L \mu\tau^c+
\Delta^{\rho}_R \mu^c\tau ) + \mathrm{h.c.},
\label{effLtau3mu1}\eea
where
\bea \mathcal{C}&\equiv&t_{\gamma}
\left(\frac{s^2_\alpha}{m^2_{\phi_{Sa36}}}
+\frac{c^2_\alpha}{m^2_{\varphi_{Sa36}}}\right).
 \label{higgfactor1} \eea Noting that factor
 $(m_{\mu}\tan\gamma)$  cannot be ignored if value of $\tan\gamma $ is large enough.
This factor causes a shift to $F^{\mu_R}_L$ and $F^{\mu_L}_R$,
explicitly
\be \delta F^{\mu_R}_L= \sqrt{2} G_Fm_{\mu}m_{\tau}
\mathcal{C}\Delta_R, \hs \delta F^{\mu_L}_R= \sqrt{2}
G_Fm_{\mu}m_{\tau} \mathcal{C}\Delta_L.\label{shiftf}\ee

The individual contribution
from Higgs exchange to $\mathrm{BR}(\tau\rightarrow
\mu^-\mu^+\mu^-)$ now is:
\be\mathrm{BR}(\tau \rightarrow \mu^-\mu^+\mu^-)_{\Phi^*} =
\frac{\left(m_{\mu}
m_{\tau}\mathcal{C}\right)^2\left(\left|\Delta^{\rho}_L\right|^2\right)
+\left|\Delta^{\rho}_R\right|^2}{32} \mathrm{BR}(\tau\rightarrow
\mu^-\bar{\nu}_{\mu}\nu_{\tau}).\label{brtau3mu2}\ee
This contribution will be suppressed in the case of small
$\tan\gamma$.  We will concentrate on this case in the numerical
calculation in the following section.

\section{\label{num}Numerical results}

In this section, let us
numerically study the
cLFV decays of the tau lepton
$\tau\rightarrow \mu \gamma,~ \tau \rightarrow \mu \mu \mu$
and $Z\rightarrow \mu \tau$.
For this purpose, we first study
some constraints on values of parameter space in the
SUSYE331 model.

\subsection{Implication on the parameter space in SUSYE331 model}

 In this section,  we pay attention to discussing on constraint of parameter
 space caused by experimental cLFV bounds (\ref{etaumugamma1})-(\ref{ztaumu1}).
  As mentioned,
this topic has been carefully studied in  many models beyond SM.
Especially ref.\cite{Anna2} not only indicated regions of
parameter space satisfying experimental bounds but also discussed
in details the correlation between dipole and non-dipole kinds of
contributions to cLFV decays in each region. The most important
assumption here is that  sources of cLFV come from only the mass
terms of sleptons in the soft term, namely only left- and
right-handed slepton mass matrices have large $\mu-\tau$ entries.
Let us compare the sources of cLFV appearing in the MSSM with
those of the SUSYE331.
 Because of the absence of the right-handed neutrinos, the MSSM
contains only three sources of cLFV,
 particularly LFV in left-handed charged sleptons, left-handed
sneutrinos  and right-handed charged leptons. But the left-handed
sleptons and their sneutrinos live in the same doublet of
$\mathrm{SU(2)_L}$ in  the MSSM, the origin of cLFV  in the
left-handed charged sleptons and left-handed sneutrinos sectors
are the same.  As a consequence, in  the MSSM there are only two
independent sources of
 cLFV. Due to the appearance of the right-handed
neutrinos in the SUSYE331, there also appears one more source of
LFV. Furthermore, there exist two Higgs multiplets   in the model,
 i.e.,  triplet $\rho$ and antitriplet $\rho'$ which
independently generate masses of charged and neutral sleptons at
tree level. These two Higgs  multiplets also create at least two
new corresponding  mass terms of sleptons in the soft term, as
shown in details in \cite{Dong1,Dong3,Giang}. In general, there
are
  at least four independent sources causing the cLFV
   in the SUSYE331 where each source is parameterized by a mixing
   angle defined in the last subsection of
   Appendix \ref{InLAno}. This parameterization creates the similarity between
   the SUSYE331 and the MSSM. We exploit this advantage to make
   some prediction for the SUSYE331 basing previous investigation
   of   the MSSM. For example,
 if  all of these sources are appeared, the
 branching ratio of decay $\tau\rightarrow \mu\gamma$
 will be much enhanced than that in  the
MSSM, even it could greatly exceed the bound of experiment given
in (\ref{etaumugamma1}).  As considered in   the MSSM,
the existence of maximal LFV mixing in the left-handed slepton
sector  means the LFV mixing in both charged and neutral sleptons,
and their contributions to cLFV decays are much larger than
contributions caused by right-handed LFV mixing. In addition, if
the superparticle spectrum  is relative light, namely the
parameter space are set below 1 TeV, there will exist very small
regions of parameter space satisfying experimental results. The
similar situation also happens for the SUSYE331.
However,  in the SUSYE331 four cLFV  sources are independent
then there
exists a situation that the model contains only one left-handed
maximal LFV source while others vanish. It is easy to  realize
that in this case, the values of branching ratios of cLFV processes in
the SUSYE331  are smaller than those of the MSSM  and the satisfied
regions of parameter space will be wider in the scale of
$\mathcal{O} (100)$ GeV.  On the other hand, if four LFV sources
are presented, the predicted results of the considered model
consistent with the experimental results at the TeV scale.
In the next subsections, we will examine
the influence of LFV sources
appearing in the soft term of the SUSYE331
 on the cLFV decays of the tau and the $Z$ boson.
In the numerical investigation, we just pay attention to the case
of soft parameters at  $ \mathcal{O}(100)$ GeV scale because this
scale allows the existence of small values of  slepton masses
which can be detected by present colliders. In fact, the detailed
investigation  to determine different
 properties of cLFV branching ratios among different regions of
 parameter space is really needed, as done in many known SUSY
 models. In the SUSYE331, this work is more complicated because of
 the addition of
 many new particles so we will come back this interesting topic in another work.

 We would like to
emphasize that the sleptons gain mass through
 main sources which come from soft terms and
 interacting terms of sleptons with Higgs
through $F$-and $D$-terms.  As mentioned in \cite{Dong2}, the
$D$-term gives contribution to mass of slepton that contribution
is proportional to the quantity  $(w^2-w^{\prime 2})$ while
$F$-term, gives contribution to mass  of slepton that is
proportional to $(\lambda_a w)$  or $(\lambda_a w^\prime)$.
Because the
 VEVs $\omega$ and $\omega'$ break down
$\mathrm{SU(3)}_\mathrm{L}$ to $\mathrm{SU(2)}_\mathrm{L}$
so these values can be set to the TeV scale.  We would like to note that  the
SUSYE331 contains the Tachyon fields, the removal Tachyon fields
leads to a condition \cite{Dong3}:
 \be |w^2-w'^2|\simeq
|v^2-v'^2|\leq 246^2 ~(\mathrm{GeV}^2).\label{tachyon1}\ee
However, in the last work \cite{Dong3}, we have ignored the
$B$-type terms, namely $B_{\rho} \rho \rho^\prime, B_{\chi }\chi
\chi^\prime$. If the $B$-type terms are included into the Higgs
potential, not only the tachyon fields are removed without any
conditions but also the stable vacuums are guaranteed. From now,
we will consider the SUSYE331  which include the above B-terms. It
is noted that the Higgs mass spectrum is different from that
presented in \cite{Dong1,Dong3}. For example, there are three
neutral massive pseudo-scalar, five neutral massive
 and four charged massive Higgs. It is interesting that
masses of light Higgs, in general, increase by the presence of
B-type terms. Especially, the charged Higgs mass of $m^2_W$ as
shown in \cite{Dong3} now changes into the new value of $(m^2_W+ 2
B_{\rho}/\sin2\gamma)$. So the $B$-terms will make masses of Higgs
satisfied with current limits of electroweak precision tests such
as LEP limit of charged Higgs boson \cite{abb}, even without loop
corrections. For more detail, the interested reader can see in
Refs. \cite{electroweakbreaking,dangchuanbi}. In the SUSYE331,
where
 the $B$-type terms are included, the $D$-term generates mass for slepton
at the $\mathrm{SU(3)_L}$ scale by sub-terms which are always
proportional to $\left(w^2(t_\beta^2-1)/t^2_{\beta}\right)$
\cite{Dong2}. So this contribution should be at the
$\mathcal{O}$(100) GeV if slepton masses are in range of
$\mathcal{O}$(100) GeV. This is similar to the condition
(\ref{tachyon1}). Furthermore, if the R-parity is imposed, the
coupling $\lambda_a$ must vanish.
 It means that the contributions from both $F$- and $D$-terms
to masses of sleptons depend on the $\mathrm{SU(3)_L}$ broken
scale and the value of $t_\beta$. So if both slepton masses are in
the range of $\mathcal{O} (1)$ TeV scale and $t_\beta$ is close to
unity, the dominant contribution to the slepton masses comes from
the soft term.

Next, let us discuss on  $\mu_{\rho}$ and $\mu_{\chi}$
 which play the same role as $\mu$ parameter in  the MSSM.
According to \cite{Dong1,Dong3},  including $B$-type terms, the
requirement of
 canceling all linear terms at tree level in the Higgs
potential leads to conditions:
\bea \mu_{\chi}^2+4m^2_{\chi} -4 \frac{B_{\chi}}{t_\beta}
&=&\frac{g^2}{1-4c_{2W}}\left[2c_{2W }\left(\frac{t^2_\beta
-1}{t_\beta^2} \right)\left(w^2+u^2\right)
+\left(\frac{t_\gamma^2-1}{t_\gamma^2}\right)v^2\right] ,\crn
\mu_{\rho}^2+4m^2_{\rho}-4 \frac{B\rho}{t_\beta}
&=&\frac{g^2}{1-4c_{2W}}\left[\left(\frac{t^2_\beta -1}{t_\beta^2}
\right)\left(w^2+u^2\right)
-2\left(\frac{t_\gamma^2-1}{t_\gamma^2}\right)v^2\right] , \crn
m_\chi^2+m_{\chi}^{\prime 2}+\frac{\mu^2_\chi}{2}&=&B_{\chi}
\frac{ t_\beta^2+1}{t_\beta},
 \crn
m_\rho^2+m_{\rho}^{\prime 2}+\frac{\mu^2_\rho}{2}&=&B_{\rho}
\frac{ t_\gamma^2+1}{t_\gamma}. \label{linear}\eea

The parameters $\mu_{\chi}^2,~\mu_{\rho}^2,~m^2_\chi$ and
 $m^2_\rho$  as well as $m_{\rho^{\prime}}^2, m_{\chi^\prime}^2$ are  positive.
 The additional $B$-terms depend on the phases of  fields. We can
redefine the phases of $ \chi, \chi^\prime, \rho$ and
$\rho^\prime$ by such way that can absorb any phase in the
$B$-terms, so we can take $B_\chi, B_\rho$ to be real and
positive. The conditions given in Eqs. (\ref{linear}) lead to the
consequences as follows: The scale of all parameters given in the
left-handed side of Eqs. (\ref{linear}) are the same order.  If
the value of $\tan \beta$ is closed to the unit, the parameters
$\mu_{\chi}^2,~\mu_{\rho}^2,~m^2_\chi, ~m^2_\rho,$
$~m_{\rho^{\prime}}^2, ~m_{\chi^\prime}^2 $ and $B_{\chi},
B_{\rho}$ are set to the scale of electroweak symmetry breaking
else they are set to the scale of $\mathrm{SU(3)_L}$ symmetry
breaking. Furthermore the Higgs mass spectrum studied in
~\cite{dangchuanbi} also leads to the conclusions on the limit of
$\tan \gamma$. Specially if the soft  given in Higgs potential are
considered at the scale of $\mathrm{SU(3)_L}$ symmetry breaking
the tree level mass of the SM Higgs boson  is $m_Z|\cos2\gamma| <
92.0$ GeV.
 This result is similar to that given in the MSSM. In this case,
  the  boundary of the SM mass can be
pushed up to 130 GeV by
 one-loop correction with large $\tan \gamma$.
 In another case if the soft
 $\mu/ B$-terms  are
 considered at the scale of electroweak symmetry breaking, there is
no constraint on the value of $\tan \gamma$.

It is known that three branching ratios
 $\tau\rightarrow\mu\gamma$, $\tau\rightarrow \mu\mu\mu$
 and $Z\rightarrow\mu\tau$ depend on some of following quantities:
 $A^{Z(Z')}_{L(R)}$,  $B^{\mu_{L(R)}}_{L(R)}$,
 $C^{\gamma(Z,Z')}_{L(R)}$ and $D^{\gamma(Z,Z')}_{L,R}$. In
different regions of parameter space, where some of these
quantities give dominant contribution, there are  precise
correlations among branching ratios. In particular, with three
considering branching ratios, we have two cases as listed below:
\begin{description}
    \item $D^{\gamma}$-{\it dominance}: It is easy to get the relation
    concerned in   \cite{Anna2}
    \be
    \frac{\mathrm{BR}(\tau\rightarrow\mu\mu\mu)}{
    \mathrm{BR}(\tau\rightarrow\mu\gamma)}\simeq
    2.2\times 10^{-3},
    \label{ftau3mutaumga1}\ee and a consequence  is
    \be \mathrm{BR}(\tau\rightarrow\mu\mu\mu) < 10^{-10}. \ee
 \item $A^Z$-{\it dominance}: In this case, we get
 \be \mathrm{BR}(\tau\rightarrow\mu\mu\mu)=\left(%
\begin{array}{c}
  0.53 \\
  0.67 \\
\end{array}%
\right) \mathrm{BR}(Z\rightarrow\mu\tau ).\label{azdomination1}\ee
Note that in the SUSYE331 model, although  $A^{Z'}_{L(R)}$
receives contribution from
 diagrams which are similar to those of $
A^{Z}_{L(R)}$,
 $A^{Z'}_R\simeq0 $. In the limit
$t_{\beta}=\omega/\omega'\simeq1~(c_{2\beta}=0)$ and $A^{Z'}_L$
depends  on only diagrams containing right-handed sneutrinos.
 \end{description}

 In next subsection, we consider
 numerical  computation for cLFV decays
of the tauon and the $Z$ boson. Remember that the effective
couplings used for numerical studying  are established in
appendices \ref{mutaugamma}, \ref{apzmutau}, \ref{apzpmutau} and
\ref{apbtaumu}. The standard loop integrals are given in
\cite{Anna2,davy}. In the following investigation, we use most of
the notations defined in \cite{Giang}. For example,  mass
parameters of gauginos and Higgsinos are listed in the formula
(B.2): $m_{\lambda}$ denotes mass of $\mathrm{SU(3)}_L$ gaugino,
$\mu_{\rho}$ and $\mu_{\chi}$ are $\mu$-terms of Higgsinos. Only
notation for mass of U(1) gaugino used in this work is $m_B$
instead of $m'$.

\subsection{In
the case of small $\tan\gamma$ and light mass spectrum}

\subsubsection{$\tau\rightarrow \mu\gamma $}
  Let us consider the numerical studying of
  branching of $\tau\rightarrow \mu\gamma
  $ decay. The analytical result given in
  Eq.(\ref{brtaumugamma1})
   depends on the effective coupling $D_{L(R)}^{\gamma}$ which can be divided
   into three parts, $D^{\gamma}_{L(R)}=
   D^{\gamma(a)}_{L(R)}+D^{\gamma(b)}_{L(R)}+D^{\gamma(c)}_{L(R)}$.
  The  analytical  formulas of $D^{\gamma(a)}_{L(R)}$,
 $D^{\gamma(b)}_{L(R)}$ and $D^{\gamma(c)}_{L(R)}$ are given in
 Appendix \ref{mutaugamma} in which only $D^{\gamma(a)}_{L(R)}$ does not
 depend on $\tan\gamma$. In addition, from  the experimental
 bound (\ref{etaumugamma1}), we can obtain the constraint on
 the effective couplings, namely
 $|D^{\gamma}_{L,R}|\leq 2.5 \times
 10^{-9}~ [\mathrm{GeV}^{-2}]$.

  We would like to remind that the diagrams which contribute to
the $D_L^\gamma$ are collected from three LFV sources: left-handed
charged slepton, left-handed sneutrinos and right-handed
sneutrinos sectors, while the diagrams contributing to
$D_R^\gamma$ are only collected from  the  charged right-handed
slepton sector. So the values of $D_L^\gamma$ are  predicted  much
larger than those of $D_R^\gamma.$ Another thing we want to
emphasize that since the SUSYE331 has many additional particles,
we expected that the additional particles can modify the predicted
results of cLFV decay in the  model under consideration. For more
details, let us consider numerical studying in this decay mode.

\begin{figure}[h]
  \centering
\begin{tabular}{cc}
\epsfig{file=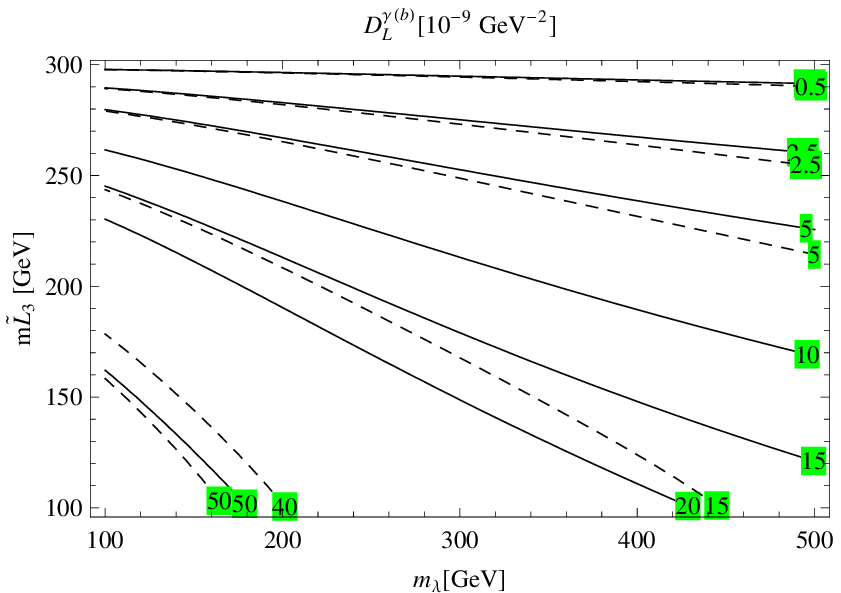,width=0.495\linewidth,clip=}
\epsfig{file=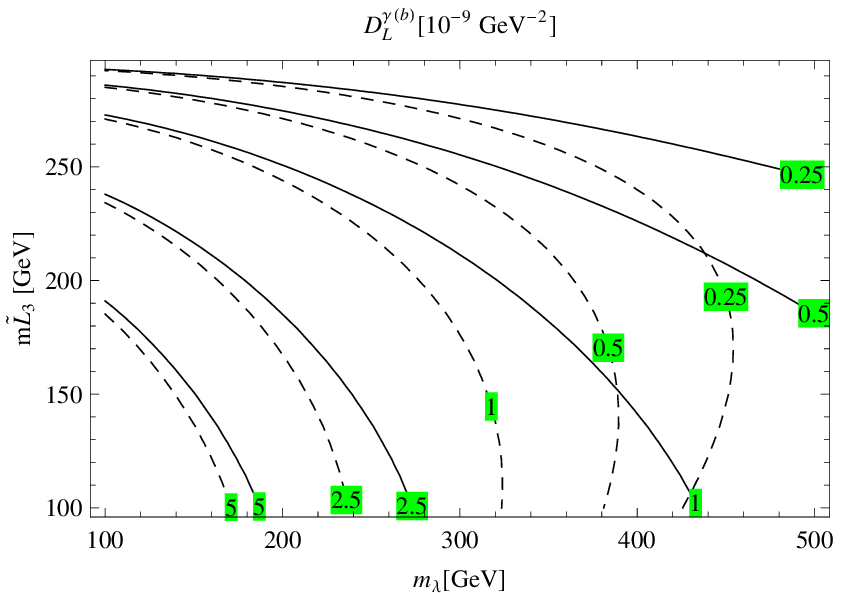,width=0.495\linewidth,clip=}
\\
\end{tabular}
  \caption{ $D^{\gamma(b)}_L$ isocontours with $\tan\gamma=3$,
  $m_{\tilde{L}_3}=m_{\tilde{\nu}_{L_3}}=m_{\tilde{\nu}_{R_3}}$
  and
  $m_{\tilde{L}_2}=m_{\tilde{\nu}_{L_2}}=m_{\tilde{\nu}_{R_2}}=300
 $ GeV,
 $\theta_L=\theta_{\tilde{\nu}_L}=\theta_{\tilde{\nu}_R}=\pi/4$
 and $\mu_{\rho}=140 $ GeV (1TeV) in the left (right) panel. The
 solid and dashed lines
 correspond to $m_B=300$ GeV and $m_B=-300$ GeV.}\label{FdgabL1huong}
\end{figure}

First we study the effects of LFV sources on $D_L^\gamma$ in the
case of small $\tan\gamma$ as well as the presence of all of the
three left-handed LFV sources, especially we fix $\tan \gamma=3$
and
$\theta_L=\theta_{\tilde{\nu}_L}=\theta_{\tilde{\nu}_R}=\pi/4$. As
in the MSSM, $D^{\gamma(b)}_L$ is  dominant contribution to
$D^{\gamma}_L$.
 Fig.\ref{FdgabL1huong} shows the dependence of
$D^{\gamma(b)}_{L}$ on soft parameters
$m_{\tilde{L}_3}=m_{\tilde{\nu}_{L_3}}=m_{\tilde{\nu}_{R_3}}$ and
$m_\lambda$, while other
 parameters  are fixed. The
predicted results given in left panel of Fig.\ref{FdgabL1huong}
are fully consistent with the experimental results if the domain
of parameter $m_{L3}$ is close to the value of $m_{\tilde{L}2}$
and all soft parameters are set at $\mathcal{O}(100)$ GeV. We also
remind that mixing mass term between left and right slepton,
 $m_{\tilde{\psi}_{\mu\tau}}$, is  small if the
 model under consideration
has maximal mixing sources. However if
 $\mu_\rho$ is set at TeV, the domain of parameter
$m_{\tilde{L}2}$ , where the values of $D_{L}^{\gamma(b)}$
 match experimental bound,
is more extensive. For more details, the reader can see in the
right panel of Fig.\ref{FdgabL1huong}.

\begin{figure}[h]
  \centering
\begin{tabular}{cc}
\epsfig{file=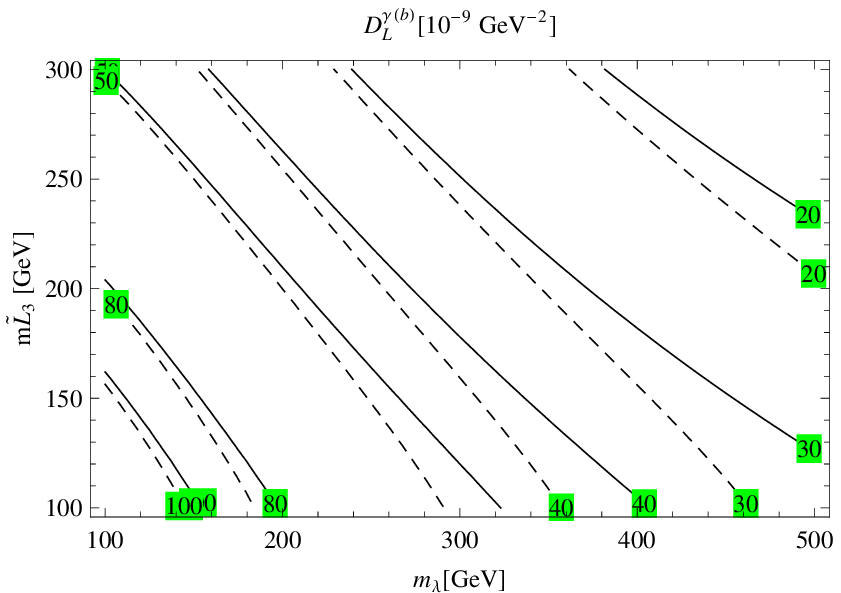,width=0.495\linewidth,clip=}
&
\epsfig{file=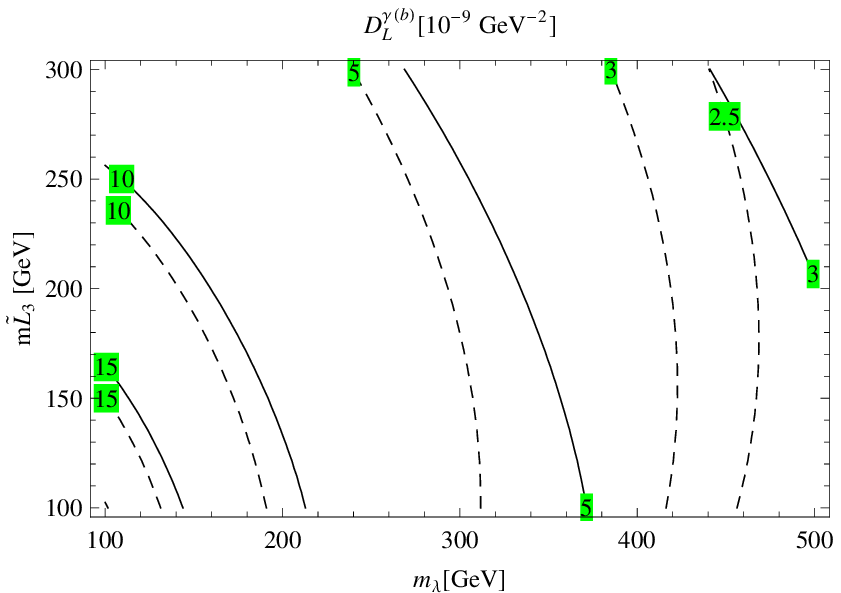,width=0.495\linewidth,clip=}\\
\end{tabular}
  \caption{ $D^{\gamma(b)}_L$ isocontours with $\tan\gamma=3$,
  $m_{\tilde{L}_2}=m_{\tilde{\nu}_{L_2}}=m_{\tilde{\nu}_{R_2}}$
  and
  $m_{\tilde{L}_2}=m_{\tilde{\nu}_{L_2}}=m_{\tilde{\nu}_{R_2}}=1
 $ TeV,
 $\theta_L=\theta_{\tilde{\nu}_L}=\theta_{\tilde{\nu}_R}=\pi/4$
 and $\mu_{\rho}=140$ GeV (1TeV) in the left (right) panel. The
 solid
  and dashed lines
 correspond to $m_B=300$ GeV and $m_B=-300$ GeV.}\label{FdgabL1}
\end{figure}

Let us consider the case $m_{\tilde{L}3}= 1$ TeV. The results
given in Fig. \ref{FdgabL1} predict that
 the values of
$D^{\gamma(b)}_{L}$ exceed the experimental bound if
 the remaining parameters are
 assumed in region of $\mathcal{O}(100)$ GeV. Even
if the values of $\mu_{\rho}$ are
  assumed as large as 1 TeV, the predicted value of
$D^{\gamma(b)}_L$ is larger than the experimental bound.
 In this
region of parameter space, the predicted results in the SUSYE331
are much similar to those in the MSSM \cite{Anna2}. As mentioned
in \cite{Anna2}, the large values of $A_{\tau}$ as well as small
values of $\mu_{\rho}$ and masses of sleptons must be required to
keep value of the total $|D^{\gamma}|$ below the experimental
bound. However, the Fig.\ref{FdgabL1} displays that the predicted
values of $D_L^{\gamma b}$ are much  greater than those in the
MSSM.  It means that in order to obtain the values of
$D^{\gamma(b)}_L$ being consistent with experimental bound, the
values of $A_{\tau}$ predicted in the SUSYE331 should be much
larger than those in the MSSM. In the limit of large values of
$A_{\tau}$, the model leads to the appearance of Tachyon sleptons.
It means that the  model under consideration does not contain the
region of parameter space such that there exists a large
difference between the values of slepton masses. The
Fig.\ref{FdgabL1} also shows that the values of $D_L^{\gamma (b)}$
exceed to the experimental results when parameter  $m_{\tilde{L}}$
expands into range of TeV.

So in the SUSYE331 model with  the case of the maximal mixing
happening in all three sources (left-handed slepton, left-handed
sneutrino and right-handed sneutrino sectors), the scale of all
slepton masses should be the same order, in range of TeV or in
range of (100) GeV.

\begin{figure}[h]
  \centering
\epsfig{file=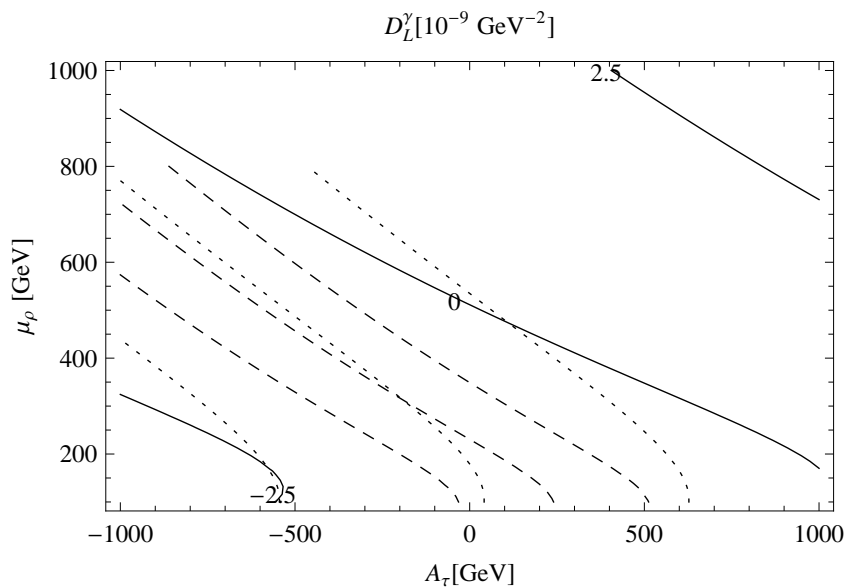,width=0.8\linewidth,clip=}
  \caption{ $D^{\gamma}_L$ isocontours with $\tan\gamma=3$, $m_{\tilde{L}_2}=1
 $ TeV,
 $\theta_L=\pi/4,~\theta_R=\theta_{\tilde{\nu}_L}=\theta_{\tilde{\nu}_R}=0$,
 $A^L_{\tau\mu}=0$
 (only LFV in $\{\tilde{m}_L,\tilde{\tau}_L\}$ sector). For illustrations, we choose
three choices of parameter space $(m_B,
~m_{\lambda},~m_{\tilde{L}_3},~m_{\tilde{R}})$ in GeV: (200, 300,
300, 200) (solid), ( 100, 400, 100, 200) (dashed), ( 100, 500,
300, 100) (dotted). For each example, the center line corresponds
 to the value of $D^{\gamma}_L=0$, two other ones limit the region
 where $|D^{\gamma}_L|\leq 2.5\times 10^{-9}~[\mathrm{GeV}^{-2}]$.
 }\label{FdgaL1}
\end{figure}

Now we consider another situation that  happens only in the
SUSYE331, not in the MSSM. It is the case of only one left-handed
LFV source appearing in the model. Looking at analytical formulas
of effective couplings, three sources contributing to left-handed
effective coupling are parameterized  by three mixing angels
$\theta_L$, $\theta_{\tilde{\nu}_L}$ and $\theta_{\tilde{\nu}_R}$.
 Two of them, $\theta_{\tilde{\nu}_L}$ and $\theta_{\tilde{\nu}_R}$, relate with
 diagrams containing
  sneutrino  propagators while the remain relates to
 diagrams with charged
 slepton propagators. Numerical
  computation  indicates that
 if the mass spectrum of superpartner particles
 is in the range of $\mathcal{O}(100)$
 GeV, contribution from
 sneutrino exchanges are larger than those
 of charged slepton ones. So if two
 sneutrino  mixing angles
 vanish, there is only one source of left-handed mixing $\theta_L$
 which generates relatively small effective couplings. The experimental
 bounds then are easily satisfied  even in regions of
 light mass spectrum. Our numerical investigation
 will focus on this case. In particular,
mixing angle parameters are fixed as
$\theta_R=\theta_{\tilde{\nu}_L}=\theta_{\tilde{\nu}_R}=0$ and
$\theta_L=\pi/4$. The Fig.\ref{FdgaL1} displays $D_L^{\gamma b}$
as function of the $A_\tau$ and $\mu_\rho$ while others are fixed:
 $m_{\tilde{L}2} =1$ TeV,
 $m_{\tilde{R}_2}=m_{\tilde{R}_3}=m_{\tilde{\nu}_{L_2}}=
 m_{\tilde{\nu}_{L_3}}=m_{\tilde{\nu}_{R_2}}=m_{\tilde{\nu}_{R_3}}\equiv
 m_{\tilde{R}}$.
The results given in the Fig.\ref{FdgaL1} illustrate that in the
considered limit, we can find the region of the small  absolute
values  of $A_{\tau}$ in which we can obtain the values of
$|D^{\gamma}_L|$ satisfying  the experimental bound, particularly:

- The value of $m_B $ should be smaller than that of $
m_{\lambda}$.

- If  the value of $m_B$ is closer to that of $ m_{\lambda}$, the
parameter space of $A_\tau $ and,  $\mu_\rho $ satisfying the
experimental bound of $|D^{\gamma}_L|$ has been expanded.

  The predicted results  given in Fig.\ref{FdgaL1} show that
if in the SUSYE331, only one source of
 lepton number violation in the charged
slepton sector, the model can contain the  region of parameter
space in
which the all soft parameters are set to
$\mathcal{O}(100)$ GeV except $m_{\tilde{L}_2}$  is set to TeV. In
this region of parameter space, the predicted results on the
$\tau\rightarrow \mu\gamma$  are matched the  experimental
 bound on the decay. The existence of the soft parameter space
below TeV scale leads to hope that the
sleptons can be detected
by LHC.

\begin{figure}[h]
  \centering
\begin{tabular}{cc}
\epsfig{file=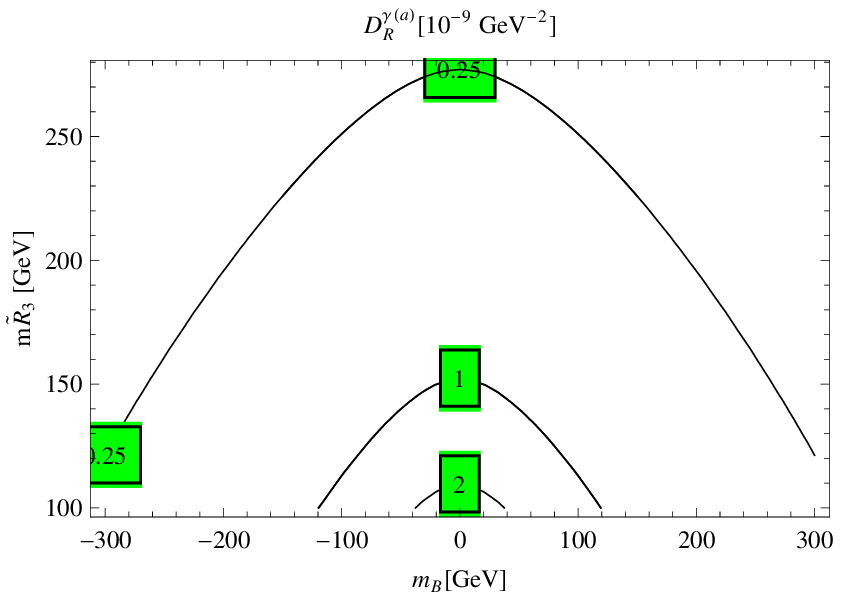,width=0.495\linewidth,clip=}
&
\epsfig{file=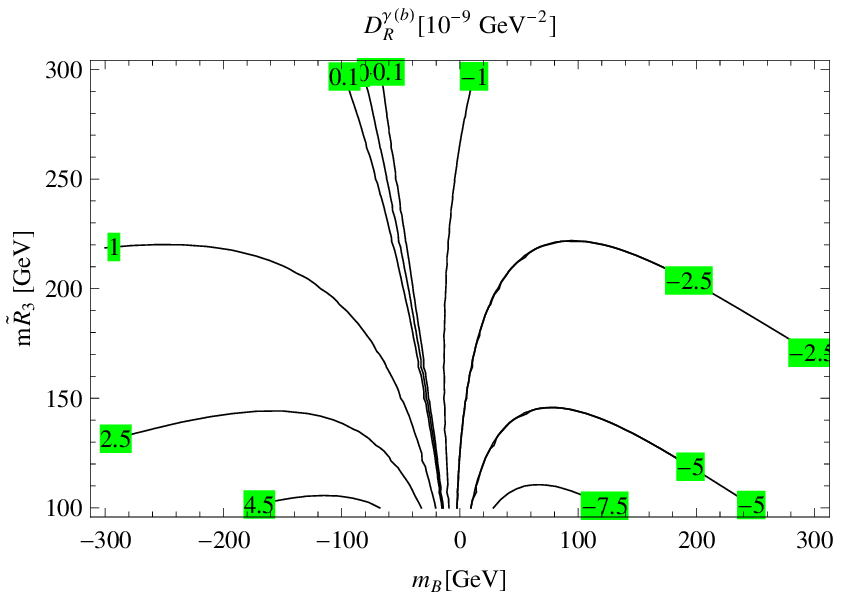,width=0.495\linewidth,clip=} \\
\end{tabular}
  \caption{Isocontours of $D^{\gamma(a)}_R$ (left panel) and isocontours of
  $D^{\gamma(b)}_R$ (right panel). The parameters are
    $\tan\gamma=3$}, $m_{\tilde{R}_2}=1$ TeV,
 $\theta_L=\theta_{\tilde{\nu}_L}=\theta_{\tilde{\nu}_R}=0$,
 $\theta_R=\pi/4$
 and $\mu_{\rho}=150$ GeV .\label{FdgabR1}
\end{figure}

Finally, we
 consider the LFV effect
 in the right-handed
 charged slepton sector to the LFV in the tauon decay. We
assume that there is only the maximal mixing of right-handed
slepton, i.e.,  $s_R=c_R=1/\sqrt{2}$ and all other mixing sources are
set to be equal to zero. Under this assumption
 the Feynman diagrams in the  model under consideration contributing to
$D^\gamma$ are exactly the same as those in the MSSM \cite{Anna2}.
In the Fig.\ref{FdgabR1}, absolute values of $D^{\gamma(a)}_R$ and
$D^{\gamma(b)}_R$ are rather small, even a bit smaller than those
in the  MSSM.
From this investigation, we see that $A_{\tau}$ should  have the
same order with the mass parameters of sleptons.

\subsubsection{Correlations}

Next, we consider the process $\tau\rightarrow \mu \mu \mu$. The
 amplitude consists of the contributions from the
effective couplings $A_{L(R)},B_{L(R)},C_{L(R)}$ and $D_{L(R)}$
which are denoted
the coupling of $Z, ~Z^\prime$ gauge bosons, photon and
Higgs to charged leptons $\mu_{L(R)},\tau_{L(R)}$.
Each type
of diagrams picking up the each particle exchange contributing to
the LFV in  the considered process depends on the region of parameter
space. In next work, we will study the contribution of each
effective coupling
to the LFV of the $\tau\rightarrow \mu \mu
\mu$ process in two typical cases.

\bc\textbf{ Only maximal mixing in the charged slepton
$(\tilde{\mu},\tilde{\tau})$ } \ec

Let us consider the process $\tau\rightarrow \mu \mu \mu$ in the
limit of small $\tan \gamma$ and in the case of existing only
maximal mixing in the charged slepton,
$s_L=c_L=\frac{1}{\sqrt{2}}$ and
$s_R=s_{\tilde{\nu}_L}=s_{\tilde{\nu}_R}=0$. This constraint leads
to $A^{2Z'}_L=A^{2Z'}_R=0$. As mentioned in the last part, it is
interesting to consider the region of parameter at
$\mathcal{O}(100)$[GeV]. One can check that $A^{Z}_{L(R)}$ is
dominated by $A^{Z(a)}_{L(R)}$ and total effective couplings
contributing to the branching ratio of the considered process are
$D^{\gamma}_{L(R)}$, $C^{\gamma}_{L(R)}$, $A^{Z(a)}_L$ and
$B^{\mu_{L(R)}}_L$. Indeed, looking at the analytical expression
of these couplings we found that only the analytical
 expression of $D^{\gamma}_L$ is affected by  left-handed
slepton parameters such as: $A_{\tau},~A^L_{\mu\tau}$.  Hence, in
order to look for the effective coupling giving dominant
contribution to the considered branching ratio, we will study
numerically two cases:

- The  $A_{\tau}$, $A_{\mu\tau}$  parameters are fixed  and
the other soft parameters are changed.

- The soft parameters are fixed and the $A_{\tau}$, $A_{\mu\tau}$
parameters are changed.

First, we  assume that $A_{\tau} =A_{\mu\tau}=0$.
 To assess the contribution of the each effective coupling $A^{Z}_L$ and
 $D^{\gamma}_L$ into the
 BR $(\tau^-\rightarrow\mu^-\mu^+\mu^-)$,  we define two factors
 $f_{A^{Z}}$ and $f_{D^{\gamma}}$  such as:
 \be f_{A^{Z}}\equiv
 \frac{g_Z^4\left[\left|A^{Z}_L\right|^2\left(\frac{1}{2}c^2_{2W}+s^4_W\right)+
 \left|A^{Z}_R\right|^2\left(\frac{1}{4}c^2_{2W}+2s^4_W\right)\right]}
 {2\left|F^{\mu_L}_L\right|^2+\left|F^{\mu_R}_L\right|^2
 +\left|F^{\mu_L}_R\right|^2+2\left|F^{\mu_R}_R\right|^2}
 \label{faz1}\ee
and
 \be f_{D^{\gamma}}\equiv
\frac{8e^4\left(\left|D^{\gamma}_L\right|^2+\left|D^{\gamma}_R\right|^2\right)
\left[\ln\left(\frac{m^2_{\tau}}{m^2_{\mu}}\right)-\frac{11}{4}\right]}
{MS}
\label{fdgamma1}\ee
where
 \bea MS &=& 8e^4\left(\left|D^{\gamma}_L\right|^2+\left|D^{\gamma}_R\right|^2\right)
\left[\ln\left(\frac{m^2_{\tau}}{m^2_{\mu}}\right)-\frac{11}{4}\right]\crn
&&
\hs  + 2\left|F^{\mu_L}_L\right|^2+\left|F^{\mu_R}_L\right|^2
+\left|F^{\mu_L}_R\right|^2+2\left|F^{\mu_R}_R\right|^2
\label{fdgamma11}\eea

The factors $f_{A^{Z}}$  and $ f_{D^{\gamma}} $ given
in Eqs. (\ref{faz1}), (\ref{fdgamma1})
quantitatively  measure contributions of
$A^{Z}_{L(R)}$ and $D^{\gamma}_{L(R)}$ to
the factor $\left|F^{\mu_{L(R)}}\right|$  and the total
branching of $(\tau^-\rightarrow\mu^-\mu^+\mu^-)$
decay, respectively.
\begin{figure}[h]
  \centering
\begin{tabular}{cc}
\epsfig{file=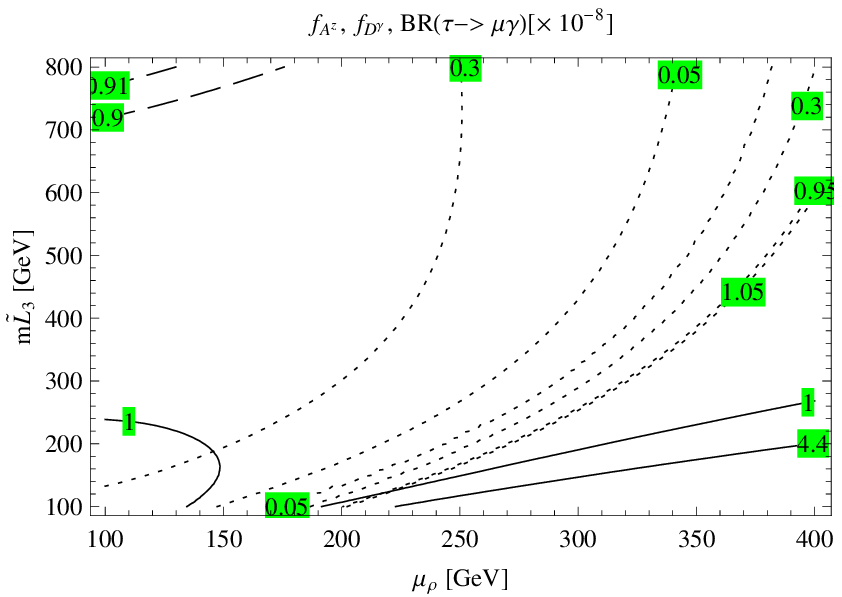,width=0.495\linewidth,clip=}
&
\epsfig{file=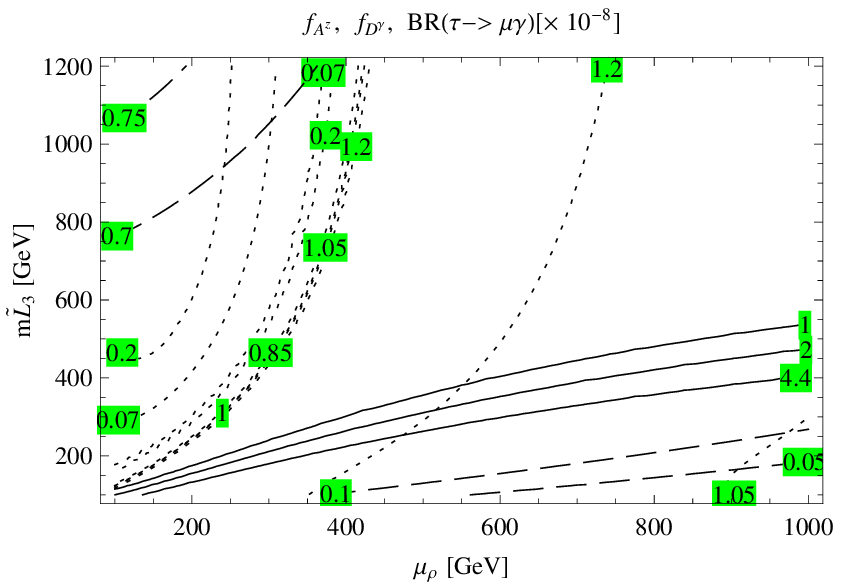,width=0.495\linewidth,clip=}
 \\
\end{tabular}
  \caption{Correlations among $ A^{Z}_{L}$, $F^{\mu_{L(R)}}_{L}$ and
  $D^{\gamma}_L$ with $A_{\tau}=0$. The contours  $f_{A^{Z}_L},~f_{D^{\gamma}_L}$
  and $BR(\tau\rightarrow\mu\gamma)$ are denoted by dashed, dotted
  and solid black lines. For illustrations, numerical values for
  parameter space
  $(m_B,~m_{\lambda},m_{\tilde{L}_2},~m_{\tilde{R}})$ are chosen
  as
  $(100,300,1000,100)$[GeV] (left panel) and $(100,500,1000,100)$ [GeV] (right
  panel).}\label{relativecon1}
\end{figure}

Looking at the Eqs.(\ref{fmul}) and (\ref{fmur}), the coefficient
$F^{\mu_{L(R)}}_{L(R)}$ depends on
 the other factors. If $A^{Z}$  gives
 dominant contribution to $F^{\mu_{L(R)}}_{L(R)}$,
 it is convenient to define regions where the factor
$ f_{A^{Z}}$
satisfies $1.05\geq f_{A^{Z}}\geq 0.95$  as the
 $A^{Z}$-domination
 regions (over $F^{\mu_{L(R)}}_{L(R)}$).  On the other hand, we also make
  convention that
   if  $f_{D^{\gamma}}\leq
 0.05$ then the dominant contribution to the
 branching ratio of $\tau\rightarrow 3\mu$
 is given by $F$-type of  couplings
 and  if $1.05\geq f_{D^{\gamma}} \geq
 0.95$ then the dominant contribution to the branching ratio of
  $\tau\rightarrow 3\mu$ is given by  $D$-type of  couplings.

  The  current experimental upper bound on the branching ratio
  of the process $BR(\tau\rightarrow\mu\gamma)$ is
  smaller than $4.4 \times 10^{-8}$.
 The results of our calculation to this process is given in Fig.
  \ref{relativecon1}.  The experimental bound of the
  branching is denoted by the solid black lines.  One can see that this
  process
 satisfies the experimental bound in
   regions of large parameter space of $\mu$ and
  $m_{\tilde{L_3}}$.

  The regions of parameter space
  where $A^{Z}_L$
  gives dominant contribution to branching ratio $\tau\rightarrow 3\mu$
  must satisfy both
  conditions: $1.05\geq |f_{A^{Z}}|\geq 0.95$ and $|f_{D^{\gamma}}|\leq
  0.05$.  The results given in
  Fig.\ref{relativecon1}
  show  that there is no region
  of $\mu_\rho$ and $m_{\tilde{L}_{2}}$ parameter space that make  $A^{Z}_L$
  given dominant contribution to branching ratio
   $\tau\rightarrow 3\mu$. If $D^{\gamma}$ gives dominant contribution to
 the considered decay mode, the region of
the  $\mu_\rho$ parameter satisfied the condition $1.05\geq
|f_{D^{\gamma}}|\geq 0.95$,  strongly depends on
 the value of charged gaugino mass, i.e.,
if the charged gaugino mass is larger, then the   value of
$\mu_\rho$ parameter  is
 larger too.

   Let us consider the effects of $A_{\tau}$
  on the  branching of the considered decay mode.
  We would like to emphasize that $A^{Z}$ does not
 depend on the  $A_{\tau}$ while the
   value of $|D^{\gamma}_L|$  depends on
the sign as well as amplitude
of $A_{\tau}$.  In particularly $D^{\gamma(c)}_L$ is proportional to
 $(A_{\tau}+\frac{1}{2}\mu_{\rho}\tan\gamma)$,  the values of $A_{\tau}$
 will affect on soft parameter space region where
 $D^{\gamma}_L$
gives dominant contribution to
 the BR$(\tau\rightarrow 3\mu)$.
 These regions of soft parameter  can be larger than that in the case of $A_\tau
  =0.$

\begin{figure}[h]
  \centering
\begin{tabular}{cc}
\epsfig{file=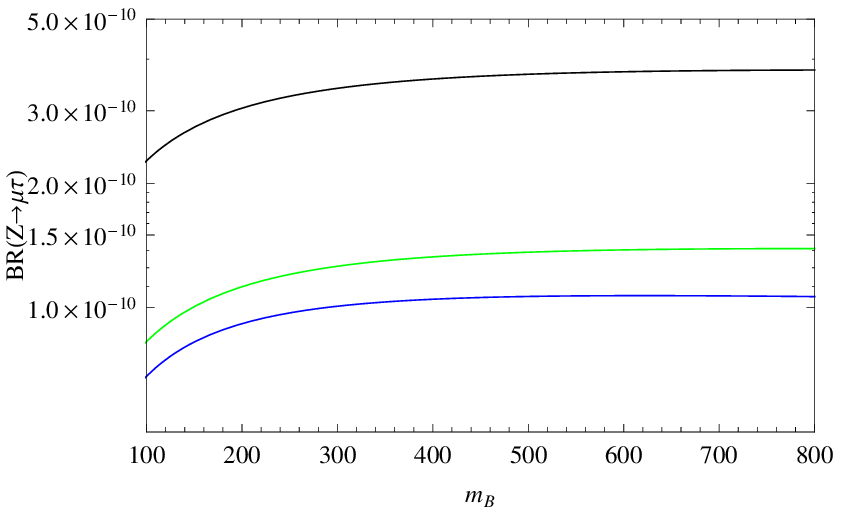,width=0.495\linewidth,clip=}
&
\epsfig{file=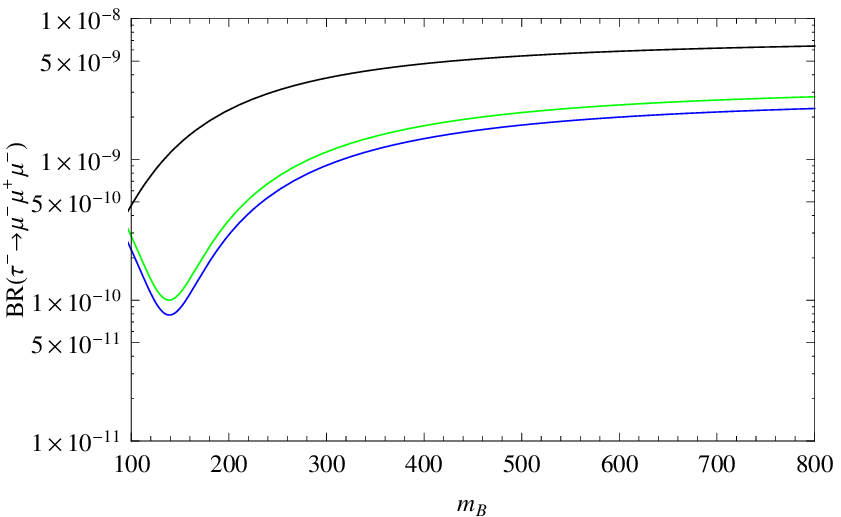,width=0.495\linewidth,clip=}
\\
\end{tabular}
  \caption{ Branching ratios $Z\rightarrow\mu\tau$ (left panel)
  and $\tau\rightarrow3\mu$ (right panel) as functions of
  $m_{B}$. Three numerical values for  parameter space
  $(m_{\lambda},\mu_{\rho},m_{\tilde{L}_2},m_{\tilde{L}_3},m_{\tilde{R}})$
  [GeV]  are chosen: (300, 150, 1000, 100, 100)-black line,  (400, 200,
  1000,
  100, 100)-green line,  (500, 150, 1000, 100, 100)-blue} line. \label{brta3muzmutau1}
\end{figure}

Now let us consider  $Z\rightarrow\mu\tau$ and
$\tau\rightarrow\mu\mu\mu$  decays.  The branching ratios of
decays $Z\rightarrow\mu\tau$ and $\tau^-\rightarrow\mu\mu\mu$ are
presented in (\ref{widthzmutau331}) and (\ref{azdomination1}). The
predicted branching is shown in Fig. \ref{brta3muzmutau1}.  The
numerical branching ratio for $Z\rightarrow\mu\tau$ has the
maximum of $5 \times 10^{-10}$ if
 the soft  parameters  of sleptons are set
 to $ \mathcal{O}(100)$ GeV and the charged gaugino mass is set to few hundreds GeV.
 This predicted result is
 very suppressed with the present experimental bound. However,
in the same region of soft parameter space, the
 predicted result for $\tau\rightarrow\mu\mu\mu$
 can reach to the experimental result and even one can find some
 regions of parameter space
 that predicted
  our  result exceed the experimental  bound.
  One can check in the left panel of the Fig. \ref{brta3muzmutau1}, the
branching values for $Z\rightarrow\mu\tau$ hardly change when we
change the value of the parameter $m_B$. However, the situation is
quite different if  the charged gaugino mass is varied, namely the
smaller value of charged gaugino mass is, the larger value of that
branching is predicted. Moreover, as we increase the values of the
 soft slepton mass parameters
these branching values decreased. Therefore, to increase
the value of the branching of $Z\rightarrow\mu\tau$, we have
to change the parameter
$\mu_{\rho}$.

\begin{figure}[h]
  \centering
\epsfig{file=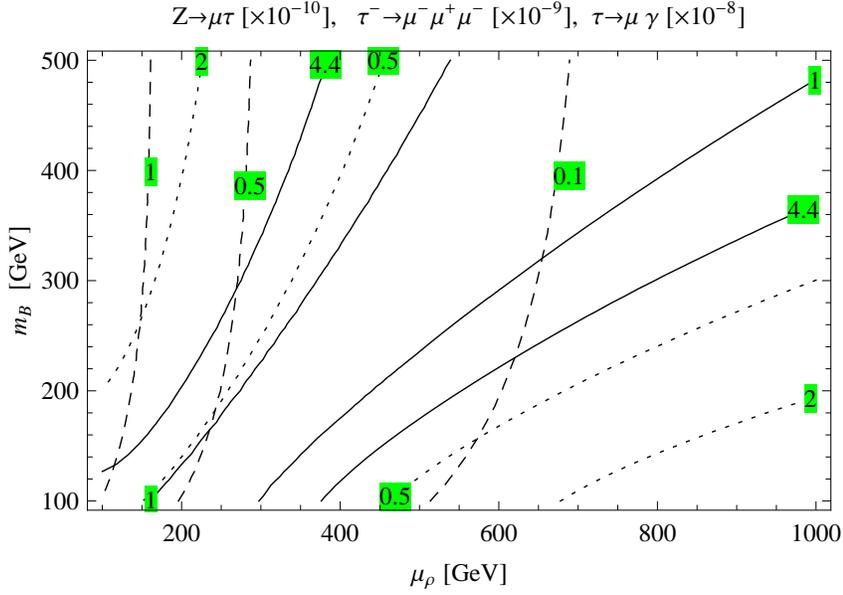,width=0.8\linewidth,clip=}
  \caption{ Contour plots for branching ratios $\tau^-\rightarrow\mu^-\mu^+\mu^-$
  (dotted), $Z\rightarrow \mu\tau$ (dashed) and
  $\tau\rightarrow\mu\gamma$ (solid) with $A_{\tau}=0$ and
  $(m_{\lambda},~m_{\tilde{L}_2},~m_{\tilde{L}_3},~m_{\tilde{R}})=(400,150,1000,100,200)$.
  }\label{brta3muzmutau2}
\end{figure}
\begin{figure}[h]
  \centering
\begin{tabular}{cc}
\epsfig{file=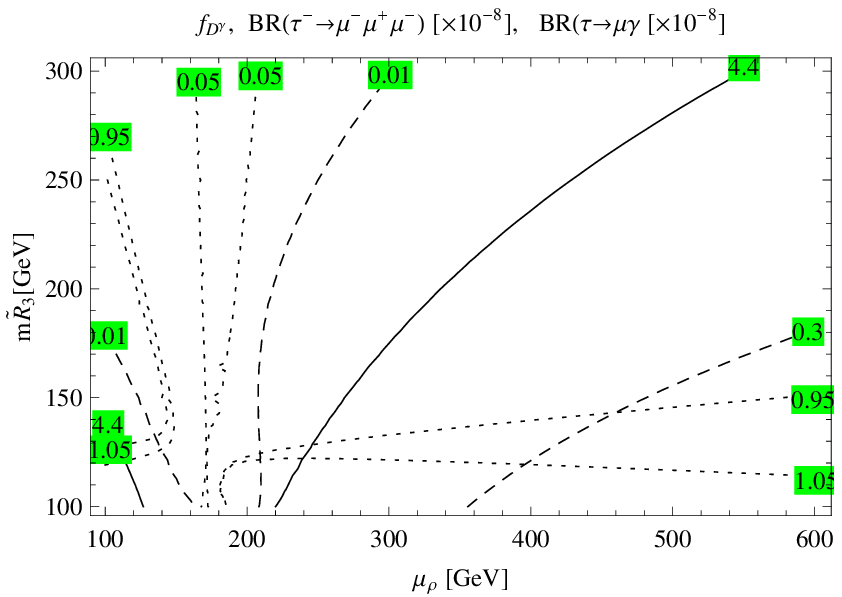,width=0.495\linewidth,clip=}
&
\epsfig{file=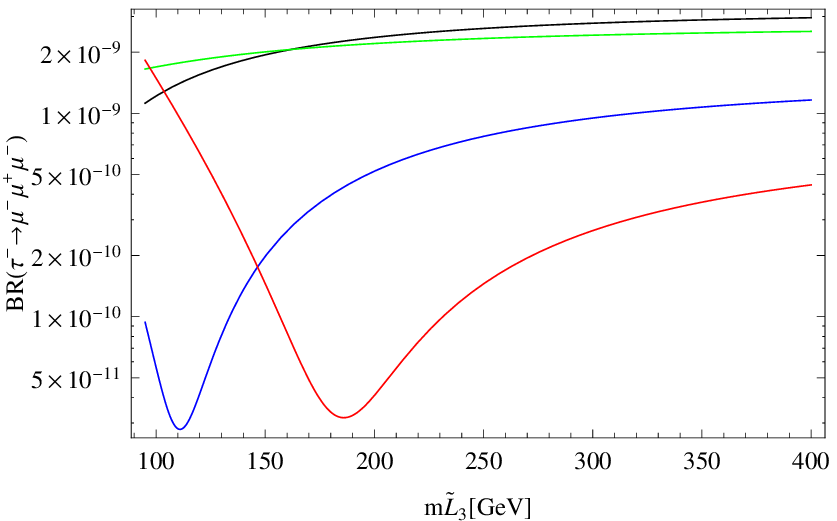,width=0.495\linewidth,clip=}
\\
\end{tabular}
  \caption{ Contours in $\mu_{\rho}-m_{\tilde{R_3}}$ plane (left
  panel) and plots of branching ratio of $\tau\rightarrow 3\mu$ (right
  panel) in the case of $\tan\gamma=3$ and $A_{\tau}=0$. The respective
   contours are BR$(\tau\rightarrow\mu\gamma)$
  (solid lines),
   $f_{D^{\gamma}}$ (dotted lines) and BR$(\tau\rightarrow3\mu)$
   (dashed lines) with numerical values of
  parameters
  $(m_B ,m_{\tilde{L}},m_{\tilde{R}_2})=(100, 100,1000)$
  ($m_{\tilde{L}_2}=m_{\tilde{L}_3}\equiv m_{\tilde{L}}$). For the plot
  of BR$(\tau\rightarrow3\mu)$ four choices of parameter space
  $(m_B,~\mu_{\rho},m_{\tilde{R}_2},m_{\tilde{R}_3})$ are:
  $(100,100,1000,100)$ (black), $(200,100,1000,100)$ (green),
  $(100,200,1000,100)$ (blue) and $(100,300,1000,100)$ (red).
  }\label{relativecon1R}
\end{figure}

The Fig. \ref{brta3muzmutau2} shows  the values  of branching
 ratios of
 the decays in the plane $m_B-\mu_{\rho}$. In this case, we
 choose $A_{\tau}=0$ and other
parameters are chosen so that the experimental branching decay of
BR$(\tau\rightarrow\mu\mu\mu)$ is satisfied. We can see that the
bounded regions of BR$(\tau\rightarrow\mu\gamma)$ supports the
small values of both remain decays. In the case $A_{\tau}\neq0$,
because only BR$(\tau\rightarrow\mu\gamma)$ depends on the values
of $A_{\tau}$  and  as we have shown in the above section, there
will exist a possibility where $D^{\gamma}_L$ vanishes. This case
allows BR$(\tau^-\rightarrow\mu^-\mu^+\mu^-)$ can reach the
 limits of experiment of order $\mathcal{O}(10^{-8})$ whilst BR$
 (Z\rightarrow\mu\tau)$ is still in maximal order of
 $\mathcal{O}(10^{-9})$.
\bc\textbf{Only maximal mixing in the right-handed charged slepton
sector $\tilde{\mu},~\tilde{\tau}$.}\ec
Now we come to consider another case, only maximal LFV in
right-handed sector of charged sleptons
($s_R=1/\sqrt{2},~s_L=s_{\tilde{\nu}_L}=s_{\tilde{\nu}_R}=0$)
where regions of parameter space can be available in range of
$\mathcal{O}(100)$[GeV].  Both   the  SUSYE331 and the MSSM models
are similar to each other in this case. So we just discuss more on
correlation among effective couplings. As shown in the left panel
of Fig. \ref{relativecon1R}, the bound of experiment of
BR$(\tau\rightarrow \mu\gamma)$ rules out the large values of
BR$(\tau\rightarrow 3\mu)$. This leads to the result
BR$(\tau\rightarrow 3\mu)\leq 10^{-9}$ if $A_{\tau}=0$. The right
panel shows that
 in the case of only large LFV in right-handed charged
 slepton sector, the $BR(\tau\rightarrow
 3\mu)$ is in maximal order of $10^{-9}$, even in the case of
  non-vanishing $A_{\tau}$
  and $A_{\tau}$ makes
  $D^{\gamma}_{R}$ suppressed.  For the
 BR$(Z\rightarrow \mu\tau)$, this case is much smaller than the
 previous case. Also we can see  a difference from the case of
 pure large LFV in left-handed charged slepton sector:
  the $D^{\gamma}$-domination
 regions now lie on the small values of $\mu_{\rho}$ while the
 large values of $\mu_{\rho}$ are ruled out
 by the condition BR$(\tau\rightarrow\mu\gamma)\leq
 4.4\times 10^{-8}$.
\section{Conclusions}

 In present paper,  we have studied the LFV   decays
 of  the tauon
 and  the $Z$ boson   in  framework of the SUSYE331 model,
and have  mainly focused on two-generations slepton mixing, namely
both left- and right-handed
  $\tilde{\mu}-\tilde{\tau}$ slepton mixings.  In order to obtain the relevant diagrams,
 we have combined the mixing of sleptons, that of charginos, Higgsinos, gauginos as
well as the interactions of gauge bosons with leptons and the
Yukawa interactions. From these diagrams, we obtained the
effective operators relating to the considered cLFV decays. This
leads to the analytical expression of the branching ratio of the
considered decay processes and the contacted relation between
decay rate of non-LFV decay mode with that of cLFV decay mode. Our
analysis is carried out in the limit of small $\tan \gamma$. The
detailed predictions in our model depend strongly on the SUSY
parameters and left- and right-handed slepton mixing.
Consequently, we have firstly considered the effects of SUSY
parameters and the mixing of left- and right-handed sleptons on
the $\tau \rightarrow \mu \gamma$ decay such as
\begin{itemize}
\item In the case of the maximal LFV  mixing, the mixing mass
terms between left- and right-handed sleptons
 $(m_{\tilde{L}_{\mu \tau}},~m_{\tilde{R}_{\mu \tau}})$ are small, our
 results  are only consistent with
 the  experimental bounds
if the domains of parameter $m_{\tilde{L}_3}$
are close to those of $m_{\tilde{L}_2}$
whenever they are set to the TeV  or $\mathcal{O}(100)$
GeV scales. It means that  in the case of maximal LFV
mixing,  the slepton mass parameters are in the same order.
\item If there is only the LFV  in  the charged left-handed
slepton sector, we can find some regions of parameter space that
allow above cLFV branching ratios matching with the experimental
 bounds. Especially
 the slepton mass $m_{\tilde{L}_3}$  is  set at 1 TeV while the
other, $m_{\tilde{L}_2}$,  is set at  $\mathcal{O}(100)$ GeV.
Noting that the value of
 parameter $m_B$ should be close to
that of $m_\lambda$
 and if the value of $m_B$
is closer to that
of $m_\lambda$, the parameter
space of $A_\tau$ has been more expanded.
In the case of left-handed LFV sector, one important
result deduced from our numerical investigation is that although
the SUSYE331 model contains much more supersymmetric particles as
well as LFV sources than the MSSM, there still exist of some wide
regions of parameter space which
allow not only masses of sleptons but also $\mu_{\rho}$
keep the small values enough to be detected by colliders.
\item If there is only the LFV in the charged right-handed slepton
sector, in order to
match the experimental  bound,
the value of $A_\tau$
 should be the same order as that of other soft parameters.
Our result is similar to the predicted result in the MSSM.
\end{itemize}
Based on the parameter space  satisfying  the experimental
bound on $\tau \rightarrow \mu \gamma$ decay rate, we consider the
branching ratios of $\tau \rightarrow \mu \mu \mu$
 and $Z\rightarrow \mu\tau$. We  have
concentrated on the LFV in the charged left-handed
  slepton sector
with  only $\theta_L=\pi/4$. In this case, there is no region of
parameter space that  $A^Z$ gives dominant contribution to the
considered decay mode, while there is region of parameter space
that $D^\gamma$ gives dominant contribution
 to the considered decay mode. The constraint on the $\mu$ parameter is
 expanded toward large values if
 the large value of the charged gaugino mass is chosen.
If we set the value of $A_\tau =0$, the constraint on the $\mu$
parameter can be expanded. Similarly, by numerical study on the
branching ratio of $\tau \rightarrow \mu \mu \mu$ decay mode in
the case  there exists LFV only in the charged right-handed
slepton sector.
 The small value of $\mu_\rho$
giving  the dominant  contribution to considered decay mode coming
from $D^\gamma$ was obtained. In both
cases, our predicted results
of
BR$(Z\rightarrow \mu \gamma)$ are very suppressed.

\section*{Acknowledgments}
L.T. Hue would like to thank Dr. A. Brignole for the explanation
of the establishing analytical formulas in \cite{Anna2}. This
research is funded by Vietnam National Foundation for Science and
Technology Development (NAFOSTED) under grant number
103.01-2011.63.
\appendix
\section{\label{InLAno} Interacting  Lagrangian and notations}
 Some of the interacting vertices  used in this work
 were given in
\cite{Giang}. In this appendix we list  the rest part of
interacting Lagrangian which is necessary  for completion our
calculation. Here we use the relation $\tilde{\ell}_R \equiv
(\tilde{\ell}^c_L)^*$ where $\tilde{\ell}^c_L$ is the superpartner
of a lepton $\ell^c_L$ in the model.
\subsection{ Interaction  between left-right slepton sector with neutral Higgs bosson}

 These terms come from two sources:
 \begin{enumerate}
    \item  from $F$-terms:
     $$ -\frac{1}{2}\left[\mu_{\rho} \rho^{0}\left(Y_{\mu}\tilde{\mu}^{*}_{L}\tilde{\mu}^{c*}_{L}
+Y_{\tau}\tilde{\tau}^{*}_{L}\tilde{\tau}^{c*}_{L}
 \right)+ \mathrm{h.c.}\right],$$
    \item  from soft-breaking term:
     \bea  \mathcal{L}^{\mathrm{soft}}_{\tilde{l}\tilde{l}H^0}
  &=& -h'_{ab}\tilde{L}_{aL}\rho'\tilde{l}^c_{bL}+
  \mathrm{h.c.}\crn&=&
  -h'_{ab} \left(\tilde{\nu}_{aL}\rho^{\prime-}_1+
  \tilde{l}_{aL}\rho^{\prime0}+
  \tilde{\nu}^c_{aL}\rho^{\prime-}_2\right)\tilde{l}^c_{bL}+ \mathrm{h.c.}
  \crn&\rightarrow& -Y_{\mu} A_{\mu}\rho^{\prime0}\tilde{\mu}_{L}
  \tilde{\mu}^c_{L}-Y_{\tau} A_{\tau}\rho^{\prime0}\tilde{\tau}_{L}
  \tilde{\tau}^c_{L}\crn&&-Y_{\tau} A^L_{\mu\tau}\rho^{\prime0}\tilde{\mu}_{L}
  \tilde{\tau}^c_{L}-Y_{\tau} A^R_{\mu\tau}\rho^{\prime0}\tilde{\tau}_{L}
  \tilde{\mu}^c_{L}+ \mathrm{h.c.}, \label{lsoftslep2h0}\eea
 where we use new notations that
 are identified with those in
 \cite{Anna2}:

 $$ h'_{22}=h'_{\mu\mu}\equiv Y_{\mu} A_{\mu}, \hs h'_{33}=
 h'_{\tau\tau}\equiv Y_{\tau} A_{\tau},
  $$
  $$ h'_{23}=h'_{\mu\tau}\equiv Y_{\tau} A^L_{\mu\tau},
   \hs h'_{32}=h'_{\tau\mu}\equiv Y_{\tau} A^R_{\mu\tau}.
  $$
 \end{enumerate}
 The total interaction part of ($H^0\tilde{l}^c_L\tilde{l}_L$) interactions is:

\bea  \mathcal{L}^{LR}_{\tilde{l}\tilde{l}H^0}
&=&-Y_{\mu}\left(\frac{1}{2}\mu_{\rho} \rho^{0*}+
A_{\mu}\rho^{\prime0}\right)
 \tilde{\mu}_{L}\tilde{\mu}^{c}_{L}
 -Y_{\tau}\left(\frac{1}{2}\mu_{\rho} \rho^{0*}+ A_{\tau}\rho^{\prime0}\right)
 \tilde{\tau}_{L}\tilde{\tau}^{c}_{L}\crn
 &&-Y_{\tau} A^L_{\mu\tau}\rho^{\prime0}\tilde{\mu}_{L}
  \tilde{\tau}^c_{L}-Y_{\tau} A^R_{\mu\tau}\rho^{\prime0}\tilde{\tau}_{L}
  \tilde{\mu}^c_{L}+ \mathrm{h.c}.\label{leftrighth0}\eea

\subsection{ \label{gaugeinteration1}Gauge boson interactions}
  This kind of vertex  is only contained in
  gauge invariant kinetics of all fields in the
  theory. In this
  work, we just study on
 cLFV in
   lepton  sector so the related part of the Lagrangian is \cite{Dong1}:
\begin{eqnarray}
\mathcal{L}_{\mathrm{kinetic}} &=& (D^\mu \rho)^{\dagger} D_\mu
 \rho + (\bar{D}^\mu \rho^{\prime})^{\dagger} \bar{D}_\mu \rho^{\prime}
 + i \bar{\tilde{\rho}}
\bar{\sigma}^{\mu}D_{\mu}\tilde{\rho}+ i
\bar{\tilde{\rho}}^{\prime}
\bar{\sigma}^{\mu}\bar{D}_{\mu}\tilde{\rho}^{\prime}\crn
 &+&
 (D^\mu \tilde{L}_{i L})^{\dagger}
 D_\mu \tilde{L}_{i L}+i\bar{L}_{iL} \bar{\sigma}^\mu D_\mu L_{iL}
 \crn&+& (D_{1\mu}l^c_{iL})^{\dagger} D_{1\mu} l^c_{iL} + i \bar{l}^c_{iL}
 \bar{\sigma}^\mu D_{1\mu} l^c_{iL} \crn &-&\fr 1 4 F_{a}^{\mu \nu}F_{a, \mu \nu}
- \fr 1 4 F^{\mu \nu}F_{\mu
 \nu} + i\bar{\lambda}^{a}_{V}\bar{\sigma}^{\mu}D^{L}_{\mu}\lambda^{a}_{V}+i
\bar{\lambda}_{B}\bar{\sigma}^{\mu}\partial_{\mu}\lambda_{B}
 \label{Lnosusy1} \end{eqnarray}
 where $i=1,2,3$ is family index, $a=1,2,...,8$  corresponds
 to eight gauge bosons of SU$(3)_L$ group.  Covariant derivatives
 $D_{\mu},~\bar{D}_{\mu}, D_{1\mu}$ and $D^{L}_{\mu}$
 correspond to triplets, anti-triplets $\mathrm{SU(3)}_L$, singlet
  $\mathrm{SU(3)}_L$ and adjoint presentation of $\mathrm{SU(3)}_L$. They are
  defined as follows:
\bea
 F_{a\mu\nu}&=&
 \partial_{\mu}V_{a\nu}-\partial_{\nu}V_{a\mu}-g f^{abc}V_{b\mu}V_{c\nu}, \hs
 F_{\mu\nu}=
 \partial_{\mu}B_{\nu}-\partial_{\nu}B_{\mu}, \crn
D_\mu &=&\partial_\mu + ig T^a V_{a\mu} +ig^\prime X T^9 B_\mu,
\crn \bar{D}_\mu &=& \partial_\mu - ig T^{a*} V_{a\mu} +ig^\prime
X T^9 B_\mu,\crn D_{1\mu} &=&\partial_\mu  +ig^\prime X T^9
B_\mu,\crn D^{L}_{\mu}\lambda^a_V&=& \partial_{\mu}\lambda^a_V-g
f^{abc}V^b\lambda^c_V. \label{dhhbien1} \eea
Here  $X$ denotes
$U(1)$ hypercharge, $f^{abc}$ is structure constant of
$\mathrm{SU(3)}$, $T^9$ is the generator of $U(1)_X$ which is
defined by $T^9= 1/\sqrt{6} ~\mathrm{diagonal} (1,1,1)$. We just
pay attention to neutral bosons in covariant derivatives so they
can be written as \cite{Dong1,e331}.
 \bea
  D^N_{\mu}&\equiv& \partial_{\mu}+i\mathcal{P}^{\mathrm{NC}}_\mu\crn&=&
  \partial_{\mu} +i g \left( T^3V_{3\mu}+T^8V_{8\mu}+t
  T^9XB_{\mu}+T^4V_{4\mu}+T^5V_{5\mu}\right),
  \crn
  \bar{D}^N_{\mu}&\equiv& \partial_{\mu}-i\mathcal{P}^{\mathrm{NC}}_\mu\crn&=&
  \partial_{\mu} -i g \left( T^3V_{3\mu}+T^8V_{8\mu}-t
  T^9XB_{\mu}\right)-i g \left(T^4V_{4\mu}-T^5V_{5\mu}\right),
  \crn
    D_{1\mu} &=&\partial_\mu  +ig^\prime X T^9
B_\mu,\crn D^{L}_{\mu}\lambda^a_V&=&
\partial_{\mu}\lambda^a_V-g\left(
f^{a3c}V^3_{\mu}+f^{a8c}V^8_{\mu}+f^{a4c}V^4_{\mu}+f^{a5c}V^5_{\mu}\right)\lambda^c_V,
\label{dhhbien2}\eea where gauge bosons $W_3,W_8, W_4$, and $B$
relate with physical states according  to the transformation:

\bea  \left(%
\begin{array}{c}
  W_3 \\
  W_8 \\
  B \\
  W_4 \\
\end{array}%
\right) = \left(%
\begin{array}{cccc}
  s_W & c_\va c_{\theta'}c_W  & s_\va c_{\theta'}c_W  & s_{\theta'}c_W \\
  -\fr{s_W}{\sqrt{3}} & \fr{c_\va \kappa_3
  -s_\va\kappa_1 \kappa_2}{\sqrt{3}c_Wc_{\theta'}} &
  \fr{s_\va \kappa_3
  +c_\va\kappa_1 \kappa_2}{\sqrt{3}c_Wc_{\theta'}} & \sqrt{3}s_{\theta'}c_W\\
  \fr{\kappa_1}{\sqrt{3}} & -\fr{t_W(c_\va \kappa_1
  +s_\va \kappa_2)}{\sqrt{3}c_{\theta'}} & -\fr{t_W(s_\va \kappa_1
  -c_\va \kappa_2)}{\sqrt{3}c_{\theta'}}  & 0 \\
  0 & -t_{\theta'}(c_\va \kappa_2
  -s_\va\kappa_1) & -t_{\theta'}(s_\va \kappa_2
  + c_\va\kappa_1) & \kappa_2 \\
\end{array}%
\right) \left(%
\begin{array}{c}
  A \\
  Z \\
  Z' \\
  W'_{4} \\
\end{array}%
\right),\crn \label{bosontrunghoa1}\eea
 where some new notations are used:
\bea t_{\theta}\equiv \tan\theta &=& \frac{u}{w},\; t_{2\theta}
\equiv \tan(2\theta),\;
 s_{\theta'}=
\fr{t_{2\theta}}{c_W\sqrt{1+4t^2_{2\theta}}}\crn t&\equiv&
\frac{g'}{g}= \fr{3\sqrt{2}s_W}{\sqrt{3-4s^2_W}},\; \kappa_1
\equiv \sqrt{4 c^2_W-1}=\frac{3 \sqrt{2}s_W}{t} \crn
 \kappa_2 &\equiv&  \sqrt{1-4 s^2_{\theta'} c^2_W}, \kappa_3
 = s^2_W-3 c^2_W s^2_{\theta'}.\eea
   In the SUSYE331 model, we have all $\theta,~ \varphi$ and $\theta'\ll 1$.
    Thus, we can use the approximation
   $\sin\theta=\sin\varphi=sin\theta'=\tan\theta=\tan\varphi=\tan\theta'=0$
   to simplify the calculation. We also take the approximation:
    $$ \kappa_2 \simeq 1, \; \kappa_3 \simeq
 s^2_W .$$
 In addition, $W_5$ and $W'_4$ make of a physical neutral
 non-Hermitian  gauge boson $X^0$ which is defined by  the
 combination:
 \be X^{0}_{\mu}\equiv \frac{W'_{4\mu}-i W_{5\mu}}{\sqrt{2}} \label{X0}\ee
So we can rewrite the above covariance derivatives in the form
below:
\bea D^{N}_{\mu}&\simeq& \partial_{\mu} + ie QA_{\mu}+
ig_Z\left( T^3-s^2_W Q\right) Z_{\mu}\crn&+&ig_{Z'}
\left[(4c^2_W-1)(T^3-Q)+3c^2_W X\right] Z'_{\mu},\crn
 \bar{D}^{N}_{\mu}&\simeq& \partial_{\mu} + ie QA_{\mu}+
ig_Z\left(-T^3-s^2_W Q\right) Z_{\mu}\crn&+&ig_{Z'}
\left[(4c^2_W-1)(-T^3-Q)+3c^2_W X\right] Z'_{\mu},\crn
D^N_{1\mu}&\simeq&\partial_{\mu} + ie QA_{\mu}- ig_Zs^2_W
Q Z_{\mu}\crn&+&ig_{Z'} \left[-(4c^2_W-1)Q+3c^2_W
X\right] Z'_{\mu}\label{Dcovariance1}\eea
  where we  have defined
 $$ g_Z \equiv \frac{gc_{\varphi}}{c_Wc_{\theta'}}\simeq \frac{g}{c_W}
 ~~~~~~~\mathrm{and}~~~~~~
 g_{Z'} \equiv \frac{gc_{\varphi}\kappa_2}{c_Wc_{\theta'}\kappa_1}\simeq
  \frac{g}{c_W\kappa_1}.$$

For the charged gauginos, we have:
\bc
 \bea \tilde{W}^{\pm}&\equiv& \frac{\lambda^1_V\mp
 i\lambda_V^2}{\sqrt{2}},\hs \tilde{Y}^{\pm}\equiv
 \frac{\lambda^6_V\pm
 i\lambda_V^7}{\sqrt{2}}.
\label{cgaugino1}\eea \ec This leads to the covariant derivative
of charged gauginos:
\bea  D^L_{\mu}\tilde{W}^{\pm}
&\simeq&\partial_{\mu}\tilde{W}^{\pm} \pm i\left(  e A_{\mu}+ gc_W
Z_{\mu}\right)\tilde{W}^{\pm} =\partial_{\mu}\tilde{W}^{\pm} \pm
i\left(  e A_{\mu}+ g_Zc^2_W Z_{\mu}\right)\tilde{W}^{\pm}\crn&=&
\partial_{\mu}\tilde{W}^{\pm} +i Q_{\tilde{W}}\left(  e A_{\mu}
+ g_Zc^2_W Z_{\mu}\right)\tilde{W}^{\pm},\crn
 D^L_{\mu}\tilde{Y}^{\pm} &=&
\partial_{\mu}\tilde{Y}^{\pm} \pm i
\left(e
A_{\mu}+g\frac{c_{\varphi}\left(c_{2W}+2c^2_Ws^2_{\theta'}\right)
+s_{\varphi}\kappa_1\kappa_2}{2c_Wc_{\theta'}}
Z_{\mu}\right.\crn&+&\left.g\frac{s_{\varphi}\left(c_{2W}+2c^2_Ws^2_{\theta'}\right)
-c_{\varphi}\kappa_1\kappa_2}{2c_Wc_{\theta'}} Z'_{\mu}
\right)\tilde{Y}^{\pm}\crn &\simeq&\partial_{\mu}\tilde{Y}^{\pm}
+i Q_{\tilde{Y}^\pm} \left(e A_{\mu}+\frac{1}{2}g_Z c_{2W}
Z_{\mu}-\frac{1}{2}g_{Z'}\kappa_1^2 Z'_{\mu}
\right)\tilde{Y}^{\pm}. \label{dcgauginos1}\eea
From these two formulas, we can deduce the  vertices of neutral
gauge boson-charged gaugino-charged gaugino.

\subsection{\label{gbslsp1} Gauge boson-slepton-slepton interactions}

This kind of vertex comes from the part \cite{Dong1}:
\bea \mathcal{L}_{\tilde{l}\tilde{l}V}&=& \frac{i
g}{2}\left[\partial^{\mu}\bar{\tilde{L}}_i
\lambda^a\tilde{L}_i-\bar{\tilde{L}}_i
\lambda^a\partial^{\mu}\tilde{L}_i\right]V^a_{\mu}\crn &+&\frac{i
g'}{\sqrt{6}}\left[-\frac{1}{3}\left(\partial^{\mu}\bar{\tilde{L}}_i
\tilde{L}_i-\bar{\tilde{L}}_i
\partial^{\mu}\tilde{L}_i\right)+
\left(\partial^{\mu}\bar{\tilde{l^c}}\tilde{l}^c-\bar{\tilde{l^c}}\partial^{\mu}\tilde{l}^c
\right)\right]B_{\mu}\nonumber\eea
 where $i=1,2,3$ is the flavor index and $a=1,2,...,8$ is generator
 index of $SU(3)$. For the $\{\tilde{\mu},\tilde{\tau}\}$ sector
 with neutral boson  we have:
\bea \mathcal{L}_{\tilde{l}\tilde{l}V}
  &\simeq& \frac{i g}{2}\left[\fr{1}{c_W} Z_{\mu}+
  \fr{c_{2W}}{\kappa_1c_W} Z'_{\mu}\right]\times
  \left( \partial^{\mu}\bar{\tilde{\nu}}_{\tau} \tilde{\nu}_{\tau}
  -\bar{\tilde{\nu}}_{\tau} \partial^{\mu}\tilde{\nu}_{\tau}\right)\crn
  &+& \frac{i g}{2}\left[-2s_W A_{\mu}-\fr{c_{2W}}{c_W}Z_{\mu}+\fr{c_{2W}}{\kappa_1c_W}Z'_{\mu}
  \right]\left( \partial^{\mu}\bar{\tilde{\tau}} \tilde{\tau}
  -\bar{\tilde{\tau}} \partial^{\mu}\tilde{\tau}\right)\crn
  &+& \frac{i g}{2}\left[-\fr{2c_W}{\kappa_1} Z'_{\mu}\right]
  \left( \partial^{\mu}\bar{\tilde{\nu^c}}_{\tau} \tilde{\nu^c}_{\tau}
  -\bar{\tilde{\nu^c}}_{\tau} \partial^{\mu}\tilde{\nu^c}_{\tau}\right)
 +(\tau \rightarrow \mu) \crn
  &+&\left[ i\left(e A_{\mu}-e ~t_WZ_{\mu}+\fr{e~t_W
   }{\kappa_1} Z'_{\mu} \right)\left(\partial^{\mu}\bar{\tilde{\tau^c}}
\tilde{\tau}^c-\bar{\tilde{\tau^c}}\partial^{\mu}\tilde{\tau}^c
\right)+ \left(\tau^c\rightarrow \mu^c\right)\right]\crn
 &=& i
\left[\frac{1}{2}g_Z Z_{\mu}+\frac{1}{2}
  g_{Z'}c_{2W}Z'_{\mu}\right]\times
  \left( \partial^{\mu}\bar{\tilde{\nu}}_{\tau} \tilde{\nu}_{\tau}
  -\bar{\tilde{\nu}}_{\tau} \partial^{\mu}\tilde{\nu}_{\tau}\right)\crn
  &-& i\left[e A_{\mu}+\frac{1}{2}g_Zc_{2W}Z_{\mu}-\frac{1}{2}g_{Z'}c_{2W}Z'_{\mu}
  \right]\left( \partial^{\mu}\bar{\tilde{\tau}} \tilde{\tau}
  -\bar{\tilde{\tau}} \partial^{\mu}\tilde{\tau}\right)\crn
  &-& i \left[g_{Z'}c^2_W Z'_{\mu}\right]
  \left( \partial^{\mu}\bar{\tilde{\nu^c}}_{\tau} \tilde{\nu^c}_{\tau}
  -\bar{\tilde{\nu^c}}_{\tau} \partial^{\mu}\tilde{\nu^c}_{\tau}\right)
 +(\tau \rightarrow \mu) \crn
  &+&\left[ i\left(e A_{\mu}-g_Z s^2_WZ_{\mu}+g_{Z'}s^2_W
   Z'_{\mu} \right)\left(\partial^{\mu}\bar{\tilde{\tau^c}}
\tilde{\tau}^c-\bar{\tilde{\tau^c}}\partial^{\mu}\tilde{\tau}^c
\right)\right.\crn&+&\left. \left(\tau^c\rightarrow
\mu^c\right)\right] \label{slep2V1}\eea
\begin{figure}[h]
\begin{center}
\begin{picture}(110,80)(-55,-40)
\DashArrowLine(-30,0)(0,0){3}\DashArrowLine(0,0)(30,0){3}
\ArrowLine(-30,-10)(-10,-10)\ArrowLine(10,-10)(30,-10)
\Photon(0,27)(0,0){2}{3} \Text(10,30)[]{$V^{\mu}$}
\Text(-30,-15)[]{$p$}\Text(30,-15)[]{$p'$}
\end{picture}
\begin{picture}(110,80)(-55,-40)
\ArrowLine(-30,0)(0,0)\ArrowLine(0,0)(30,0)
\ArrowLine(-30,-10)(-10,-10)\ArrowLine(10,-10)(30,-10)
 \Photon(0,25)(0,0){2}{3} \Text(10,30)[]{$V^{\mu}$}
\Text(-30,-15)[]{$p$}\Text(30,-15)[]{$p'$}
\end{picture}
\end{center}
\caption{\ftsz Notations of directions of scalars and
fermions. Here $V^{\mu}$ denotes a photon $A$,
$Z$ or $Z'$ boson.} \label{dinhvohuong}
\end{figure}
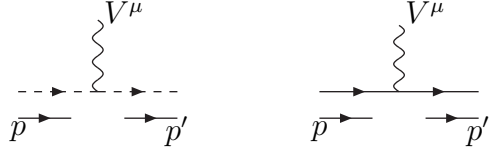

 Interaction vertices of photon, $Z$ and $Z'$ bosons relating with our calculation are summarized in
  tables \ref{photon1v}, \ref{zvertex} and \ref{zpvertex},
  respectively. We denote directions of momentums in
  Fig.\ref{dinhvohuong}. For simplicity, we omit spinor index in
  the formulas of boson-fermion-fermion vertices. The  precise formulas of
   this kind of vertices is easily deduced using rules concerned in
   \cite{SpMartin1}.
\begin{table}[h]
  \centering
  \caption{Photon vertices.}
  \begin{tabular}{|c|c|c|c|}
    \hline
    vertex & factor & vertex & factor \\
    \hline
   Photon-scalar-scalar &  &  &  \\
    (scalar $\varphi: H,\tilde{f}$) &$-ie Q_{\varphi} (p+p')^{\mu}$  &  &  \\
    \hline
    Photon-spinor-spinor & & &\\
    (spinor $\psi$: fermion, Higgsino) & $-ie Q_{\psi} \bar{\sigma}^{\mu}$ ($ie Q_{\psi} \sigma^{\mu}$)
    & $\gamma\psi^c\psi^c$  & $ie Q_{\psi} \bar{\sigma}^{\mu}$ ($-ie Q_{\psi} \sigma^{\mu}$) \\
    \hline
Photon-boson-boson & & & \\
 $W^{+\rho}W^{-\mu}A^{\nu}$ &$ie[p_{+\rho},~p_{-\mu},~ p_{A\nu}]$ & $Y^{+\rho}Y^{-\mu}A^{\nu}$
 &$-ie[p_{+\rho},~p_{-\mu},~p_{A\nu}]$\\
    \hline
Photon-Higgs-gauge boson & &  &  \\
$A^{\mu} W^{\nu} \rho_1$ & $ \frac{1}{2}(ie g) g_{\mu\nu}$  &
$A^{\mu} Y^{\nu} \rho_2$ & $ \frac{1}{2}(i e g) g_{\mu\nu}$ \\
 \hline
Photon-gaugino-gaugino &  &  &  \\
$\overline{\tilde{W}^+} A_{\mu}\tilde{W}^{+}$
&$-ie\bar{\sigma}^{\mu}$ (or $ie\sigma^{\mu}$) &
$\overline{\tilde{W}^-} A_{\mu}\tilde{W}^{-}$&$ie\bar{\sigma}^{\mu}$ (or $-ie\sigma^{\mu}$)\\
$\overline{\tilde{Y}^+} A_{\mu}\tilde{Y}^{+}$
&$-ie\bar{\sigma}^{\mu}$ (or $ie\sigma^{\mu}$) &
$\overline{\tilde{Y}^-} A_{\mu}\tilde{Y}^{-}$&$ie\bar{\sigma}^{\mu}$ (or $-ie\sigma^{\mu}$)\\
    \hline
  \end{tabular}
    \label{photon1v}
\end{table}
\begin{table}[h]
  \centering
   \caption{$Z$ boson vertices}
 \begin{tabular}{|c|c|c|c|}
   \hline
   Vertex & Factor & Vertex & Factor \\
   \hline
   $Z_{\mu}\tilde{\nu}^*_{L}\tilde{\nu}_L$ & $-\frac{i}{2}~g_{Z}(p+p')^{\mu}$
   &  &  \\
   $Z_{\mu}\tilde{\ell}^*_L\tilde{\ell}_L$ &$ \frac{i}{2}c_{2W}g_{Z}(p+p')^{\mu}$ &
    $Z_{\mu}\tilde{\ell}_R^*\tilde{\ell}_R$ & $ig_{Z}s^2_W(p+p')^{\mu} $\\
   \hline
  $\rho^{0*}\rho^0 Z_{\mu}$ &$\frac{i}{2}g_{Z}(p+p')^{\mu}$&$\rho^{\prime0*}\rho^{\prime0} Z_{\mu}$&
  $-\frac{i}{2}g_{Z}(p+p')^{\mu}$\\
  $\chi^{0*}_1\chi^0_1 Z_{\mu}$ &$-\frac{i}{2}g_{Z}(p+p')^{\mu}$&$\chi^{\prime0*}_1\chi^{\prime0}_1 Z_{\mu}$&
  $\frac{i}{2}g_{Z}(p+p')^{\mu}$\\
   \hline
   $\overline{\tilde{\rho}^{0}}\tilde{\rho}^0 Z_{\mu}$ &$\frac{i}{2}g_{Z}\bar{\sigma}^{\mu}$
   (or $-\frac{i}{2}g_{Z}\sigma^{\mu}$)&
   $\overline{\tilde{\rho}^{\prime0}}\tilde{\rho}^{\prime0} Z_{\mu}$&
  $-\frac{i}{2}g_{Z}\sigma^{\mu}$  (or $\frac{i}{2}g_{Z}\sigma^{\mu}$)\\
   $\overline{\tilde{\rho}^{+}_1}\tilde{\rho}^+_1 Z_{\mu}$ &$-\frac{i}{2}g_{Z}c_{2W}\bar{\sigma}^{\mu}$
   (or $\frac{i}{2}g_{Z}c_{2W}\sigma^{\mu}$)&
   $\overline{\tilde{\rho}^{\prime-}_1}\tilde{\rho}^{\prime-}_1 Z_{\mu}$&
  $\frac{i}{2}g_{Z}c_{2W}\sigma^{\mu}$  (or $-\frac{i}{2}g_{Z}c_{2W}\sigma^{\mu}$)\\
  $\overline{\tilde{\rho}^{+}_2}\tilde{\rho}^+_2 Z_{\mu}$ &$\frac{i}{2}g_{Z}s^2_{W}\bar{\sigma}^{\mu}$
   (or $-\frac{i}{2}g_{Z}s^2_{W}\sigma^{\mu}$)&
   $\overline{\tilde{\rho}^{\prime-}_2}\tilde{\rho}^{\prime-}_2 Z_{\mu}$&
  $-\frac{i}{2}g_{Z}s^2_{W}\sigma^{\mu}$  (or $\frac{i}{2}g_{Z}s^2_{W}\sigma^{\mu}$)\\
   \hline
   $\overline{\tilde{W}^+} Z_{\mu}\tilde{W}^{+}$&$-ig_{Z}c^2_W\bar{\sigma}^{\mu}$
   (or $ig_{Z}c^2_W\sigma^{\mu}$) & $\overline{\tilde{W}^-} Z_{\mu}\tilde{W}^{-}$
   &$ig_{Z}c^2_W\bar{\sigma}^{\mu}$
   (or $-ig_{Z}c^2_W\sigma^{\mu}$)\\
   $\overline{\tilde{Y}^+} Z_{\mu}\tilde{Y}^{+}$&$-\frac{i}{2}g_{Z}c_{2W}\bar{\sigma}^{\mu}$
   (or $\frac{i}{2}g_{Z}c_{2W}\sigma^{\mu}$)&$\overline{\tilde{Y}^-} Z_{\mu}\tilde{Y}^{-}$&
   $\frac{i}{2}g_{Z}c_{2W}\bar{\sigma}^{\mu}$
   (or $-\frac{i}{2}g_{Z}c_{2W}\sigma^{\mu}$)\\
   \hline
 \end{tabular}
 \label{zvertex}
\end{table}
\begin{table}[h]
  \centering
    \caption{$Z'$ boson vertices}\label{zpvertex}
 \begin{tabular}{|c|c|c|c|}
   \hline
   Vertex & Factor & Vertex & Factor \\
   \hline
   $Z'_{\mu}\tilde{\nu}^*_L\tilde{\nu}_L$ &$-\frac{i}{2}~g_{Z'}c_{2W}(p+p')^{\mu}$  &
    $Z'_{\mu}\tilde{\nu}_R^*\tilde{\nu}_R$  & $-g_{Z'}c^2_W(p+p')^{\mu}$ \\
    $Z'_{\mu}\tilde{\ell}^*_L\tilde{\ell}_L$ &$ -\frac{i}{2}c_{2W}g_{Z'}(p+p')^{\mu}$ &
    $Z'_{\mu}\tilde{\ell}_R^*\tilde{\ell}_R$ & $-ig_{Z'}s^2_W(p+p')^{\mu} $\\
    \hline
  $\rho^{0*}\rho^0 Z'_{\mu}$ &$-\frac{i}{2}g_{Z}(p+p')^{\mu}$&$\rho^{\prime0*}\rho^{\prime0} Z'_{\mu}$&
  $\frac{i}{2}g_{Z}(p+p')^{\mu}$\\
  $\chi^{0*}_1\chi^0_1 Z'_{\mu}$ &$-\frac{i}{2}g_{Z'}c_{2W}(p+p')^{\mu}$&$\chi^{\prime0*}_1\chi^{\prime0}_1 Z'_{\mu}$&
  $\frac{i}{2}g_{Z'}c_{2W}(p+p')^{\mu}$\\
  $\chi^{0*}_2\chi^0_2 Z'_{\mu}$ &$-\frac{i}{2}g_{Z'}c^2_{W}(p+p')^{\mu}$&$\chi^{\prime0*}_2\chi^{\prime0}_2 Z'_{\mu}$&
  $\frac{i}{2}g_{Z'}c^2_{W}(p+p')^{\mu}$\\
\hline
   $\overline{\tilde{\rho}^{0}}\tilde{\rho}^0 Z'_{\mu}$ &$-\frac{i}{2}g_{Z'}\bar{\sigma}^{\mu}$
   (or $\frac{i}{2}g_{Z'}\sigma^{\mu}$)&
   $\overline{\tilde{\rho}^{\prime0}}\tilde{\rho}^{\prime0} Z_{\mu}$&
  $\frac{i}{2}g_{Z'}\sigma^{\mu}$  (or $-\frac{i}{2}g_{Z'}\sigma^{\mu}$)\\
   $\overline{\tilde{\rho}^{+}_1}\tilde{\rho}^+_1 Z'_{\mu}$ &$-\frac{i}{2}g_{Z'}\bar{\sigma}^{\mu}$
   (or $\frac{i}{2}g_{Z'}\sigma^{\mu}$)&
   $\overline{\tilde{\rho}^{\prime-}_1}\tilde{\rho}^{\prime-}_1 Z'_{\mu}$&
  $\frac{i}{2}g_{Z'}\sigma^{\mu}$  (or $-\frac{i}{2}g_{Z'}\sigma^{\mu}$)\\
  $\overline{\tilde{\rho}^{+}_2}\tilde{\rho}^+_2 Z'_{\mu}$ &$\frac{i}{2}g_{Z'}\bar{\sigma}^{\mu}$
   (or $-\frac{i}{2}g_{Z}\sigma^{\mu}$)&
   $\overline{\tilde{\rho}^{\prime-}_2}\tilde{\rho}^{\prime-}_2 Z'_{\mu}$&
  $-\frac{i}{2}g_{Z'}c_{2W}\sigma^{\mu}$  (or $\frac{i}{2}g_{Z}c_{2W}\sigma^{\mu}$)\\

   \hline
    $\overline{\tilde{Y}^+} Z'_{\mu}\tilde{Y}^{+}$&$\frac{i}{2}g_{Z'}\kappa_1^2\bar{\sigma}^{\mu}$
   (or $-\frac{i}{2}g_{Z'}\kappa_1^2\sigma^{\mu}$)&$\overline{\tilde{Y}^-} Z'_{\mu}\tilde{Y}^{-}$&
   $-\frac{i}{2}g_{Z'}\kappa_1^2\bar{\sigma}^{\mu}$
   (or $\frac{i}{2}g_{Z'}\kappa_1^2\sigma^{\mu}$)\\
   \hline
 \end{tabular}
\end{table}

\subsection{Mixing in the slepton sector}
As we know, in supersymmetric models, in order to keep the
conversation of LFV in the lepton sector at tree level, the
sources of LFV are assumed to be  from the slepton mass terms in
the soft-breaking part of the Lagrangian \cite{Anna2,hisano}. For
the SUSYE331, there are three  mass terms of left-handed slepton,
right-handed slepton and sneutrinos which may independently be
sources of LFV. In addition, there exists another LFV source
original from the Yukawa couplings between Higgs and neutrinos.
Thus in the SUSYE331 model, there are at least four independent
 sources of LFV and we will parameterize
them as follows. In each case of supersymmetric particle
(sleptons) $\tilde{\psi}$ ($\psi= l_{L},~l_{R},~\nu_L,\nu_R $), we
define a corresponding mixing angle $\theta_{\tilde{\varphi}}$
which was defined in \cite{Giang}. In what follows we just remind
some general formulas for the review. The mass mixing matrices of
smuon and stau as well as their sneutrinos can be written in the
general form of:
\be\mathcal{M}^2_{\tilde{\psi}} =\left(%
 \begin{array}{cc}
  m^2_{\tilde{\psi}_{\mu\mu}} & m^2_{\tilde{\psi}_{\mu\tau}} \\
 m^2_{\tilde{\psi}_{\mu\tau}} & m^2_{\tilde{\psi}_{\tau\tau}} \\
\end{array}%
\right).\label{mixingm} \ee
Mixing angles then can be determined as
\be s_{\tilde{\psi}}\equiv \sin\theta_{\tilde{\psi}}, \hs
c_{\tilde{\psi}}\equiv \cos\theta_{\tilde{\psi}}\hs
\mathrm{where}\hs
s_{\tilde{\psi}}c_{\tilde{\psi}}=\frac{m^2_{\tilde{\psi}_{\mu\tau}}}{m^2_{\psi_2}-m^2_{\psi_3}},
\label{mixingangles}\ee
 where $s_{\tilde{\psi}}=\{ s_L, s_R, s_{\tilde{\nu}_L},
s_{\tilde{\nu}_R}\}$ and $\{m^2_{\psi_2},~m^2_{\psi_3}\}$ are
eigenvalues of $\mathcal{M}^2_{\tilde{\psi}} $, according to
notations in \cite{Giang}. In addition, for convenience  we denote
$m^2_{\tilde{\psi}}$ instead of $\tilde{m}^2_{\psi}$.
We always choose $ m^2_{\psi_3}<m^2_{\psi_2}$ to take the positive
values of $s_{\tilde{\psi}}$ and $c_{\tilde{\psi}}$. The
mass-eigenstates of sleptons are denoted as $\{ \tilde{\psi}_2,~
\tilde{\psi}_3 \}$ while the flavor-eigenstates are $\{
\tilde{\psi}_{\mu},~ \tilde{\psi}_{\tau} \}$. The relation between
two
bases are: \be \tilde{\psi}_{\mu}=c_{\tilde{\psi}}
\tilde{\psi}_2-s_{\tilde{\psi}} \tilde{\psi}_3\hs \mathrm{and} \hs
\tilde{\psi}_{\tau}=s_{\tilde{\psi}}
\tilde{\psi}_2+c_{\tilde{\psi}} \tilde{\psi}_3.\ee

\section{\label{mutaugamma} Contribution to $\tau \rightarrow \mu\gamma$}

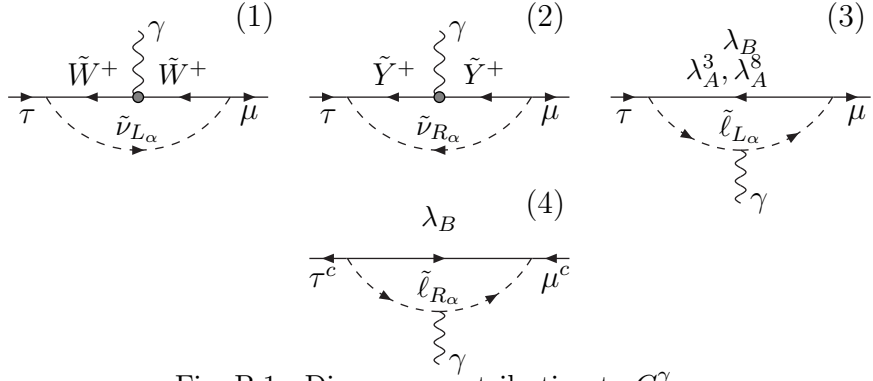
\begin{figure}[h]
\begin{center}
\begin{picture}(110,60)(-55,-30)
\ArrowLine(-49,0)(-35,0) \ArrowLine(0,0)(-35,0)
\ArrowLine(35,0)(0,0) \ArrowLine(35,0)(49,0)
\Photon(0,0)(0,25){2}{3} \GCirc(0,0){2}{0.5}
\DashArrowArc(0,20)(40,210,330){3} \Text(-42,-7)[]{$\tau$}
\Text(42,-7)[]{$\mu$} \Text(-17,10)[]{$\tilde{W}^+$}
\Text(17,10)[]{{\small$\tilde{W}^+$}} \Text(7,25)[]{$\gamma$}
\Text(0,-12)[]{\small{$\tilde{\nu}_{L_\alpha}$}}
\Text(44,30)[]{(1)}
\end{picture}
\begin{picture}(110,60)(-55,-30)
\ArrowLine(-49,0)(-35,0) \ArrowLine(0,0)(-35,0)
\ArrowLine(35,0)(0,0) \ArrowLine(35,0)(49,0)
\Photon(0,0)(0,25){2}{3} \GCirc(0,0){2}{0.5}
\DashArrowArcn(0,20)(40,330,210){3} \Text(-42,-7)[]{$\tau$}
\Text(42,-7)[]{$\mu$} \Text(-17,10)[]{$\tilde{Y}^+$}
\Text(17,10)[]{{\small$\tilde{Y}^+$}} \Text(7,25)[]{$\gamma$}
\Text(0,-12)[]{\small{$\tilde{\nu}_{R_\alpha}$}}
\Text(40,30)[]{$(2)$}
\end{picture}
\begin{picture}(110,60)(-55,-30)
\ArrowLine(-49,0)(-35,0) \ArrowLine(35,0)(-35,0)
\ArrowLine(35,0)(49,0) \DashArrowArc(0,20)(40,210,270){3}
\DashArrowArc(0,20)(40,270,330){3} \Photon(0,-20)(0,-40){2}{3}
\Text(-44,-7)[]{$\tau$} \Text(44,-7)[]{$\mu$}
\Text(0,21)[]{\small{$\lambda_B$}}
\Text(-5,10)[]{\small{$\lambda^3_A,\lambda^8_A$}}
\Text(7,-40)[]{$\gamma$}
\Text(0,-11)[]{\small{$\tilde{\ell}_{L_\alpha}$}}
\Text(40,30)[]{$(3)$}
\end{picture}
\end{center}
\begin{center}
%
\begin{picture}(110,60)(-55,-30)
\ArrowLine(-35,0)(-49,0) \ArrowLine(-35,0)(35,0)
\ArrowLine(49,0)(35,0) \DashArrowArc(0,20)(40,210,270){3}
\DashArrowArc(0,20)(40,270,330){3} \Photon(0,-20)(0,-40){2}{3}
\Text(-44,-7)[]{$\tau^c$} \Text(44,-8)[]{$\mu^c$}
\Text(0,15)[]{\small{$\lambda_B$}} \Text(7,-40)[]{$\gamma$}
\Text(0,-11)[]{\small{$\tilde{\ell}_{R_\alpha}$}}
\Text(40,20)[]{$(4)$}
\end{picture}
\end{center}
\caption{ Diagrams contributing to $C^{\gamma}_{L,R}$.}
\label{cga1}
\end{figure}

Diagrams relating to  $C^{\gamma}_{L,R}$ are drawn in Fig.
\ref{cga1} with no line of Higgs insertion.

Formulas  of $C^{\gamma}_{L,R}$ are:
 \bea
 C^{\gamma}_{L}&=&
\frac{(g^2 c_Ls_L)}{16\pi^2}\times \frac{1}{9} \left[-K_5
(m^2_{\lambda},m^2_{\tilde{l}_{L2}},m^2_{\tilde{l}_{L2}},m^2_{\tilde{l}_{L2}},m^2_{\tilde{l}_{L2}})
\right]\crn&+&\frac{(g^2 c_{\nu_L}s_{\nu_L})}{16\pi^2}\times
\frac{1}{6}\left[-2K_{5}(m^2_{\lambda},m^2_{\lambda},m^2_{\lambda},m^2_{\lambda},
\tilde{m}^2_{\nu_{L2}})\right.\crn&+&\left. 3 m^2_{\lambda}
J_5(m^2_{\lambda},m^2_{\lambda},m^2_{\lambda},m^2_{\lambda},
\tilde{m}^2_{\nu_{L2}})\right]\crn&+& \frac{(g^2
c_{\nu_R}s_{\nu_R})}{16\pi^2}\times \frac{1}{6}
\left[-2K_{5}(m^2_{\lambda},m^2_{\lambda},m^2_{\lambda},m^2_{\lambda}
,\tilde{m}^2_{\nu_{R2}})\right.\crn&+&\left. 3 m^2_{\lambda}
J_5(m^2_{\lambda},m^2_{\lambda},m^2_{\lambda},m^2_{\lambda}
,\tilde{m}^2_{\nu_{R2}})\right]\crn&+&\frac{g^{\prime2}c_Ls_L }{16
\pi^2}\times\frac{1}{162}\left[-K_5
    (m^2_B,m^2_{\tilde{l}_{L2}},m^2_{\tilde{l}_{L2}},m^2_{\tilde{l}_{L2}},m^2_{\tilde{l}_{L2}})\right]\crn&-&
     (L_2\rightarrow L_3, R_2\rightarrow R_3),\crn
  C^{\gamma}_{R} &=& \frac{g^{\prime2}c_Rs_R}{16\pi^2}\times\frac{1}{18}\left[-K_5
    (m^{2}_B,m^2_{\tilde{l}_{R2}},m^2_{\tilde{l}_{R2}},
    m^2_{\tilde{l}_{R2}},m^2_{\tilde{l}_{R2}})\right]-[R_2\rightarrow R_3].\label{cgamma1}\eea

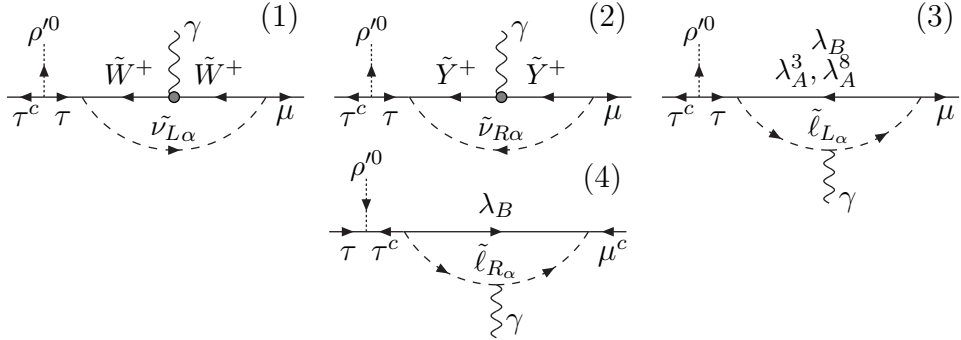
\begin{figure}[h]
\vspace{0.4cm}
\begin{center}
\begin{picture}(120,60)(-65,-30)
\ArrowLine(-49,0)(-63,0) \ArrowLine(-49,0)(-35,0)
\ArrowLine(0,0)(-35,0) \ArrowLine(35,0)(0,0)
\ArrowLine(35,0)(49,0) \DashArrowLine(-49,0)(-49,20){1}
\Photon(0,0)(0,25){2}{3} \GCirc(0,0){2}{0.5}
\DashArrowArc(0,20)(40,210,330){3} \Text(-56,-7)[]{$\tau^c$}
\Text(-42,-8)[]{$\tau$} \Text(42,-7)[]{$\mu$}
\Text(-49,25)[]{\small{$\rho^{\prime0}$}}
\Text(-17,10)[]{\small{$\tilde{W}^+$}}
\Text(17,10)[]{\small{$\tilde{W}^+$}} \Text(7,25)[]{$\gamma$}
\Text(0,-12)[]{\small{$\tilde{\nu
_L}_{\alpha}$}}\Text(40,30)[]{$(1)$}
\end{picture}
\begin{picture}(120,60)(-65,-30)
\ArrowLine(-49,0)(-63,0) \ArrowLine(-49,0)(-35,0)
\ArrowLine(0,0)(-35,0) \ArrowLine(35,0)(0,0)
\ArrowLine(35,0)(49,0) \DashArrowLine(-49,0)(-49,20){1}
\Photon(0,0)(0,25){2}{3} \GCirc(0,0){2}{0.5}
\DashArrowArcn(0,20)(40,330,210){3} \Text(-56,-7)[]{$\tau^c$}
\Text(-42,-8)[]{$\tau$} \Text(42,-7)[]{$\mu$}
\Text(-49,25)[]{\small{$\rho^{\prime0}$}}
\Text(-17,10)[]{\small{$\tilde{Y}^+$}}
\Text(17,10)[]{\small{$\tilde{Y}^+$}} \Text(7,25)[]{$\gamma$}
\Text(0,-12)[]{\small{$\tilde{\nu}_{R\alpha}$}}\Text(40,30)[]{$(2)$}
\end{picture}
\begin{picture}(120,60)(-65,-30)
\ArrowLine(-49,0)(-63,0) \ArrowLine(-49,0)(-35,0)
\ArrowLine(35,0)(-35,0) \ArrowLine(35,0)(49,0)
\DashArrowLine(-49,0)(-49,20){1}
\DashArrowArc(0,20)(40,210,270){3}
\DashArrowArc(0,20)(40,270,330){3} \Photon(0,-20)(0,-40){2}{3}
\Text(-56,-7)[]{$\tau^c$} \Text(-42,-8)[]{$\tau$}
\Text(44,-7)[]{$\mu$} \Text(-49,25)[]{\small{$\rho^{\prime0}$}}
\Text(0,21)[]{\small{$\lambda_B$}}
\Text(-5,10)[]{\small{$\lambda^3_A,\lambda^8_A$}}
\Text(7,-40)[]{$\gamma$}
\Text(0,-11)[]{\small{$\tilde{\ell}_{L_\alpha}$}}\Text(40,30)[]{$(3)$}
\end{picture}
\begin{picture}(120,50)(-65,-30)
\ArrowLine(-63,0)(-49,0) \ArrowLine(-35,0)(-49,0)
\ArrowLine(-35,0)(35,0) \ArrowLine(49,0)(35,0)
\DashArrowLine(-49,20)(-49,0){1}
\DashArrowArc(0,20)(40,210,270){3}
\DashArrowArc(0,20)(40,270,330){3} \Photon(0,-20)(0,-40){2}{3}
\Text(-56,-8)[]{$\tau$} \Text(-42,-7)[]{$\tau^c$}
\Text(44,-8)[]{$\mu^c$} \Text(-49,25)[]{\small{$\rho^{\prime 0}$}}
\Text(0,10)[]{\small{$\lambda_B$}} \Text(7,-35)[]{$\gamma$}
\Text(0,-11)[]{\small{$\tilde{\ell}_{R_\alpha}$}}\Text(40,20)[]{$(4)$}
\end{picture}
\vspace{0.5cm}
 \caption{ Contribution to $D^{\gamma(a)}_L$ [1-3] and $D^{\gamma(a)}_R$
  [4]. } \label{fdgaa1}
 \end{center}
\end{figure}

\begin{figure}[h]
\begin{center}
\begin{picture}(110,80)(-55,-40)
\ArrowLine(-35,0)(-49,0) \ArrowLine(-35,0)(0,0)
\ArrowLine(35,0)(0,0) \ArrowLine(35,0)(49,0)
\Photon(-32,-35)(-20,-15){2}{3} \DashArrowLine(0,-20)(0,-40){1}
\DashArrowArc(0,20)(40,210,240){3}
\DashArrowArc(0,20)(40,240,270){3}
\DashArrowArc(0,20)(40,270,330){3} \Text(-42,-7)[]{$\tau^c$}
\Text(44,-7)[]{$\mu$} \Text(0,8)[]{\small{$\lambda_B$}}
\Text(-20,-35)[]{$\gamma$} \Text(10,-35)[]{\small{$\rho^{\prime
0}$}} \Text(-11,-9)[]{\small{$\tilde{\ell}_{R_\alpha}$}}
\Text(8,-10)[]{\small{$\tilde{\ell}_{L_\beta}$}}
\Text(40,20)[]{$(1)$}
\end{picture}
\begin{picture}(110,80)(-55,-40)
\ArrowLine(-35,0)(-49,0) \ArrowLine(-35,0)(0,0)
\ArrowLine(35,0)(0,0) \ArrowLine(35,0)(49,0)
\Photon(-32,-35)(-20,-15){2}{3} \DashArrowLine(0,-40)(0,-20){1}
\DashArrowArc(0,20)(40,210,240){3}
\DashArrowArc(0,20)(40,240,270){3}
\DashArrowArc(0,20)(40,270,330){3} \Text(-42,-7)[]{$\tau^c$}
\Text(44,-7)[]{$\mu$} \Text(0,8)[]{\small{$\lambda_B$}}
\Text(-20,-35)[]{$\gamma$} \Text(10,-35)[]{\small{$\rho^0$}}
\Text(-11,-9)[]{\small{$\tilde{\ell}_{R_\alpha}$}}
\Text(8,-10)[]{\small{$\tilde{\ell}_{L_\beta}$}}
\Text(40,20)[]{$(2)$}
\end{picture}
\begin{picture}(110,80)(-55,-40)
\ArrowLine(-35,0)(-49,0) \ArrowLine(-35,0)(0,0)
\ArrowLine(35,0)(0,0) \ArrowLine(35,0)(49,0)
\Photon(32,-35)(20,-15){2}{3} \DashArrowLine(0,-20)(0,-40){1}
\DashArrowArc(0,20)(40,210,270){3}
\DashArrowArc(0,20)(40,270,300){3}
\DashArrowArc(0,20)(40,300,330){3} \Text(-42,-7)[]{$\tau^c$}
\Text(44,-7)[]{$\mu$} \Text(0,8)[]{\small{$\lambda_B$}}
\Text(20,-35)[]{$\gamma$} \Text(-10,-35)[]{\small{$\rho^{\prime
0}$}} \Text(-11,-9)[]{\small{$\tilde{\ell}_{R_\alpha}$}}
\Text(10,-9)[]{\small{$\tilde{\ell}_{L_\beta}$}}\Text(40,20)[]{$(3)$}
\end{picture}
\end{center}
\begin{center}
\begin{picture}(110,80)(-55,-40)
\ArrowLine(-35,0)(-49,0) \ArrowLine(-35,0)(0,0)
\ArrowLine(35,0)(0,0) \ArrowLine(35,0)(49,0)
\Photon(32,-35)(20,-15){2}{3} \DashArrowLine(0,-40)(0,-20){1}
\DashArrowArc(0,20)(40,210,270){3}
\DashArrowArc(0,20)(40,270,300){3}
\DashArrowArc(0,20)(40,300,330){3} \Text(-42,-7)[]{$\tau^c$}
\Text(44,-7)[]{$\mu$} \Text(0,8)[]{\small{$\lambda_B$}}
\Text(20,-35)[]{$\gamma$} \Text(-10,-35)[]{\small{$\rho^0$}}
\Text(-11,-9)[]{\small{$\tilde{\ell}_{R_\alpha}$}}
\Text(10,-9)[]{\small{$\tilde{\ell}_{L_\beta}$}}\Text(40,20)[]{$(4)$}
\end{picture}
\begin{picture}(110,80)(-55,-40)
\ArrowLine(-49,0)(-35,0) \ArrowLine(0,0)(-35,0)
\ArrowLine(0,0)(35,0) \ArrowLine(49,0)(35,0)
\Photon(-32,-35)(-20,-15){2}{3} \DashArrowLine(0,-40)(0,-20){1}
\DashArrowArc(0,20)(40,210,240){3}
\DashArrowArc(0,20)(40,240,270){3}
\DashArrowArc(0,20)(40,270,330){3} \Text(-42,-7)[]{$\tau$}
\Text(44,-8)[]{$\mu^c$} \Text(0,8)[]{\small{$\lambda_B$}}
\Text(-20,-35)[]{$\gamma$} \Text(10,-35)[]{\small{$\rho^{\prime
0}$}} \Text(-11,-9)[]{\small{$\tilde{\ell}_{L_\alpha}$}}
\Text(8,-10)[]{\small{$\tilde{\ell}_{R_\beta}$}}\Text(40,20)[]{$(5)$}
\end{picture}
\begin{picture}(110,80)(-55,-40)
\ArrowLine(-49,0)(-35,0) \ArrowLine(0,0)(-35,0)
\ArrowLine(0,0)(35,0) \ArrowLine(49,0)(35,0)
\Photon(-32,-35)(-20,-15){2}{3} \DashArrowLine(0,-20)(0,-40){1}
\DashArrowArc(0,20)(40,210,240){3}
\DashArrowArc(0,20)(40,240,270){3}
\DashArrowArc(0,20)(40,270,330){3} \Text(-42,-7)[]{$\tau$}
\Text(44,-8)[]{$\mu^c$} \Text(0,8)[]{\small{$\lambda_B$}}
\Text(-20,-35)[]{$\gamma$} \Text(10,-35)[]{\small{$\rho^0$}}
\Text(-11,-9)[]{\small{$\tilde{\ell}_{L_\alpha}$}}
\Text(8,-10)[]{\small{$\tilde{\ell}_{R_\beta}$}}\Text(40,20)[]{$(6)$}
\end{picture}
\end{center}
\bc
\begin{picture}(110,80)(-55,-40)
\ArrowLine(-49,0)(-35,0) \ArrowLine(0,0)(-35,0)
\ArrowLine(0,0)(35,0) \ArrowLine(49,0)(35,0)
\Photon(32,-35)(20,-15){2}{3} \DashArrowLine(0,-40)(0,-20){1}
\DashArrowArc(0,20)(40,210,270){3}
\DashArrowArc(0,20)(40,270,300){3}
\DashArrowArc(0,20)(40,300,330){3} \Text(-42,-7)[]{$\tau$}
\Text(44,-8)[]{$\mu^c$} \Text(0,8)[]{\small{$\lambda_B$}}
\Text(20,-35)[]{$\gamma$} \Text(-10,-35)[]{\small{$\rho^{\prime
0}$}} \Text(-11,-9)[]{\small{$\tilde{\ell}_{L_\alpha}$}}
\Text(10,-9)[]{\small{$\tilde{\ell}_{R_\beta}$}}\Text(40,20)[]{$(7)$}
\end{picture}
\begin{picture}(110,80)(-55,-40)
\ArrowLine(-49,0)(-35,0) \ArrowLine(0,0)(-35,0)
\ArrowLine(0,0)(35,0) \ArrowLine(49,0)(35,0)
\Photon(32,-35)(20,-15){2}{3} \DashArrowLine(0,-20)(0,-40){1}
\DashArrowArc(0,20)(40,210,270){3}
\DashArrowArc(0,20)(40,270,300){3}
\DashArrowArc(0,20)(40,300,330){3} \Text(-42,-7)[]{$\tau$}
\Text(44,-8)[]{$\mu^c$} \Text(0,8)[]{\small{$\lambda_B$}}
\Text(20,-35)[]{$\gamma$} \Text(-10,-35)[]{\small{$\rho^0$}}
\Text(-11,-9)[]{\small{$\tilde{\ell}_{L_\alpha}$}}
\Text(10,-9)[]{\small{$\tilde{\ell}_{R_\beta}$}}\Text(40,20)[]{$(8)$}
\end{picture}
  \caption{Contribution to $D^{\gamma(c)}_L$ [1-6] and $D^{\gamma}_R$
  [7,8].}\label{dgac1}
  \ec
\end{figure}
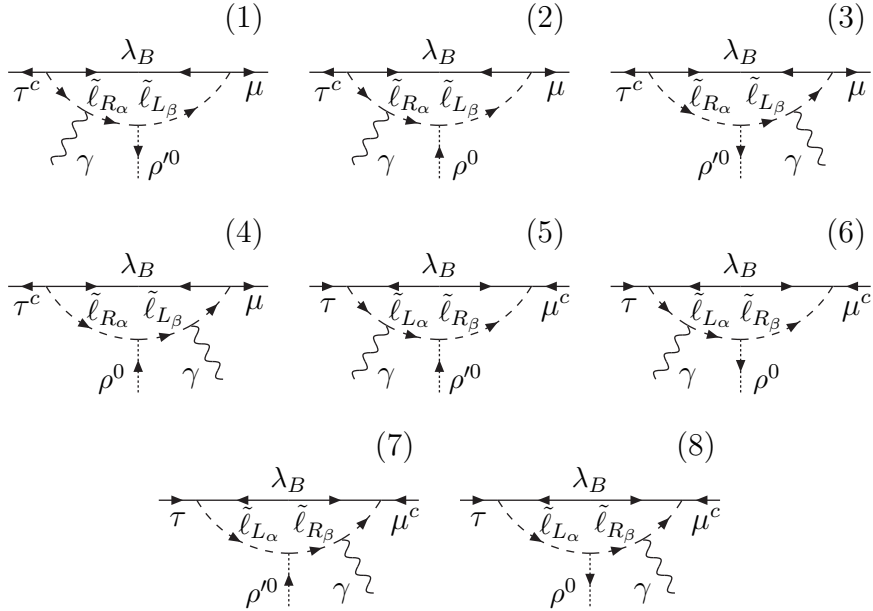

 On the
other hand, $D^{\gamma}$ gets contributions from diagrams with one
line Higgs insertion, figs. \ref{fdgaa1},\ref{dgab1} and
\ref{dgac1}. There is another class of LFV sources relating with
neutrino-mediation in which their contributions
  are very
small \cite{cheng} so we will ignore  them in our
investigation.
 The  $D^{\gamma}_{L,R}$ can be separated into three parts:
 $$ D^{\gamma}_{L,R}=
 D^{\gamma(a)}_{L,R}+D^{\gamma(b)}_{L,R}+D^{\gamma(c)}_{L,R},$$
 where diagrams involving each part are expressed in
  three figs.\ref{fdgaa1}, \ref{dgab1} and
 \ref{dgac1}.

 For $D^{\gamma(a)}$:
\bea
D^{\gamma(a)}_L&=&\frac{g^2c_Ls_L}{16\pi^2}\times\frac{1}{3}\left[
m^2_{\tilde{l}_{L2}}
 J_5(m^2_{\lambda},m^2_{\tilde{l}_{L2}},m^2_{\tilde{l}_{L2}}
 ,m^2_{\tilde{l}_{L2}},m^2_{\tilde{l}_{L2}} )\right]\crn&-&
\frac{g^2c_{\nu_L}s_{\nu_L}}{16\pi^2}\times\frac{1}{2} \left[
m^2_{\lambda}~J_5(m^2_{\lambda},m^2_{\lambda},m^2_{\lambda},
 m^2_{\lambda},m^2_{\tilde{\nu}_{L2}})\right]\crn&-&
 \frac{g^2c_{\nu_R}s_{\nu_R}}{16\pi^2}\times\frac{1}{2}\left[
  m^2_{\lambda}~J_5(m^2_{\lambda},m^2_{\lambda},m^2_{\lambda},
 m^2_{\lambda},m^2_{\tilde{\nu}_{R2}})\right]
\crn&+&\frac{g^{\prime2}c_Ls_L}{16\pi^2}m^2_{\tilde{l}_{L2}}\left[\frac{1}{54}
 J_5(m^2_B,m^2_{\tilde{l}_{L2}},m^2_{\tilde{l}_{L2}}
 ,m^2_{\tilde{l}_{L2}},m^2_{\tilde{l}_{L2}} )\right]\crn&-&[L_2\rightarrow
 L_3,R_2\rightarrow R_3], \crn
 D^{\gamma(a)}_R&=&\frac{g^{\prime2}c_Rs_R}{16\pi^2}m^2_{\tilde{l}_{R2}}\left[\frac{1}{6}
 J_5(m^2_B,m^2_{\tilde{l}_{R2}},m^2_{\tilde{l}_{R2}}
 ,m^2_{\tilde{l}_{R2}},m^2_{\tilde{l}_{R2}} )\right]-[R_2\rightarrow
 R_3].
 \label{dgab2}\eea

\begin{figure}[h]
\vspace{0.5cm}
\begin{center}
\begin{picture}(110,60)(-55,-30)
\ArrowLine(-35,0)(-49,0) \ArrowLine(-35,0)(0,0)
\ArrowLine(20,0)(0,0) \ArrowLine(35,0)(20,0)
\ArrowLine(35,0)(49,0) \DashLine(0,0)(0,15){1}
\DashArrowLine(0,15)(0,30){1} \Photon(20,0)(20,30){2}{3}
\GCirc(20,0){2}{0.5} \DashArrowArc(0,20)(40,210,330){3}
\Text(-42,-8)[]{$\tau^c$} \Text(42,-7)[]{$\mu$}
\Text(-10,25)[]{\small{$\rho^{\prime 0}$}}
\Text(-15,8)[]{\small{$\tilde{\rho}^{\prime -}_1$}}
\Text(10,9)[]{\small{$\tilde{W}^+$}}
\Text(32,9)[]{\small{$\tilde{W}^+$}} \Text(28,25)[]{$\gamma$}
\Text(0,-12)[]{\small{$\tilde{\nu}_{L\alpha}$}}\Text(45,30)[]{$(1)$}
\end{picture}
\begin{picture}(110,60)(-55,-30)
\ArrowLine(-35,0)(-49,0) \ArrowLine(-35,0)(-18,0)
\ArrowLine(-18,0)(0,0) \ArrowLine(35,0)(0,0)
\ArrowLine(35,0)(49,0) \DashLine(0,0)(0,15){1}
\DashArrowLine(0,15)(0,30){1} \Photon(-20,0)(-20,30){2}{3}
\GCirc(-20,0){2}{0.5} \DashArrowArc(0,20)(40,210,330){3}
\Text(-42,-8)[]{$\tau^c$} \Text(42,-7)[]{$\mu$}
\Text(10,25)[]{\small{$\rho^{\prime 0}$}}
\Text(-29,8)[]{\small{$\tilde{\rho}^{\prime -}_1$}}
\Text(-9,8)[]{\small{$\tilde{\rho}^{\prime -}_1$}}
\Text(20,9)[]{\small{$\tilde{W}^+$}} \Text(-28,25)[]{$\gamma$}
\Text(0,-12)[]{\small{$\tilde{\nu}_{L\alpha}$}}\Text(45,30)[]{$(2)$}
\end{picture}
\begin{picture}(110,60)(-55,-30)
\ArrowLine(-35,0)(-49,0) \ArrowLine(-35,0)(0,0)
\ArrowLine(20,0)(0,0) \ArrowLine(35,0)(20,0)
\ArrowLine(35,0)(49,0) \DashLine(0,0)(0,15){1}
\DashArrowLine(0,15)(0,30){1} \Photon(20,0)(20,30){2}{3}
 \GCirc(20,0){2}{0.5}
\DashArrowArcn(0,20)(40,330,210){3} \Text(-42,-8)[]{$\tau^c$}
\Text(42,-7)[]{$\mu$} \Text(-10,25)[]{\small{$\rho^{\prime 0}$}}
\Text(-15,8)[]{\small{$\tilde{\rho}^{\prime -}_2$}}
\Text(10,9)[]{\small{$\tilde{Y}^+$}}
\Text(32,9)[]{\small{$\tilde{Y}^+$}} \Text(28,25)[]{$\gamma$}
\Text(0,-12)[]{\small{$\tilde{\nu}_{R\alpha}$}}\Text(45,30)[]{$(3)$}
\end{picture}
\end{center}
\vspace{0.5 cm} 
\begin{center}
\begin{picture}(110,60)(-55,-30)
\ArrowLine(-35,0)(-49,0) \ArrowLine(-35,0)(-18,0)
\ArrowLine(-18,0)(0,0) \ArrowLine(35,0)(0,0)
\ArrowLine(35,0)(49,0) \DashLine(0,0)(0,15){1}
\DashArrowLine(0,15)(0,30){1} \Photon(-20,0)(-20,30){2}{3}
\GCirc(-20,0){2}{0.5} \DashArrowArcn(0,20)(40,330,210){3}
\Text(-42,-8)[]{$\tau^c$} \Text(42,-7)[]{$\mu$}
\Text(10,25)[]{\small{$\rho^{\prime 0}$}}
\Text(-29,8)[]{\small{$\tilde{\rho}^{\prime -}_2$}}
\Text(-9,8)[]{\small{$\tilde{\rho}^{\prime -}_2$}}
\Text(20,9)[]{\small{$\tilde{Y}^+$}} \Text(-28,25)[]{$\gamma$}
\Text(0,-12)[]{\small{$\tilde{\nu}_{R\alpha}$}}\Text(40,40)[]{$(4)$}
\end{picture}
\begin{picture}(110,60)(-55,-30)
\ArrowLine(-35,0)(-49,0) \ArrowLine(-35,0)(-18,0)
\ArrowLine(0,0)(-18,0) \ArrowLine(0,0)(20,0)
\ArrowLine(35,0)(20,0) \ArrowLine(35,0)(49,0)
\DashLine(0,0)(0,15){1} \DashArrowLine(0,30)(0,15){1}
\Photon(20,0)(20,30){2}{3} \GCirc(20,0){2}{0.5}
\DashArrowArc(0,20)(40,210,330){3} \Text(-42,-8)[]{$\tau^c$}
\Text(42,-7)[]{$\mu$} \Text(-10,25)[]{\small{$\rho^0$}}
\Text(-28,8)[]{\small{$\tilde{\rho}^{\prime -}_1$}}
\Text(-13,8)[]{\small{$\tilde{\rho}^+_1$}}
\Text(10,9)[]{\small{$\tilde{W}^-$}}
\Text(32,9)[]{\small{$\tilde{W}^+$}} \Text(28,25)[]{$\gamma$}
\Text(0,-12)[]{\small{$\tilde{\nu}_{L\alpha}$}}
\Text(40,40)[]{$(5)$}
\end{picture}
\begin{picture}(110,60)(-55,-30)
\ArrowLine(-35,0)(-49,0) \ArrowLine(-35,0)(-20,0)
\ArrowLine(0,0)(-20,0) \ArrowLine(0,0)(20,0)
\ArrowLine(35,0)(20,0) \ArrowLine(35,0)(49,0)
\DashLine(0,0)(0,15){1} \DashArrowLine(0,30)(0,15){1}
\Photon(-20,0)(-20,30){2}{3} \GCirc(-20,0){2}{0.5}
\DashArrowArc(0,20)(40,210,330){3} \Text(-42,-8)[]{$\tau^c$}
\Text(42,-7)[]{$\mu$} \Text(10,25)[]{\small{$\rho^0$}}
\Text(-30,8)[]{\small{$\tilde{\rho}^{\prime -}_1$}}
\Text(-8,8)[]{\small{$\tilde{\rho}^+_1$}}
\Text(12,9)[]{\small{$\tilde{W}^-$}}
\Text(30,9)[]{\small{$\tilde{W}^+$}} \Text(-28,25)[]{$\gamma$}
\Text(0,-12)[]{\small{$\tilde{\nu}_{L\alpha}$}}
\Text(40,40)[]{$(6)$}
\end{picture}
\end{center}
\vspace{0.5 cm}
\begin{center}
\begin{picture}(110,60)(-55,-30)
\ArrowLine(-35,0)(-49,0) \ArrowLine(-35,0)(-18,0)
\ArrowLine(0,0)(-18,0) \ArrowLine(0,0)(20,0)
\ArrowLine(35,0)(20,0) \ArrowLine(35,0)(49,0)
\DashLine(0,0)(0,15){1} \DashArrowLine(0,30)(0,15){1}
\Photon(20,0)(20,30){2}{3} \GCirc(20,0){2}{0.5}
\DashArrowArcn(0,20)(40,330,210){3} \Text(-42,-8)[]{$\tau^c$}
\Text(42,-7)[]{$\mu$} \Text(-10,25)[]{\small{$\rho^0$}}
\Text(-28,8)[]{\small{$\tilde{\rho}^{\prime -}_2$}}
\Text(-13,8)[]{\small{$\tilde{\rho}^+_2$}}
\Text(10,9)[]{\small{$\tilde{Y}^-$}}
\Text(32,9)[]{\small{$\tilde{Y}^+$}} \Text(28,25)[]{$\gamma$}
\Text(0,-12)[]{\small{$\tilde{\nu}_{R\alpha}$}}
\Text(40,40)[]{$(7)$}
\end{picture}
\begin{picture}(110,60)(-55,-30)
\ArrowLine(-35,0)(-49,0) \ArrowLine(-35,0)(-20,0)
\ArrowLine(0,0)(-20,0) \ArrowLine(0,0)(20,0)
\ArrowLine(35,0)(20,0) \ArrowLine(35,0)(49,0)
\DashLine(0,0)(0,15){1} \DashArrowLine(0,30)(0,15){1}
\Photon(-20,0)(-20,30){2}{3} \GCirc(-20,0){2}{0.5}
\DashArrowArcn(0,20)(40,330,210){3} \Text(-42,-8)[]{$\tau^c$}
\Text(42,-7)[]{$\mu$} \Text(10,25)[]{\small{$\rho^0$}}
\Text(-30,8)[]{\small{$\tilde{\rho}^{\prime -}_2$}}
\Text(-8,8)[]{\small{$\tilde{\rho}^+_2$}}
\Text(12,9)[]{\small{$\tilde{Y}^-$}}
\Text(30,9)[]{\small{$\tilde{Y}^+$}} \Text(-28,25)[]{$\gamma$}
\Text(0,-12)[]{\small{$\tilde{\nu}_{R\alpha}$}}
\Text(40,40)[]{$(8)$}
\end{picture}
\begin{picture}(110,70)(-55,-40)
\ArrowLine(-35,0)(-49,0) \ArrowLine(-35,0)(0,0)
\ArrowLine(35,0)(0,0) \ArrowLine(35,0)(49,0)
\DashLine(0,0)(0,15){1} \DashArrowLine(0,15)(0,30){1}
\Photon(0,-20)(0,-40){2}{3} \DashArrowArc(0,20)(40,210,270){3}
\DashArrowArc(0,20)(40,270,330){3} \Text(-42,-8)[]{$\tau^c$}
\Text(42,-7)[]{$\mu$} \Text(-10,25)[]{\small{$\rho^{\prime 0}$}}
\Text(-15,8)[]{\small{$\tilde{\rho}^{\prime 0}$}}
\Text(18,9)[]{\small{$\lambda^3_A,\lambda ^ 8_A $}}
\Text(18,20)[]{\small{$\lambda_B$}} \Text(8,-40)[]{$\gamma$}
\Text(0,-11)[]{\small{$\tilde{\ell}_{L_\alpha}$}}
\Text(40,40)[]{$(9)$}
\end{picture}
\end{center}
\vspace{0.5 cm}
\begin{center}
\begin{picture}(110,70)(-55,-40)
\ArrowLine(-35,0)(-49,0) \ArrowLine(-35,0)(-18,0)
\ArrowLine(0,0)(-18,0) \ArrowLine(0,0)(18,0)
\ArrowLine(35,0)(18,0) \ArrowLine(35,0)(49,0)
\DashLine(0,0)(0,15){1} \DashArrowLine(0,30)(0,15){1}
\Photon(0,-20)(0,-40){2}{3} \DashArrowArc(0,20)(40,210,270){3}
\DashArrowArc(0,20)(40,270,330){3} \Text(-42,-8)[]{$\tau^c$}
\Text(42,-7)[]{$\mu$} \Text(-10,25)[]{\small{$\rho^0$}}
\Text(-27,8)[]{\small{$\tilde{\rho}^{\prime 0}$}}
\Text(-8,8)[]{\small{$\tilde{\rho}^0$}}
\Text(18,9)[]{\small{$\lambda^3_A, \lambda^8_A $}}
\Text(18,20)[]{\small{$\lambda_B$}} \Text(8,-40)[]{$\gamma$}
\Text(0,-11)[]{\small{$\tilde{\ell}_{L_\alpha}$}}
\Text(40,40)[]{$(10)$}
\end{picture}
\begin{picture}(110,70)(-55,-40)
\ArrowLine(-49,0)(-35,0) \ArrowLine(0,0)(-35,0)
\ArrowLine(0,0)(35,0) \ArrowLine(49,0)(35,0)
\DashLine(0,0)(0,15){1} \DashArrowLine(0,30)(0,15){1}
\Photon(0,-20)(0,-40){2}{3} \DashArrowArc(0,20)(40,210,270){3}
\DashArrowArc(0,20)(40,270,330){3} \Text(-42,-7)[]{$\tau$}
\Text(44,-8)[]{$\mu^c$} \Text(-10,25)[]{\small{$\rho^{\prime 0}$}}
\Text(-15,8)[]{\small{$\tilde{\rho}^{\prime 0}$}}
\Text(18,9)[]{\small{$\lambda_B$}} \Text(8,-40)[]{$\gamma$}
\Text(0,-11)[]{\small{$\tilde{\ell}_{R_\alpha}$}}\Text(40,40)[]{$(11)$}
\end{picture}
\begin{picture}(110,70)(-55,-40)
\ArrowLine(-49,0)(-35,0) \ArrowLine(-18,0)(-35,0)
\ArrowLine(-18,0)(0,0) \ArrowLine(18,0)(0,0)
\ArrowLine(18,0)(35,0) \ArrowLine(49,0)(35,0)
\DashLine(0,0)(0,15){1} \DashArrowLine(0,15)(0,30){1}
\Photon(0,-20)(0,-40){2}{3} \DashArrowArc(0,20)(40,210,270){3}
\DashArrowArc(0,20)(40,270,330){3} \Text(-42,-7)[]{$\tau$}
\Text(44,-8)[]{$\mu^c$} \Text(-10,25)[]{\small{$\rho^0$}}
\Text(-26,8)[]{\small{$\tilde{\rho}^{\prime 0}$}}
\Text(-9,8)[]{\small{$\tilde{\rho}^0$}}
\Text(18,9)[]{\small{$\lambda_B$}} \Text(8,-40)[]{$\gamma$}
\Text(0,-11)[]{\small{$\tilde{\ell}_{R_\alpha}$}}\Text(40,40)[]{$(12)$}
\end{picture}
\end{center}
 \caption{\ftsz Contribution to $D^{\gamma(b)}_L$ [1-10] and $D^{\gamma(b)}_R$
  [11,12].}\label{dgab1}
\end{figure}
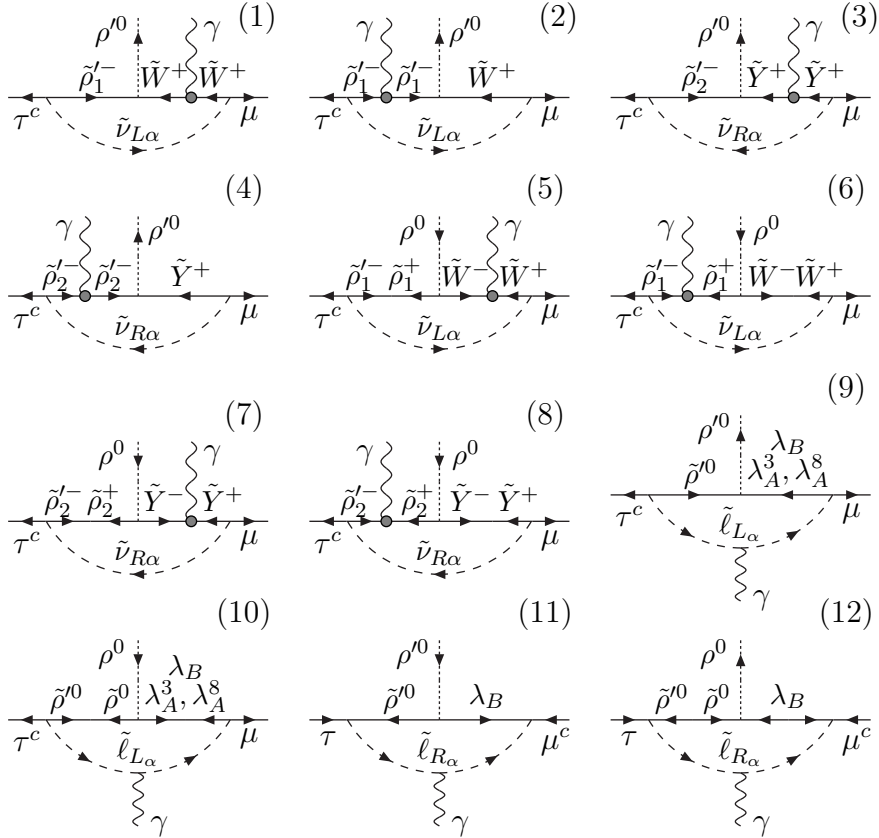
For $D^{\gamma(b)}$:
\bea D^{\gamma(b)}_L&=&
-\frac{g^{2}s_{\nu_L}c_{\nu_L}}{16 \pi^2}
m^4_{\tilde{\nu}_{L2}}I_5(m_{\lambda}^2,\mu^2_{\rho},
m^2_{\tilde{\nu}_{L2}},m^2_{\tilde{\nu}_{L2}},m^2_{\tilde{\nu}_{L2}})
\crn&-& \frac{g^{2}s_{\nu_R}c_{\nu_R}}{16 \pi^2}
m^4_{\tilde{\nu}_{R2}}I_5(m_{\lambda}^2,\mu^2_{\rho},
m^2_{\tilde{\nu}_{R2}},m^2_{\tilde{\nu}_{R2}},m^2_{\tilde{\nu}_{R2}})
\crn&+& \frac{g^2s_{\nu_{L2}}c_{\nu_{L2}}}{16\pi^2}\times
m_{\lambda}~\mu~\tan\gamma
\left[J_5(m_{\lambda}^2,m_{\lambda}^2,\mu^2_{\rho},\mu^2_{\rho},m^2_{\tilde{\nu}_{L2}})
\right.\crn&+&\left.J_5(m_{\lambda}^2,m_{\lambda}^2,m_{\lambda}^2,\mu^2_{\rho},m^2_{\tilde{\nu}_{L2}})
+J_5(m_{\lambda}^2,\mu^2_{\rho},\mu^2_{\rho},\mu^2_{\rho},m^2_{\tilde{\nu}_{L2}})
\right] \crn
&+& \frac{g^2s_{\nu_{R2}}c_{\nu_{R2}}}{16\pi^2}\times
m_{\lambda}~\mu~\tan\gamma
\left[J_5(m_{\lambda}^2,m_{\lambda}^2,\mu^2_{\rho},\mu^2_{\rho},m^2_{\tilde{\nu}_{R2}})
\right.\crn&+&\left.J_5(m_{\lambda}^2,m_{\lambda}^2,m_{\lambda}^2,\mu^2_{\rho},m^2_{\tilde{\nu}_{L2}})
+J_5(m_{\lambda}^2,\mu^2_{\rho},\mu^2_{\rho},\mu^2_{\rho},m^2_{\tilde{\nu}_{R2}})
\right]
\crn&-&
 \frac{g^{2}s_Lc_L}{16 \pi^2}~m^2_{\tilde{l}_{L2}}\times\frac{2}{3} \left[
J_5(m_{\lambda}^2,\mu^2_{\rho},
m^2_{\tilde{l}_{L2}},m^2_{\tilde{l}_{L2}},m^2_{\tilde{l}_{L2}})\right.\crn
&-&
\left.m_{\lambda}~\mu_{\rho}\tan\gamma~I_5(m_{\lambda}^2,\mu^2_{\rho},
m^2_{\tilde{l}_{L2}},m^2_{\tilde{l}_{L2}},m^2_{\tilde{l}_{L2}})\right]
 \crn &+& \frac{g^{\prime2}s_Lc_L}{16 \pi^2}~m^2_{\tilde{l}_{L2}}
\times\frac{2}{27}
\left[J_5(m_B^2,\mu^2_{\rho},m^2_{\tilde{l}_{L2}},m^2_{\tilde{l}_{L2}},m^2_{\tilde{l}_{L2}}
)\right.\crn
&-&\left. m_B~\mu~\tan\gamma~
 I_5(m_B^2,\mu^2_{\rho},m^2_{\tilde{l}_{L2}},m^2_{\tilde{l}_{L2}},m^2_{\tilde{l}_{L2}})
  \right]\crn&-&[L_2\rightarrow L_3],\crn
 D^{\gamma(b)}_R&=&\frac{g^{\prime2}s_Rc_R}{16
\pi^2} m^2_{\tilde{l}_{R2}} \times\frac{2}{9}
\left[-J_5(m_B^2,\mu^2_{\rho},m^2_{\tilde{l}_{R2}},m^2_{\tilde{l}_{R2}},m^2_{\tilde{l}_{R2}}
)\right.\crn&+& m_B~\mu_{\rho}\tan\gamma~
I_5(m_B^2,\mu^2_{\rho},m^2_{\tilde{l}_{R2}},m^2_{\tilde{l}_{R2}},m^2_{\tilde{l}_{R2}}
)\left. \right]\crn&-&[R_2\rightarrow R_3].\label{dgab1}\eea

For $D^{\gamma(c)}$:
 \bea
 D^{\gamma(c)}_L= -\frac{g'}{16\pi^2}~\frac{m^3_B}{9}&\times&
 \left\{ \left[ s_Lc_L\left(s^2_R[A_{\tau}+\frac{1}{2}\mu_{\rho}
 \tan\gamma]+s_Rc_RA^R_{\mu\tau}\right)+ c^2_Ls^2_RA^L_{\mu\tau}\right]\right.\crn
 &\times& \left.I_5(m_B^2,m_B^2,m_B^2,m^2_{\tilde{l}_{L2}},
 m^2_{\tilde{l}_{R2}}) \right.\crn
 &-&\left.
 \left[ s_Lc_L\left(s^2_R[A_{\tau}+\frac{1}{2}\mu_{\rho}
 \tan\gamma]+s_Rc_RA^R_{\mu\tau}\right)-s^2_Ls^2_RA^L_{\mu\tau}\right]\right.\crn
 &\times& \left.I_5(m_B^2,m_B^2,m_B^2,m^2_{\tilde{l}_{L3}},
 m^2_{\tilde{l}_{R2}}) \right.\crn
 &+&\left.\left[ s_Lc_L\left(c^2_R[A_{\tau}+\frac{1}{2}\mu_{\rho}
 \tan\gamma]-s_Rc_RA^R_{\mu\tau}\right)+c^2_Lc^2_RA^L_{\mu\tau}\right]\right.\crn
 &\times& \left.I_5(m_B^2,m_B^2,m_B^2,m^2_{\tilde{l}_{L2}},
 m^2_{\tilde{l}_{R3}}) \right.\crn
 &-&\left. \left[ s_Lc_L\left(c^2_R[A_{\tau}+\frac{1}{2}\mu_{\rho}
 \tan\gamma]-s_Rc_RA^R_{\mu\tau}\right)-s^2_Lc^2_RA^L_{\mu\tau}\right]\right.\crn
 &\times& \left.I_5(m_B^2,m_B^2,m_B^2,m^2_{\tilde{l}_{L3}},
 m^2_{\tilde{l}_{R3}})\right\},
 \crn
 D^{\gamma(c)}_R= -\frac{g'}{16\pi^2}~\frac{m^3_B}{9}&\times&
 \left\{ \left[ s_Rc_R\left(s^2_L[A_{\tau}+\frac{1}{2}\mu_{\rho}
 \tan\gamma]+s_Lc_LA^L_{\mu\tau}\right)+ c^2_Rs^2_LA^R_{\mu\tau}\right]\right.\crn
 &\times& \left.I_5(m_B^2,m_B^2,m_B^2,m^2_{\tilde{l}_{L2}},
 m^2_{\tilde{l}_{R2}}) \right.\crn
 &-&\left.
 \left[ s_Rc_R\left(s^2_L[A_{\tau}+\frac{1}{2}\mu_{\rho}
 \tan\gamma]+s_Lc_LA^L_{\mu\tau}\right)-s^2_Rs^2_LA^R_{\mu\tau}\right]\right.\crn
 &\times& \left.I_5(m_B^2,m_B^2,m_B^2,m^2_{\tilde{l}_{L2}},
 m^2_{\tilde{l}_{R3}}) \right.\crn
 &+&\left.\left[ s_Rc_R\left(c^2_L[A_{\tau}+\frac{1}{2}\mu_{\rho}
 \tan\gamma]-s_Lc_LA^L_{\mu\tau}\right)+c^2_Rc^2_LA^R_{\mu\tau}\right]\right.\crn
 &\times& \left.I_5(m_B^2,m_B^2,m_B^2,m^2_{\tilde{l}_{L3}},
 m^2_{\tilde{l}_{R2}}) \right.\crn
 &-&\left. \left[ s_Rc_R\left(c^2_L[A_{\tau}+\frac{1}{2}\mu_{\rho}
 \tan\gamma]-s_Lc_LA^L_{\mu\tau}\right)-s^2_Rc^2_LA^R_{\mu\tau}\right]\right.\crn
 &\times& \left.I_5(m_B^2,m_B^2,m_B^2,m^2_{\tilde{l}_{L3}},
 m^2_{\tilde{l}_{R3}})\right\}.\label{dga(c)1}\eea

\section{\label{apzmutau}Contributions to $Z\rightarrow \mu\tau$}
In this appendix, we draw all the possible diagrams which
contribute to the effective operator $Z\rightarrow \mu\tau$ in the
limit of assumption given out in \cite{Anna2}. All of these
diagrams can be applied to  the case of $Z'$ boson.
\subsection{\label{apazmutau}Contributions to $A^Z_{L,R}$ \label{a3}}
\begin{figure}[h]
\begin{center}
\begin{picture}(200,70)(-10,-30)
\ArrowLine(-49,0)(-35,0) \ArrowLine(-21,0)(-35,0)
\ArrowLine(-21,0)(-7,0) \ArrowLine(7,0)(-7,0)
\ArrowLine(7,0)(21,0) \ArrowLine(35,0)(21,0)
\ArrowLine(35,0)(49,0) \DashArrowLine(-21,30)(-21,15){1}
\DashLine(-21,15)(-21,0){1} \DashArrowLine(7,30)(7,15){1}
\DashLine(7,15)(7,0){1} \DashArrowArc(0,20)(40,210,330){3}
\Text(-42,-7)[]{$\tau$} \Text(42,-7)[]{$\mu$}
\Text(-29,10)[]{\ftsz{$\tilde{W}^+$}}
\Text(-13,9)[]{\ftsz{$\tilde{\rho}_1^{\prime-}$}}
\Text(0,9)[]{\ftsz{$\tilde{\rho}_1^+$}}
\Text(16,10)[]{\ftsz{$\tilde{W}^-$}}
\Text(32,10)[]{\ftsz{$\tilde{W}^+$}}
\Text(-21,32)[b]{\small{$\rho{\prime^0}$}}
\Text(7,32)[b]{\small{$\rho^0$}}
\Text(0,-12)[]{\small{$\tilde{\nu}_{L\alpha}$}}
\Text(42,30)[]{(1)}
\end{picture}
\begin{picture}(180,70)(80,-30)
\ArrowLine(-49,0)(-35,0) \ArrowLine(-21,0)(-35,0)
\ArrowLine(-21,0)(-7,0) \ArrowLine(7,0)(-7,0)
\ArrowLine(7,0)(21,0) \ArrowLine(35,0)(21,0)
\ArrowLine(35,0)(49,0) \DashArrowLine(-7,15)(-7,30){1}
\DashLine(-7,15)(-7,0){1} \DashArrowLine(21,15)(21,30){1}
\DashLine(21,15)(21,0){1} \DashArrowArc(0,20)(40,210,330){3}
\Text(-42,-7)[]{$\tau$} \Text(42,-7)[]{$\mu$}
\Text(-30,10)[]{\ftsz{$\tilde{W}^+$}}
\Text(-14,10)[]{\ftsz{$\tilde{W}^-$}}
\Text(1,9)[]{\ftsz{$\tilde{\rho}_1^+$}}
\Text(15,9)[]{\ftsz{$\tilde{\rho}_1^{\prime-}$}}
\Text(32,10)[]{\ftsz{$\tilde{W}^+$}}
\Text(-7,32)[b]{\small{$\rho^0$}}
\Text(21,32)[b]{\small{$\rho^{\prime0}$}}
\Text(0,-12)[]{\small{$\tilde{\nu}_{L_\alpha}$}}\Text(42,30)[]{(2)}
\end{picture}
\begin{picture}(0,70)(150,-30)
\ArrowLine(-49,0)(-35,0) \ArrowLine(-9,0)(-35,0)
\ArrowLine(-9,0)(9,0) \ArrowLine(35,0)(9,0) \ArrowLine(35,0)(49,0)
\DashArrowLine(-9,30)(-9,15){1} \DashLine(-9,15)(-9,0){1}
\DashArrowLine(9,15)(9,30){1} \DashLine(9,15)(9,0){1}
\DashArrowArc(0,20)(40,210,330){3} \Text(-42,-7)[]{$\tau$}
\Text(42,-7)[]{$\mu$} \Text(-21,10)[]{\ftsz{$\tilde{W}^+$}}
\Text(0,9)[]{\ftsz{$\tilde{\rho}_1^{\prime-}$}}
\Text(24,10)[]{\ftsz{$\tilde{W}^+$}}
\Text(-9,32)[b]{\small{$\rho^{\prime0}$}}
\Text(9,32)[b]{\small{$\rho^{\prime0}$}}
\Text(0,-12)[]{\small{$\tilde{\nu}_{L_\alpha}$}}
\Text(42,30)[]{(3)}
\end{picture}
\begin{picture}(-50,70)(40,-30)
\ArrowLine(-49,0)(-35,0) \ArrowLine(-22,0)(-35,0)
\ArrowLine(-22,0)(-9,0) \ArrowLine(9,0)(-9,0)
\ArrowLine(9,0)(22,0) \ArrowLine(35,0)(22,0)
\ArrowLine(35,0)(49,0) \DashArrowLine(-9,15)(-9,30){1}
\DashLine(-9,15)(-9,0){1} \DashArrowLine(9,30)(9,15){1}
\DashLine(9,15)(9,0){1} \DashArrowArc(0,20)(40,210,330){3}
\Text(-42,-7)[]{$\tau$} \Text(42,-7)[]{$\mu$}
\Text(-32,10)[]{\ftsz{$\tilde{W}^+$}}
\Text(-16,10)[]{\ftsz{$\tilde{W}^-$}}
\Text(0,9)[]{\ftsz{$\tilde{\rho}_1^+$}}
\Text(18,10)[]{\ftsz{$\tilde{W}^-$}}
\Text(34,10)[]{\ftsz{$\tilde{W}^+$}}
\Text(-9,32)[b]{\small{$\rho^0$}} \Text(9,32)[b]{\small{$\rho^0$}}
\Text(0,-12)[]{\small{$\tilde{\nu}_{L_\alpha}$}}
\Text(42,30)[]{(4)}
\end{picture}
\end{center}
\begin{center}
\begin{picture}(200,70)(-10,-30)
\ArrowLine(-49,0)(-35,0) \ArrowLine(-21,0)(-35,0)
\ArrowLine(-21,0)(-7,0) \ArrowLine(7,0)(-7,0)
\ArrowLine(7,0)(21,0) \ArrowLine(35,0)(21,0)
\ArrowLine(35,0)(49,0) \DashArrowLine(-21,30)(-21,15){1}
\DashLine(-21,15)(-21,0){1} \DashArrowLine(7,30)(7,15){1}
\DashLine(7,15)(7,0){1} \DashArrowArcn(0,20)(40,330,210){3}
\Text(-42,-7)[]{$\tau$} \Text(42,-7)[]{$\mu$}
\Text(-29,10)[]{\ftsz{$\tilde{Y}^+$}}
\Text(-13,9)[]{\ftsz{$\tilde{\rho}_2^{\prime-}$}}
\Text(0,9)[]{\ftsz{$\tilde{\rho}_2^+$}}
\Text(16,10)[]{\ftsz{$\tilde{Y}^-$}}
\Text(32,10)[]{\ftsz{$\tilde{Y}^+$}}
\Text(-21,32)[b]{\small{$\rho{\prime^0}$}}
\Text(7,32)[b]{\small{$\rho^0$}}
\Text(0,-12)[]{\small{$\tilde{\nu}_{R_\alpha}$}}
\Text(42,30)[]{(5)}
\end{picture}
\begin{picture}(180,70)(80,-30)
\ArrowLine(-49,0)(-35,0) \ArrowLine(-21,0)(-35,0)
\ArrowLine(-21,0)(-7,0) \ArrowLine(7,0)(-7,0)
\ArrowLine(7,0)(21,0) \ArrowLine(35,0)(21,0)
\ArrowLine(35,0)(49,0) \DashArrowLine(-7,15)(-7,30){1}
\DashLine(-7,15)(-7,0){1} \DashArrowLine(21,15)(21,30){1}
\DashLine(21,15)(21,0){1} \DashArrowArcn(0,20)(40,330,210){3}
\Text(-42,-7)[]{$\tau$} \Text(42,-7)[]{$\mu$}
\Text(-30,10)[]{\ftsz{$\tilde{Y}^+$}}
\Text(-14,10)[]{\ftsz{$\tilde{Y}^-$}}
\Text(1,9)[]{\ftsz{$\tilde{\rho}_2^+$}}
\Text(15,9)[]{\ftsz{$\tilde{\rho}_2^{\prime-}$}}
\Text(32,10)[]{\ftsz{$\tilde{Y}^+$}}
\Text(-7,32)[b]{\small{$\rho^0$}}
\Text(21,32)[b]{\small{$\rho^{\prime0}$}}
\Text(0,-12)[]{\small{$\tilde{\nu}_{R_\alpha}$}}
\Text(42,30)[]{(6)}
\end{picture}
\begin{picture}(0,70)(150,-30)
\ArrowLine(-49,0)(-35,0) \ArrowLine(-9,0)(-35,0)
\ArrowLine(-9,0)(9,0) \ArrowLine(35,0)(9,0) \ArrowLine(35,0)(49,0)
\DashArrowLine(-9,30)(-9,15){1} \DashLine(-9,15)(-9,0){1}
\DashArrowLine(9,15)(9,30){1} \DashLine(9,15)(9,0){1}
\DashArrowArcn(0,20)(40,330,210){3} \Text(-42,-7)[]{$\tau$}
\Text(42,-7)[]{$\mu$} \Text(-21,10)[]{\ftsz{$\tilde{Y}^+$}}
\Text(0,9)[]{\ftsz{$\tilde{\rho}_2^{\prime-}$}}
\Text(24,10)[]{\ftsz{$\tilde{Y}^+$}}
\Text(-9,32)[b]{\small{$\rho^{\prime0}$}}
\Text(9,32)[b]{\small{$\rho^{\prime0}$}}
\Text(0,-12)[]{\small{$\tilde{\nu}_{R_\alpha}$}}
\Text(42,30)[]{(7)}
\end{picture}
\begin{picture}(-50,70)(40,-30)
\ArrowLine(-49,0)(-35,0) \ArrowLine(-22,0)(-35,0)
\ArrowLine(-22,0)(-9,0) \ArrowLine(9,0)(-9,0)
\ArrowLine(9,0)(22,0) \ArrowLine(35,0)(22,0)
\ArrowLine(35,0)(49,0) \DashArrowLine(-9,15)(-9,30){1}
\DashLine(-9,15)(-9,0){1} \DashArrowLine(9,30)(9,15){1}
\DashLine(9,15)(9,0){1} \DashArrowArcn(0,20)(40,330,210){3}
\Text(-42,-7)[]{$\tau$} \Text(42,-7)[]{$\mu$}
\Text(-32,10)[]{\ftsz{$\tilde{Y}^+$}}
\Text(-16,10)[]{\ftsz{$\tilde{Y}^-$}}
\Text(0,9)[]{\ftsz{$\tilde{\rho}_2^+$}}
\Text(18,10)[]{\ftsz{$\tilde{Y}^-$}}
\Text(34,10)[]{\ftsz{$\tilde{Y}^+$}}
\Text(-9,32)[b]{\small{$\rho^0$}} \Text(9,32)[b]{\small{$\rho^0$}}
\Text(0,-12)[]{\small{$\tilde{\nu}_{R_\alpha}$}}
\Text(42,30)[]{(8)}
\end{picture}
\end{center}

\begin{center}
\begin{picture}(200,70)(0,-30)
\ArrowLine(-49,0)(-35,0) \ArrowLine(-9,0)(-35,0)
\ArrowLine(-9,0)(9,0) \ArrowLine(35,0)(9,0) \ArrowLine(35,0)(49,0)
\DashArrowLine(-9,30)(-9,15){1} \DashLine(-9,15)(-9,0){1}
\DashArrowLine(9,15)(9,30){1} \DashLine(9,15)(9,0){1}
\DashArrowArc(0,20)(40,210,330){3} \Text(-42,-7)[]{$\tau$}
\Text(42,-7)[]{$\mu$} \Text(-21,10)[]{\ftsz{$\lambda_i$}}
\Text(0,9)[]{\ftsz{$\tilde{H}^0_k$}}
\Text(24,10)[]{\ftsz{$\lambda_i$}}
\Text(-9,32)[b]{\small{$H^0_k$}} \Text(9,32)[b]{\small{$H^0_k$}}
\Text(0,-11)[]{\small{$\tilde{\ell}_{L_\alpha}$}}
\Text(42,30)[]{(9)}
\end{picture}
\begin{picture}(180,70)(80,-30)
\ArrowLine(-49,0)(-35,0) \ArrowLine(-22,0)(-35,0)
\ArrowLine(-22,0)(-9,0) \ArrowLine(9,0)(-9,0)
\ArrowLine(9,0)(22,0) \ArrowLine(35,0)(22,0)
\ArrowLine(35,0)(49,0) \DashArrowLine(-9,15)(-9,30){1}
\DashLine(-9,15)(-9,0){1} \DashArrowLine(9,30)(9,15){1}
\DashLine(9,15)(9,0){1} \DashArrowArc(0,20)(40,210,330){3}
\Text(-42,-7)[]{$\tau$} \Text(42,-7)[]{$\mu$}
\Text(-21,10)[]{\ftsz{$\lambda_i$}}
\Text(0,9)[]{\ftsz{$\tilde{H}^0_k$}}
\Text(24,10)[]{\ftsz{$\lambda_i$}}
\Text(-9,32)[b]{\small{$H^0_k$}} \Text(9,32)[b]{\small{$H^0_k$}}
\Text(0,-11)[]{\small{$\tilde{\ell}_{L_\alpha}$}}
\Text(42,30)[]{(10)}
\end{picture}
\begin{picture}(0,70)(150,-30)
\ArrowLine(-49,0)(-35,0) \ArrowLine(-9,0)(-35,0)
\ArrowLine(-9,0)(9,0) \ArrowLine(35,0)(9,0) \ArrowLine(35,0)(49,0)
\DashArrowLine(-9,30)(-9,15){1} \DashLine(-9,15)(-9,0){1}
\DashArrowLine(9,15)(9,30){1} \DashLine(9,15)(9,0){1}
\DashArrowArc(0,20)(40,210,330){3} \Text(-42,-7)[]{$\tau$}
\Text(42,-7)[]{$\mu$} \Text(-21,10)[]{\ftsz{$\lambda_i$}}
\Text(0,9)[]{\ftsz{$\tilde{H}^0_k$}}
\Text(24,10)[]{\ftsz{$\lambda_j$}}
\Text(-9,32)[b]{\small{$H^0_k$}} \Text(9,32)[b]{\small{$H^0_k$}}
\Text(0,-11)[]{\small{$\tilde{\ell}_{L_\alpha}$}}
\Text(42,30)[]{(11)}
\end{picture}
\begin{picture}(-50,70)(40,-30)
\ArrowLine(-49,0)(-35,0) \ArrowLine(-22,0)(-35,0)
\ArrowLine(-22,0)(-9,0) \ArrowLine(9,0)(-9,0)
\ArrowLine(9,0)(22,0) \ArrowLine(35,0)(22,0)
\ArrowLine(35,0)(49,0) \DashArrowLine(-9,15)(-9,30){1}
\DashLine(-9,15)(-9,0){1} \DashArrowLine(9,30)(9,15){1}
\DashLine(9,15)(9,0){1} \DashArrowArc(0,20)(40,210,330){3}
\Text(-42,-7)[]{$\tau$} \Text(42,-7)[]{$\mu$}
\Text(-21,10)[]{\ftsz{$\lambda_i$}}
\Text(0,9)[]{\ftsz{$\tilde{H}^0_k$}}
\Text(24,10)[]{\ftsz{$\lambda_j$}}
\Text(-9,32)[b]{\small{$H^0_k$}} \Text(9,32)[b]{\small{$H^0_k$}}
\Text(0,-11)[]{\small{$\tilde{\ell}_{L_\alpha}$}}
\Text(42,30)[]{(12)}
\end{picture}
\end{center}
%
%
\begin{center}
\begin{picture}(110,70)(-55,-30)
\ArrowLine(-35,0)(-49,0) \ArrowLine(-35,0)(-9,0)
\ArrowLine(9,0)(-9,0) \ArrowLine(9,0)(35,0) \ArrowLine(49,0)(35,0)
\DashArrowLine(-9,15)(-9,30){1} \DashLine(-9,15)(-9,0){1}
\DashArrowLine(9,30)(9,15){1} \DashLine(9,15)(9,0){1}
\DashArrowArc(0,20)(40,210,330){3} \Text(-42,-7)[]{$\tau^c$}
\Text(44,-8)[]{$\mu^c$} \Text(-21,10)[]{\ftsz{$\lambda_B$}}
\Text(0,9)[]{\ftsz{$\tilde{H}^0_k$}}
\Text(24,10)[]{\ftsz{$\lambda_B$}}
\Text(-9,32)[b]{\small{$H^0_k$}} \Text(9,32)[b]{\small{$H^0_k$}}
\Text(0,-11)[]{\small{$\tilde{\ell}_{R_\alpha}$}}
\Text(42,30)[]{(13)}
\end{picture}
\begin{picture}(110,70)(-55,-30)
\ArrowLine(-35,0)(-49,0) \ArrowLine(-35,0)(-22,0)
\ArrowLine(-9,0)(-22,0) \ArrowLine(-9,0)(9,0)
\ArrowLine(22,0)(9,0) \ArrowLine(22,0)(35,0)
\ArrowLine(49,0)(35,0) \DashArrowLine(-9,30)(-9,15){1}
\DashLine(-9,15)(-9,0){1} \DashArrowLine(9,15)(9,30){1}
\DashLine(9,15)(9,0){1} \DashArrowArc(0,20)(40,210,330){3}
\Text(-42,-7)[]{$\tau^c$} \Text(44,-8)[]{$\mu^c$}
\Text(-21,10)[]{\ftsz{$\lambda_B$}}
\Text(0,9)[]{\ftsz{$\tilde{H}^0_k$}}
\Text(24,10)[]{\ftsz{$\lambda_B$}}
\Text(-9,32)[b]{\small{$H^0_k$}} \Text(9,32)[b]{\small{$H^0_k$}}
\Text(0,-11)[]{\small{$\tilde{\ell}_{R_\alpha}$}}
\Text(42,30)[]{(14)}
\end{picture}
\end{center}
\caption{\ftsz Diagrams contributing to $A^{Z (a)}_L$ (or $A^{1Z'
(a)}_L$) (first, second and third rows) and $A^{Z (a)}_R$
 (or  $A^{1Z' (a)}_R$) (fourth row). Here we denote
$H^0_k\in\{\rho^0,~\rho^{\prime0}\}$ while $\lambda_{i,j}$ implies
$i,j= \{B,3,8\}$ and $i\neq j$.}
\label{alra}
\end{figure}
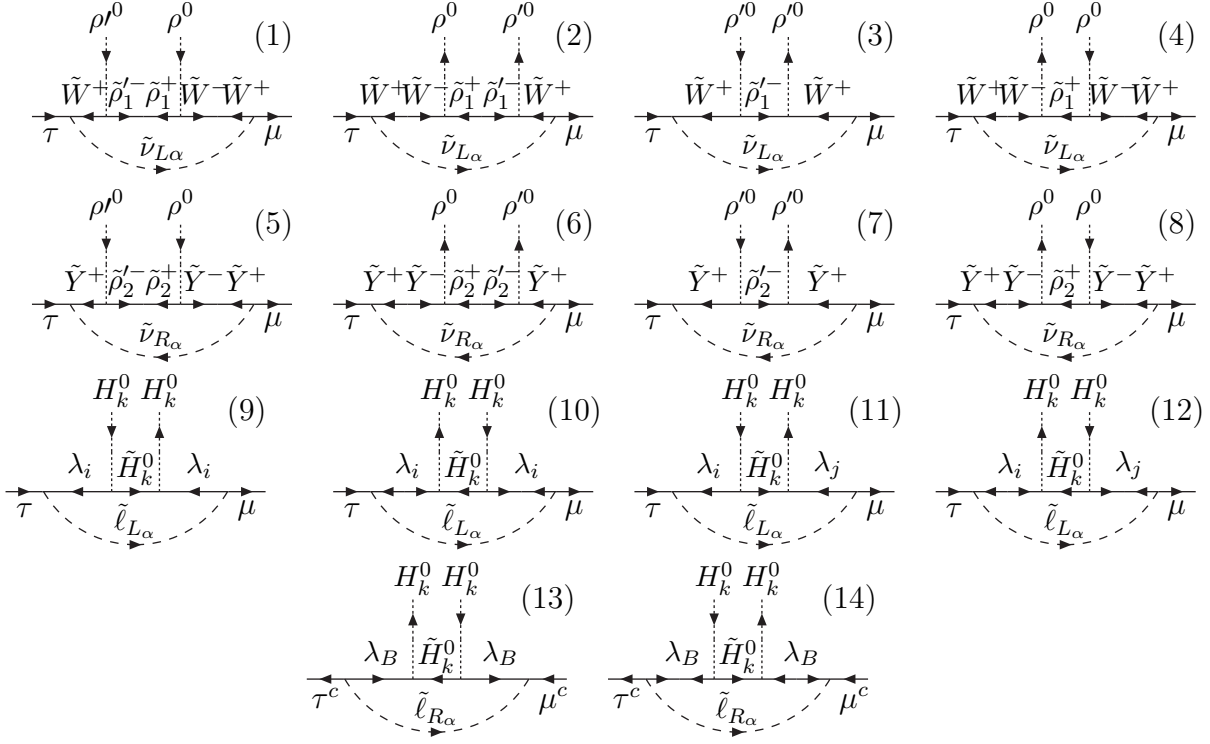

Diagrams  contributing to $A^{Z(a)}_{L,R}$ are shown in
Fig.\ref{alra}.
 The formulas are:
\bea  A^{Z(a)}_{L}&=&
 (s_{\nu_L}c_{\nu_L})\times \frac{g^2c^2_W}{16\pi^2}  \times
\frac{1}{4} (1+c_{2\gamma}) \crn&\times&\left[-\mu^2_{\rho}
J_5(m^2_{\lambda},m^2_{\lambda},\mu^2_{\rho},\mu^2_{\rho},m^2_{\tilde{\nu}_{L2}})-
2
J_4(m^2_{\lambda},m^2_{\lambda},\mu^2_{\rho},m^2_{\tilde{\nu}_{L2}})\right]
\crn&+& (s_{\nu_R}c_{\nu_R})\times \frac{g^2c^2_W}{16\pi^2} \times
\frac{1}{4}(1+c_{2\gamma})\crn&\times&\left[-\mu^2_{\rho}
J_5(m^2_{\lambda},m^2_{\lambda},\mu^2_{\rho},\mu^2_{\rho},m^2_{\tilde{\nu}_{R2}})-
2J_4(m^2_{\lambda},m^2_{\lambda},\mu^2_{\rho},m^2_{\tilde{\nu}_{R2}})\right]
\crn&+&(s_Lc_L)\times\frac{g^2 c^2_W}{16\pi^2} \times
\frac{11}{36}c_{2\gamma}\crn&\times&\left[-\mu^2_{\rho}
J_5(m^2_\lambda,m^2_\lambda,\mu_\rho^2,\mu_\rho^2,m^2_{\tilde{l}_{L2}})
- 2
J_4(m^2_\lambda,m^2_\lambda,\mu_\rho^2,m^2_{\tilde{l}_{L2}})\right.\crn
&+&\left. m_\lambda^2\left(I_4(m_\lambda^2,m_\lambda^2,
\mu_\rho^2, m^2_{\tilde{l}_{L2}}) -\mu_\rho^2
I_5(m_\lambda^2,m_\lambda^2,\mu_\rho^2,\mu_\rho^2,
m^2_{\tilde{l}_{L2}})\right) \right]
\crn &-&(s_{\nu_L}c_{\nu_L})\times\frac{g^2c^2_W
m^2_{\lambda}}{16\pi^2}\times \frac{1}{4}(1-c_{2\gamma})
\crn&\times&\left[\mu_{\rho}^2
I_5(m_{\lambda}^2,m_{\lambda}^2,\mu_{\rho}^2,\mu_{\rho}^2,
m^2_{\tilde{\nu}_{L2}})-
I_4(m_{\lambda}^2,m_{\lambda}^2,\mu_{\rho}^2,
m^2_{\tilde{\nu}_{L2}})\right]
\crn &-& (s_{\nu_R}c_{\nu_R})\times\frac{g^2c^2_W
m^2_{\lambda}}{16\pi^2}\times \frac{1}{4}(1-c_{2\gamma})
\crn&\times&\left[\mu_{\rho}^2
I_5(m_{\lambda}^2,m_{\lambda}^2,\mu_{\rho}^2,\mu_{\rho}^2,
m^2_{\tilde{\nu}_{R2}})-
I_4(m_{\lambda}^2,m_{\lambda}^2,\mu_{\rho}^2,
m^2_{\tilde{\nu}_{R2}})\right]
\crn &+&  (s_Lc_L)\times\frac{g^{\prime2}c^2_W}{16\pi^2}\times
\frac{8}{81}c_{2\gamma}\crn &\times& \left\{\mu^2_{\rho}
J_5(m^2_B,m^2_\lambda,\mu_\rho^2,\mu_\rho^2,m^2_{\tilde{l}_{L2}})
+ 2 J_4(m^2_B,m^2_\lambda,\mu_\rho^2,m^2_{\tilde{l}_{L2}}) \right.
\crn &-& \left. m_B m_\lambda \left[\mu_\rho^2
I_5(m_B^2,m_\lambda^2,\mu_\rho^2,\mu_\rho^2,
m^2_{\tilde{l}_{L2}})-I_4(m_B^2,m_\lambda^2,\mu_\rho^2,
m_{\tilde{l}_{L2}})\right]\right\}
\crn &+&(s_{\nu_L}c_{\nu_L})\times \frac{g^2c^2_W}{16\pi^2}
 \times\frac{1}{2}  s_{2\gamma}\left[\mu_{\rho}~ m_{\lambda}~
 J_5(m_{\lambda}^2,m_{\lambda}^2,\mu^2_{\rho},\mu^2_{\rho},m^2_{\tilde{\nu}_{L2}})\right]
\crn &+&(s_{\nu_R}c_{\nu_R})
 \times\frac{g^2c^2_W}{16\pi^2}\times
 \frac{1}{2} s_{2\gamma}\left[\mu_{\rho}~ m_{\lambda}~
  J_5(m_{\lambda}^2,m_{\lambda}^2,\mu^2_{\rho},\mu^2_{\rho},m^2_{\tilde{\nu}_{R2}})\right]
\crn &+&(s_Lc_L)\times \frac{g^{\prime2}t^2c^2_W}{16\pi^2} \times
\frac{2}{729}c_{2\gamma}
  \crn&\times& \left[-\mu^2_{\rho}
J_5(m_B^2,m_B^2,\mu_\rho^2,\mu_\rho^2,m^2_{\tilde{l}_{L2}}) - 2
J_4(m_B^2,m_B^2,\mu_\rho^2,m^2_{\tilde{l}_{L2}})\right.\crn&+&\left.
m^2_{B}\left(\mu_\rho^2 I_5(m_B^2,m_B^2,\mu_\rho^2,\mu_\rho^2,
m^2_{\tilde{l}_{L2}})-I_4(m_B^2,m_B^2, \mu_\rho^2,
m^2_{\tilde{l}_{L2}}) \right)\right]\crn &-& (L_2\rightarrow L_3,
R_2\rightarrow R_3),\label{azL1} \eea
\bea A^{Z(a)}_{R}&=& (s_Rc_R) \frac{g^{\prime2}t^2c^2_W
}{16\pi^2}\times \frac{2}{81}c_{2\gamma} \crn&\times&
    \left\{\mu^2_{\rho}
J_5(m_B^2,m_B^2,\mu_{\rho}^2,\mu_{\rho}^2,m^2_{\tilde{l}_{R2}}) +
2 J_4(m_B^2,m_B^2,\mu_{\rho}^2,m^2_{\tilde{l}_{R2}}) \right. \crn
&-& \left. m_B^2 \left[\mu_{\rho}^2
I_5(m_B^2,m_B^2,\mu_{\rho}^2,\mu_{\rho}^2,
m^2_{\tilde{l}_{R2}})\right.
\right.\crn&-&\left.\left.I_4(m_B^2,m_B^2,\mu_{\rho}^2,
m^2_{\tilde{l}_{R2}}) \right]\right\}-\left[R_2\rightarrow
R_3\right]. \label{azr1}\eea
For $A^{Z(b,c)}_{L,R}$, see Fig. \ref{alrbc331}:

\begin{figure}[h]
\begin{center}
\begin{picture}(200,80)(-10,-40)
\ArrowLine(-49,0)(-35,0) \ArrowLine(-17,0)(-35,0)
\ArrowLine(-17,0)(0,0) \ArrowLine(35,0)(0,0)
\ArrowLine(35,0)(49,0) \DashArrowLine(0,15)(0,30){1}
\DashLine(0,15)(0,0){1} \DashArrowLine(0,-40)(0,-20){1}
\DashArrowArc(0,20)(40,210,270){3}
\DashArrowArc(0,20)(40,270,330){3} \Text(-42,-7)[]{$\tau$}
\Text(44,-7)[]{$\mu$}
\Text(-27,9)[]{\ftsz{$\tilde{\rho}^{\prime0}$}}
\Text(-9,9)[]{\ftsz{$\tilde{\rho}^0$}}
\Text(20,21)[]{\ftsz{$(\lambda_3,\lambda_8)$}}
\Text(17,10)[]{\ftsz{$\lambda_B$}}
\Text(-10,30)[t]{\small{$\rho^0$}}
\Text(-10,-40)[b]{\small{$\rho^0$}}
\Text(-25,-20)[]{\small{$\tilde{\ell}_{R_\alpha}$}}
\Text(35,-20)[]{\small{$\tilde{\ell}_{L_\beta}$}}
\Text(42,33)[]{(1)}
\end{picture}
\begin{picture}(180,80)(80,-40)
\ArrowLine(-35,0)(-49,0) \ArrowLine(-35,0)(-17,0)
\ArrowLine(0,0)(-17,0) \ArrowLine(0,0)(35,0)
\ArrowLine(49,0)(35,0) \DashArrowLine(0,30)(0,15){1}
\DashLine(0,15)(0,0){1} \DashArrowLine(0,-20)(0,-40){1}
\DashArrowArc(0,20)(40,210,270){3}
\DashArrowArc(0,20)(40,270,330){3} \Text(-42,-7)[]{$\tau^c$}
\Text(44,-8)[]{$\mu^c$}
\Text(-27,9)[]{\ftsz{$\tilde{\rho}^{\prime0}$}}
\Text(-9,9)[]{\ftsz{$\tilde{\rho}^0$}}
\Text(17,10)[]{\ftsz{$\lambda_B$}}
\Text(-10,30)[t]{\small{$\rho^0$}}
\Text(-10,-40)[b]{\small{$\rho^0$}}
\Text(-25,-20)[]{\small{$\tilde{\ell}_{L_\alpha}$}}
\Text(35,-20)[]{\small{$\tilde{\ell}_{R_\beta}$}}
\Text(42,33)[]{(2)}
\end{picture}
%
%
\begin{picture}(0,80)(150,-40)
\ArrowLine(-49,0)(-35,0) \ArrowLine(35,0)(-35,0)
\ArrowLine(35,0)(49,0) \DashArrowLine(-20,-15)(-32,-35){1}
\DashArrowLine(32,-35)(20,-15){1}
\DashArrowArc(0,20)(40,210,240){3}
\DashArrowArc(0,20)(40,240,300){3}
\DashArrowArc(0,20)(40,300,330){3} \Text(-44,-7)[]{$\tau$}
\Text(44,-7)[]{$\mu$} \Text(0,10)[]{\ftsz{$\lambda_3$}}
\Text(0,21)[]{\ftsz{$(\lambda_B,\lambda_8)$}}
\Text(-20,-35)[]{\small{$\rho^0$}}
\Text(20,-35)[]{\small{$\rho^0$}}
\Text(-30,-12)[]{\small{$\tilde{\ell}_{L_\alpha}$}}
\Text(36,-17)[]{\small{$\tilde{\ell}_{L_\gamma}$}}
\Text(0,-11)[]{\small{$\tilde{\ell}_{R_\beta}$}}
\Text(42,33)[]{(3)}
\end{picture}
\begin{picture}(-50,80)(40,-40)
\ArrowLine(-35,0)(-49,0) \ArrowLine(-35,0)(35,0)
\ArrowLine(49,0)(35,0) \DashArrowLine(-32,-35)(-20,-15){1}
\DashArrowLine(20,-15)(32,-35){1}
\DashArrowArc(0,20)(40,210,240){3}
\DashArrowArc(0,20)(40,240,300){3}
\DashArrowArc(0,20)(40,300,330){3} \Text(-44,-7)[]{$\tau^c$}
\Text(44,-8)[]{$\mu^c$} \Text(0,10)[]{\ftsz{$\lambda_B$}}
\Text(-20,-35)[]{\small{$\rho^0$}}
\Text(20,-35)[]{\small{$\rho^0$}}
\Text(-30,-12)[]{\small{$\tilde{\ell}_{R_\alpha}$}}
\Text(36,-17)[]{\small{$\tilde{\ell}_{R_\gamma}$}}
\Text(0,-11)[]{\small{$\tilde{\ell}_{L_\beta}$}}
\Text(42,33)[]{(4)}
\end{picture}
\end{center}
\caption{\ftsz Diagrams contributing to $A^{Z (b)}_{L,R}$ (left
side) and $A^{Z (c)}_{L,R}$ (right side) in SUSYE331.}
\label{alrbc331}
\end{figure}
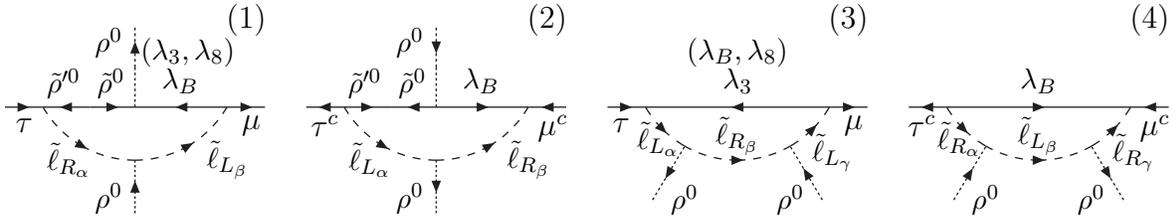
Formula for $A^{Z(b)}_L$:
 \bea A^{Z(b)}_L &=&(s_Lc_L)\times\frac{m^2_{\tau}c^2_W}{16\pi^2 V^2}\times\frac{1}{3}
   ( t^2_{\gamma}~\mu^2_{\rho})\left[s^2_R\left(J_5(m^2_\lambda,\mu^2_{\rho},\mu^2_{\rho},m^2_{\tilde{l}_{L_2}},
m^2_{\tilde{l}_{R_2}})\right.\right.\crn&+&\left.
\left.J_5(m^2_\lambda,\mu_{\rho}^2,m^2_{\tilde{l}_{L_2}},m^2_{\tilde{l}_{R_2}},
m^2_{\tilde{l}_{R_2}})\right)+c^2_R\left(J_5(m^2_\lambda,\mu^2_{\rho},\mu^2_{\rho},m^2_{\tilde{l}_{L_2}},
m^2_{\tilde{l}_{R_3}})\right.\right.\crn&+&\left.
\left.J_5(m^2_\lambda,\mu_{\rho}^2,m^2_{\tilde{l}_{L_2}},m^2_{\tilde{l}_{R_3}},
m^2_{\tilde{l}_{R_3}})\right)\right]
 \crn&+&(s_Lc_L)\times\frac{m^2_{\tau}t^2c^2_W }{16\pi^2~V^2}\times\frac{1}{27}
     (t^2_{\gamma}~\mu^2_{\rho})\left[-s^2_R\left(J_5(m^2_B,\mu^2_{\rho},\mu^2_{\rho},m^2_{\tilde{l}_{L2}},
m^2_{\tilde{l}_{R2}})\right.\right.\crn&+&\left.\left.
J_5(m^2_B,\mu_{\rho}^2,m^2_{\tilde{l}_{L2}},m^2_{\tilde{l}_{R2}},
m^2_{\tilde{l}_{R2}})\right)-c^2_R\left(J_5(m^2_B,\mu^2_{\rho},\mu^2_{\rho},m^2_{\tilde{l}_{L2}},
m^2_{\tilde{l}_{R3}})\right.\right.\crn&+&\left.\left.
J_5(m^2_B,\mu_{\rho}^2,m^2_{\tilde{l}_{L2}},m^2_{\tilde{l}_{R3}},
m^2_{\tilde{l}_{R3}})\right) \right]-(L_2\rightarrow L_3),\eea
where  $m_{\tau}= Y_{\tau}\times v'/\sqrt{2}$ is mass of the tau
and $V\equiv v_{\mathrm{weak}}= \sqrt{v^2+v^{'2}}$ in the SUSYE331
model. We also have formula of $ A^{Z(b)}_R$:
\bea
 A^{Z(b)}_R
&=&(s_Rc_R)\times\frac{m_{\tau}^2t^2c^2_W}{16\pi^2~V^2} \times
\frac{1}{9}
t^2_{\gamma}\mu^2_{\rho}\crn&\times&\left[-s^2_L\left(J_5(m^2_B,\mu^2_{\rho},\mu^2_{\rho},m^2_{\tilde{l}_{L_2}},
m^2_{\tilde{l}_{R_2}})\right.\right.\crn&&~~~~~~+
\left.\left.J_5(m^2_B,\mu_{\rho}^2,m^2_{\tilde{l}_{L_2}},m^2_{\tilde{l}_{L_2}},
m^2_{\tilde{l}_{R_2}})\right)\right.
\crn&&~-\left.c^2_L\left(J_5(m^2_B,\mu^2_{\rho},\mu^2_{\rho},m^2_{\tilde{l}_{L_3}},
m^2_{\tilde{l}_{R_2}})\right.\right.\crn&&~~~~~~+
\left.\left.J_5(m^2_B,\mu_{\rho}^2,m^2_{\tilde{l}_{L_3}},m^2_{\tilde{l}_{L_3}},
m^2_{\tilde{l}_{R_2}})\right)\right]-\left(R_2\rightarrow
R_3\right). \label{AzbR1}\eea

Formulas for $A^{Z(c)}$:
\bea  A^{Z(c)}_L&=& (s_Lc_L)\times
\frac{m^2_{\tau}t^2c^2_W}{16\pi^2~V^2}\times
\frac{1}{6}(t^2_{\gamma}\mu^2_{\rho})\crn&\times& \left\{
s^2_L\left[s^2_RJ_5(m^2_\lambda,
m^2_{\tilde{l}_{R_2}},m^2_{\tilde{l}_{R_2}},
 m^2_{\tilde{l}_{L_2}},m^2_{\tilde{l}_{L_2}})\right.\right.\crn
 &+&\left.\left.c^2_RJ_5(m^2_\lambda, m^2_{\tilde{l}_{R_3}},m^2_{\tilde{l}_{R_3}},
 m^2_{\tilde{l}_{L_2}},m^2_{\tilde{l}_{L_2}})\right]\right. \crn &-&\left.
 c^2_L\left[s^2_RJ_5(m^2_\lambda, m^2_{\tilde{l}_{R_2}},m^2_{\tilde{l}_{R_2}},
 m^2_{\tilde{l}_{L_3}},m^2_{\tilde{l}_{L_3}})\right.\right.\crn
 &+&\left.\left.c^2_RJ_5(m^2_\lambda, m^2_{\tilde{l}_{R_3}},m^2_{\tilde{l}_{R_3}},
 m^2_{\tilde{l}_{L_3}},m^2_{\tilde{l}_{L_3}})\right]\right. \crn &-&\left.
 \left(s^2_L-c^2_L\right)\left[s^2_RJ_5(m^2_\lambda, m^2_{\tilde{l}_{R_2}},m^2_{\tilde{l}_{R_2}},
 m^2_{\tilde{l}_{L_2}},m^2_{\tilde{l}_{L_3}})\right.\right.\crn
 &+&\left.\left.c^2_RJ_5(m^2_\lambda, m^2_{\tilde{l}_{R_3}},m^2_{\tilde{l}_{R_3}},
 m^2_{\tilde{l}_{L_2}},m^2_{\tilde{l}_{L_3}})\right]\right\}
 \crn&+&(s_Lc_L)\times \frac{m^2_{\tau}t^2c^2_W}{16\pi^2~V^2}\times
\frac{1}{108}(t^2_{\gamma}\mu^2_{\rho})\crn&\times& \left\{
s^2_L\left[s^2_RJ_5(m^2_B,
m^2_{\tilde{l}_{R_2}},m^2_{\tilde{l}_{R_2}},
 m^2_{\tilde{l}_{L_2}},m^2_{\tilde{l}_{L_2}})\right.\right.\crn
 &+&\left.\left.c^2_RJ_5(m^2_B, m^2_{\tilde{l}_{R_3}},m^2_{\tilde{l}_{R_3}},
 m^2_{\tilde{l}_{L_2}},m^2_{\tilde{l}_{L_2}})\right]\right. \crn &-&\left.
 c^2_L\left[s^2_RJ_5(m^2_B, m^2_{\tilde{l}_{R_2}},m^2_{\tilde{l}_{R_2}},
 m^2_{\tilde{l}_{L_3}},m^2_{\tilde{l}_{L_3}})\right.\right.\crn
 &+&\left.\left.c^2_RJ_5(m^2_B, m^2_{\tilde{l}_{R_3}},m^2_{\tilde{l}_{R_3}},
 m^2_{\tilde{l}_{L_3}},m^2_{\tilde{l}_{L_3}})\right]\right. \crn &-&\left.
 \left(s^2_L-c^2_L\right)\left[s^2_RJ_5(m^2_B, m^2_{\tilde{l}_{R_2}},m^2_{\tilde{l}_{R_2}},
 m^2_{\tilde{l}_{L_2}},m^2_{\tilde{l}_{L_3}})\right.\right.\crn
 &+&\left.\left.c^2_RJ_5(m^2_B, m^2_{\tilde{l}_{R_3}},m^2_{\tilde{l}_{R_3}},
 m^2_{\tilde{l}_{L_2}},m^2_{\tilde{l}_{L_3}})\right]\right\}\eea
\bea A^{Z(c)}_R&=& (s_Rc_R)\times
\frac{m^2_{\tau}t^2c^2_W}{16\pi^2~V^2}\times
\frac{1}{12}(t^2_{\gamma}\mu^2_{\rho})\crn&\times& \left\{ -
s^2_R\left[s^2_LJ_5(m^2_B,
m^2_{\tilde{l}_{R_2}},m^2_{\tilde{l}_{R_2}},
 m^2_{\tilde{l}_{L_2}},m^2_{\tilde{l}_{L_2}})\right.\right.\crn
 &+&\left.\left.c^2_LJ_5(m^2_B, m^2_{\tilde{l}_{R_2}},m^2_{\tilde{l}_{R_2}},
 m^2_{\tilde{l}_{L_3}},m^2_{\tilde{l}_{L_3}})\right]\right. \crn &+&\left.
 c^2_R\left[s^2_LJ_5(m^2_B, m^2_{\tilde{l}_{R_3}},m^2_{\tilde{l}_{R_3}},
 m^2_{\tilde{l}_{L_2}},m^2_{\tilde{l}_{L_2}})\right.\right.\crn
 &+&\left.\left.c^2_LJ_5(m^2_B, m^2_{\tilde{l}_{R_3}},m^2_{\tilde{l}_{R_3}},
 m^2_{\tilde{l}_{L_3}},m^2_{\tilde{l}_{L_3}})\right]\right. \crn &+&\left.
 \left(s^2_R-c^2_R\right)\left[s^2_LJ_5(m^2_B, m^2_{\tilde{l}_{R_2}},m^2_{\tilde{l}_{R_3}},
 m^2_{\tilde{l}_{L_2}},m^2_{\tilde{l}_{L_2}})\right.\right.\crn
 &+&\left.\left.c^2_LJ_5(m^2_B, m^2_{\tilde{l}_{R_2}},m^2_{\tilde{l}_{R_3}},
 m^2_{\tilde{l}_{L_3}},m^2_{\tilde{l}_{L_3}})\right]\right\}.\label{azc1}\eea

\subsection{\label{apczmutau}Contributions to $C^Z_{L,R}$}

\begin{figure}[h]
\begin{center}
\begin{picture}(110,80)(-55,-40)
\ArrowLine(-49,0)(-35,0) \ArrowLine(0,0)(-35,0)
\ArrowLine(35,0)(0,0) \ArrowLine(35,0)(49,0)
\Photon(0,0)(0,25){2}{3} \GCirc(0,0){2}{0.5}
\DashArrowArc(0,20)(40,210,330){3} \Text(-42,-7)[]{$\tau$}
\Text(42,-7)[]{$\mu$} \Text(-17,10)[]{\ftsz{$\tilde{W}^+$}}
\Text(17,10)[]{\ftsz{$\tilde{W}^+$}} \Text(8,25)[]{$Z$}
\Text(0,-12)[]{\small{$\tilde{\nu}_{\alpha}$}} \Text(30,30)[]{(1)}
\end{picture}
\begin{picture}(110,80)(-55,-40)
\ArrowLine(-49,0)(-35,0) \ArrowLine(35,0)(-35,0)
\ArrowLine(35,0)(49,0) \DashArrowArc(0,20)(40,210,270){3}
\DashArrowArc(0,20)(40,270,330){3} \Photon(0,-20)(0,-40){2}{3}
\Text(-44,-7)[]{$\tau$} \Text(44,-7)[]{$\mu$}
\Text(0,10)[]{\ftsz{$\tilde{W}^+$}} \Text(7,-40)[]{$Z$ }
\Text(20,-40)[]{($Z'$)}
\Text(0,-12)[]{\small{$\tilde{\nu}_{\alpha}$}} \Text(30,25)[]{(2)}
\end{picture}
\begin{picture}(110,80)(-55,-40)
\ArrowLine(-49,0)(-35,0) \ArrowLine(35,0)(-35,0)
\ArrowLine(35,0)(49,0) \DashArrowArc(0,20)(40,210,270){3}
\DashArrowArc(0,20)(40,270,330){3} \Photon(0,-20)(0,-40){2}{3}
\Text(-44,-7)[]{$\tau$} \Text(44,-7)[]{$\mu$}
\Text(0,21)[]{\ftsz{$(\lambda_3,\lambda_8)$}}
\Text(0,10)[]{\ftsz{$\lambda_B$}}
\Text(7,-40)[]{$Z$}\Text(20,-40)[]{($Z'$)}
\Text(0,-11)[]{\small{$\tilde{\ell}_{L_\alpha}$}}\Text(30,25)[]{(3)}
\end{picture}
\end{center}
\begin{center}
\begin{picture}(110,80)(-55,-40)
\ArrowLine(-49,0)(-35,0) \ArrowLine(0,0)(-35,0)
\ArrowLine(35,0)(0,0) \ArrowLine(35,0)(49,0)
\Photon(0,0)(0,25){2}{3} \GCirc(0,0){2}{0.5}
\DashArrowArcn(0,20)(40,330,210){3} \Text(-42,-7)[]{$\tau$}
\Text(42,-7)[]{$\mu$} \Text(-17,10)[]{\ftsz{$\tilde{Y}^+$}}
\Text(17,10)[]{\ftsz{$\tilde{Y}^+$}}
\Text(8,25)[]{$Z$}\Text(8,35)[]{($Z'$)}
\Text(0,-12)[]{\small{$\tilde{\nu}_{R\alpha}$}}
\Text(30,25)[]{(4)}
\end{picture}
\begin{picture}(110,80)(-55,-40)
\ArrowLine(-49,0)(-35,0) \ArrowLine(35,0)(-35,0)
\ArrowLine(35,0)(49,0) \DashArrowArcn(0,20)(40,270,210){3}
\DashArrowArcn(0,20)(40,330,270){3} \Photon(0,-20)(0,-40){2}{3}
\Text(-44,-7)[]{$\tau$} \Text(44,-7)[]{$\mu$}
\Text(0,10)[]{\ftsz{$\tilde{Y}^+$}} \Text(7,-40)[]{$Z'$}
\Text(0,-12)[]{\small{$\tilde{\nu}_{R\alpha}$}}\Text(30,25)[]{(5)}
\end{picture}
\begin{picture}(110,80)(-55,-40)
\ArrowLine(-35,0)(-49,0) \ArrowLine(-35,0)(35,0)
\ArrowLine(49,0)(35,0) \DashArrowArc(0,20)(40,210,270){3}
\DashArrowArc(0,20)(40,270,330){3} \Photon(0,-20)(0,-40){2}{3}
\Text(-44,-7)[]{$\tau^c$} \Text(44,-8)[]{$\mu^c$}
\Text(0,10)[]{\ftsz{$\lambda_B$}} \Text(7,-40)[]{$Z$}
\Text(23,-40)[]{$(Z')$}
\Text(0,-11)[]{\small{$\tilde{\ell}_{R_\alpha}$}}\Text(30,30)[]{(6)}
\end{picture}
\hspace{1cm}
\end{center}
\caption{\ftsz Diagrams contributing to $C^Z_{L,R}$
($C^{Z'}_{L,R}$ ). Only the last gives contribution to $C^Z_{R}$
($C^{Z'}_{R}$). The first diagram only contributes to $C^Z_{L}$
while the fifth only contributes to $C^{Z'}_{L}$.}
\label{cz}
\end{figure}
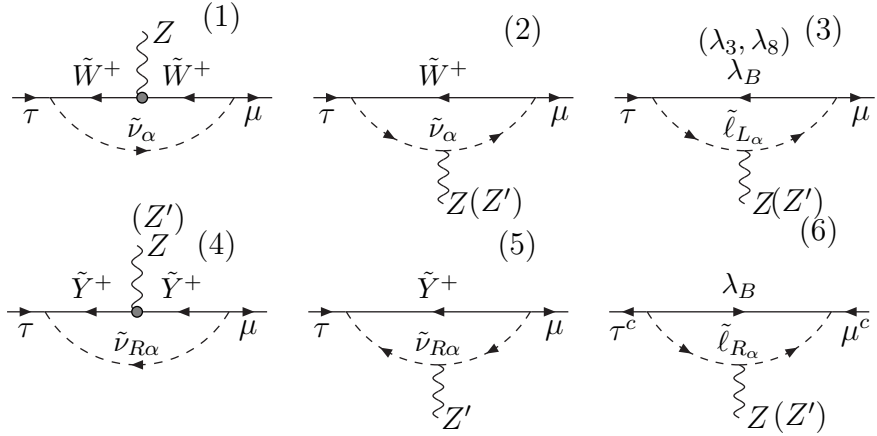

For $C^Z_{L,R}$, in Fig. \ref{cz}.
The formulas for these two quantities are written as below: \bea
    C^{Z}_{L} &=&
    (c_{\nu_L}s_{\nu_L})\times \frac{g^{2}}{16\pi^2}\times\frac{1}{12}\left[-K_5
    (m^{2}_{\lambda},m^2_{\tilde{\nu}_{L2}},m^2_{\tilde{\nu}_{L2}},
    m^2_{\tilde{\nu}_{L2}},m^2_{\tilde{\nu}_{L2}})\right] \crn
    &+& (c_Ls_L)\times\frac{g^2 }{16\pi^2}\times \frac{1}{18}c_{2W}
\left[-K_5
(m^2_{\lambda},m^2_{\tilde{l}_{L2}},m^2_{\tilde{l}_{L2}},m^2_{\tilde{l}_{L2}},m^2_{\tilde{l}_{L2}})
\right]\crn
&+&(c_{\nu_L}s_{\nu_L})\times\frac{g^2}{16\pi^2}\times
\frac{1}{12}c^2_W\left[-2K_{5}(m^2_{\lambda},m^2_{\lambda},m^2_{\lambda},m^2_{\lambda},
\tilde{m}^2_{\nu_{L2}})\right.\crn&+&\left. 3 m^2_{\lambda}
J_5(m^2_{\lambda},m^2_{\lambda},m^2_{\lambda},m^2_{\lambda},
\tilde{m}^2_{\nu_{L2}})\right]\crn
&+&(c_{\nu_R}s_{\nu_R})\times\frac{g^2 }{16\pi^2}\times
\frac{c_{2W}}{12}
\left[-2K_{5}(m^2_{\lambda},m^2_{\lambda},m^2_{\lambda},m^2_{\lambda}
,\tilde{m}^2_{\nu_{R2}})\right.\crn&+&\left.3 m^2_{\lambda}
J_5(m^2_{\lambda},m^2_{\lambda},m^2_{\lambda},m^2_{\lambda}
,\tilde{m}^2_{\nu_{R2}})\right]\crn
 &+&(c_Ls_L )\times\frac{g^{\prime2}}{16
\pi^2}\times\frac{1}{324}(1-2s^2_W) \left[K_5
    (m^2_B,m^2_{\tilde{l}_{L2}},m^2_{\tilde{l}_{L2}},
    m^2_{\tilde{l}_{L2}},m^2_{\tilde{l}_{L2}})\right]\crn
    &-&\left[L_2\rightarrow L_3,~R_2 \rightarrow R_3\right],\eea
 \bea
     C^{Z}_{R} &=& (c_Rs_R)\times\frac{g^{\prime2}}{16\pi^2}\times\frac{1}{36}s^2_W\left[K_5
    (m^{2}_B,m^2_{\tilde{l}_{R2}},m^2_{\tilde{l}_{R2}},
    m^2_{\tilde{l}_{R2}},m^2_{\tilde{l}_{R2}})\right]\crn&-&[R_2\rightarrow R_3].\label{czlr1}\eea
We note that because $Z$ boson couples much weakly to right-handed
neutrinos so the diagram 5 in the Fig.\ref{cz} give suppressed
contribution to $C^Z_L$. In contrast, the case of $Z'$ boson is
different, it
 weakly couples
 with $\tilde{W}^\pm$ but non-negligible to right-handed
neutrinos. So for the $C^{Z'}_L$, we neglect the first diagram and
keep the fifth. This conclusion is held in  the case of $D^{Z}$
and $D^{Z'}$.

\subsection{\label{apdzmutau}Contributions to $D^Z_{L,R}$}

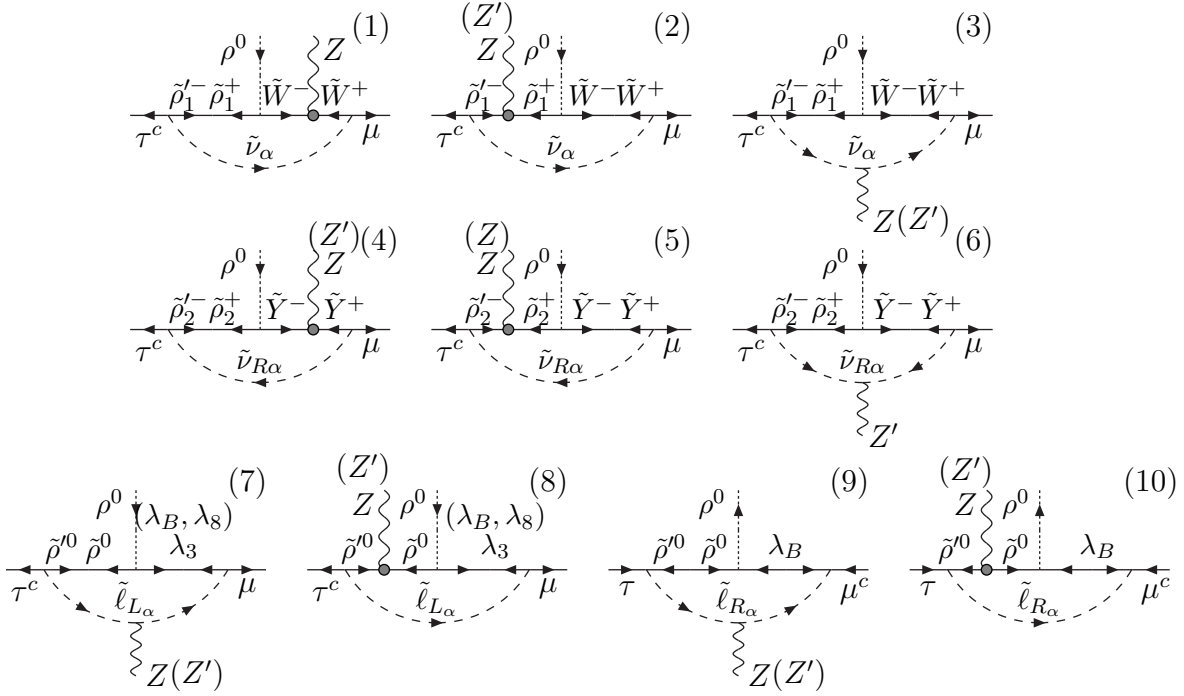
\begin{figure}[h]
\vspace{1cm}
\begin{center}
\begin{picture}(110,60)(-55,-50)
\ArrowLine(-35,0)(-49,0) \ArrowLine(-35,0)(-18,0)
\ArrowLine(0,0)(-18,0) \ArrowLine(0,0)(20,0)
\ArrowLine(35,0)(20,0) \ArrowLine(35,0)(49,0)
\DashLine(0,0)(0,15){1} \DashArrowLine(0,30)(0,15){1}
\Photon(20,0)(20,30){2}{3} \GCirc(20,0){2}{0.5}
\DashArrowArc(0,20)(40,210,330){3} \Text(-42,-8)[]{$\tau^c$}
\Text(42,-7)[]{$\mu$} \Text(-10,25)[]{\small{$\rho^0$}}
\Text(-27,8)[]{\ftsz{$\tilde{\rho}^{\prime-}_1$}}
\Text(-13,8)[]{\ftsz{$\tilde{\rho}^{+}_1$}}
\Text(10,9)[]{\ftsz{$\tilde{W}^-$}}
\Text(32,9)[]{\ftsz{$\tilde{W}^+$}} \Text(28,25)[]{$Z$}
\Text(0,-12)[]{\small{$\tilde{\nu}_{\alpha}$}} \Text(42,33)[]{(1)}
\end{picture}
\begin{picture}(110,60)(-55,-50)
\ArrowLine(-35,0)(-49,0) \ArrowLine(-35,0)(-20,0)
\ArrowLine(0,0)(-20,0) \ArrowLine(0,0)(20,0)
\ArrowLine(35,0)(20,0) \ArrowLine(35,0)(49,0)
\DashLine(0,0)(0,15){1} \DashArrowLine(0,30)(0,15){1}
\Photon(-20,0)(-20,30){2}{3} \GCirc(-20,0){2}{0.5}
\DashArrowArc(0,20)(40,210,330){3} \Text(-42,-8)[]{$\tau^c$}
\Text(42,-7)[]{$\mu$} \Text(-9,25)[]{\small{$\rho^0$}}
\Text(-29,8)[]{\ftsz{$\tilde{\rho}^{\prime-}_1$}}
\Text(-8,8)[]{\ftsz{$\tilde{\rho}^+_1$}}
\Text(12,9)[]{\ftsz{$\tilde{W}^-$}}
\Text(30,9)[]{\ftsz{$\tilde{W}^+$}}
\Text(-28,25)[]{$Z$}\Text(-28,37)[]{$(Z')$}
\Text(0,-12)[]{\small{$\tilde{\nu}_{\alpha}$}}\Text(42,33)[]{(2)}
\end{picture}
\begin{picture}(110,60)(-55,-50)
\ArrowLine(-35,0)(-49,0) \ArrowLine(-35,0)(-18,0)
\ArrowLine(0,0)(-18,0) \ArrowLine(0,0)(18,0)
\ArrowLine(35,0)(18,0) \ArrowLine(35,0)(49,0)
\DashLine(0,0)(0,15){1} \DashArrowLine(0,30)(0,15){1}
\Photon(0,-20)(0,-40){2}{3} \DashArrowArc(0,20)(40,210,270){3}
\DashArrowArc(0,20)(40,270,330){3} \Text(-42,-8)[]{$\tau^c$}
\Text(42,-7)[]{$\mu$} \Text(-10,25)[]{\small{$\rho^0$}}
\Text(-27,8)[]{\ftsz{$\tilde{\rho}^{\prime-}_1$}}
\Text(-13,8)[]{\ftsz{$\tilde{\rho}^+_1$}}
\Text(12,9)[]{\ftsz{$\tilde{W}^-$}}
\Text(30,9)[]{\ftsz{$\tilde{W}^+$}}
\Text(8,-40)[]{$Z$}\Text(23,-40)[]{$(Z')$}
\Text(0,-12)[]{\small{$\tilde{\nu}_{\alpha}$}} \Text(42,33)[]{(3)}
\end{picture}
\end{center}
\begin{center}
\begin{picture}(110,80)(-55,-50)
\ArrowLine(-35,0)(-49,0) \ArrowLine(-35,0)(-18,0)
\ArrowLine(0,0)(-18,0) \ArrowLine(0,0)(20,0)
\ArrowLine(35,0)(20,0) \ArrowLine(35,0)(49,0)
\DashLine(0,0)(0,15){1} \DashArrowLine(0,30)(0,15){1}
\Photon(20,0)(20,30){2}{3} \GCirc(20,0){2}{0.5}
\DashArrowArcn(0,20)(40,330,210){3} \Text(-42,-8)[]{$\tau^c$}
\Text(42,-7)[]{$\mu$} \Text(-10,25)[]{\small{$\rho^0$}}
\Text(-27,8)[]{\ftsz{$\tilde{\rho}^{\prime-}_2$}}
\Text(-13,8)[]{\ftsz{$\tilde{\rho}^+_2$}}
\Text(10,9)[]{\ftsz{$\tilde{Y}^-$}}
\Text(32,9)[]{\ftsz{$\tilde{Y}^+$}}
\Text(28,25)[]{$Z$}\Text(28,35)[]{$(Z')$}
\Text(0,-12)[]{\small{$\tilde{\nu}_{R\alpha}$}}
\Text(45,33)[]{(4)}
\end{picture}
\begin{picture}(110,80)(-55,-50)
\ArrowLine(-35,0)(-49,0) \ArrowLine(-35,0)(-20,0)
\ArrowLine(0,0)(-20,0) \ArrowLine(0,0)(20,0)
\ArrowLine(35,0)(20,0) \ArrowLine(35,0)(49,0)
\DashLine(0,0)(0,15){1} \DashArrowLine(0,30)(0,15){1}
\Photon(-20,0)(-20,30){2}{3} \GCirc(-20,0){2}{0.5}
\DashArrowArcn(0,20)(40,330,210){3} \Text(-42,-8)[]{$\tau^c$}
\Text(42,-7)[]{$\mu$} \Text(-9,25)[]{\small{$\rho^0$}}
\Text(-29,8)[]{\ftsz{$\tilde{\rho}^{\prime-}_2$}}
\Text(-8,8)[]{\ftsz{$\tilde{\rho}^+_2$}}
\Text(12,9)[]{\ftsz{$\tilde{Y}^-$}}
\Text(30,9)[]{\ftsz{$\tilde{Y}^+$}} \Text(-28,25)[]{$Z$}
\Text(-28,35)[]{$(Z)$}
\Text(0,-12)[]{\small{$\tilde{\nu}_{R\alpha}$}}
\Text(42,33)[]{(5)}
\end{picture}
\begin{picture}(110,80)(-55,-50)
\ArrowLine(-35,0)(-49,0) \ArrowLine(-35,0)(-18,0)
\ArrowLine(0,0)(-18,0) \ArrowLine(0,0)(18,0)
\ArrowLine(35,0)(18,0) \ArrowLine(35,0)(49,0)
\DashLine(0,0)(0,15){1} \DashArrowLine(0,30)(0,15){1}
\Photon(0,-20)(0,-40){2}{3} \DashArrowArc(0,20)(40,210,270){3}
\DashArrowArcn(0,20)(40,330,270){3} \Text(-42,-8)[]{$\tau^c$}
\Text(42,-7)[]{$\mu$} \Text(-10,25)[]{\small{$\rho^0$}}
\Text(-27,8)[]{\ftsz{$\tilde{\rho}^{\prime-}_2$}}
\Text(-13,8)[]{\ftsz{$\tilde{\rho}^+_2$}}
\Text(12,9)[]{\ftsz{$\tilde{Y}^-$}}
\Text(30,9)[]{\ftsz{$\tilde{Y}^+$}} \Text(8,-40)[]{$Z'$}
\Text(0,-12)[]{\small{$\tilde{\nu}_{R\alpha}$}}
\Text(42,33)[]{(6)}
\end{picture}
\end{center}
\begin{center}
%
\begin{picture}(200,80)(-10,-40)
\ArrowLine(-35,0)(-49,0) \ArrowLine(-35,0)(-18,0)
\ArrowLine(0,0)(-18,0) \ArrowLine(0,0)(18,0)
\ArrowLine(35,0)(18,0) \ArrowLine(35,0)(49,0)
\DashLine(0,0)(0,15){1} \DashArrowLine(0,30)(0,15){1}
\Photon(0,-20)(0,-40){2}{3} \DashArrowArc(0,20)(40,210,270){3}
\DashArrowArc(0,20)(40,270,330){3} \Text(-42,-8)[]{$\tau^c$}
\Text(42,-7)[]{$\mu$} \Text(-10,25)[]{\small{$\rho^0$}}
\Text(-28,8)[]{\ftsz{$\tilde{\rho}^{\prime0}$}}
\Text(-14,8)[]{\ftsz{$\tilde{\rho}^0$}}
\Text(18,9)[]{\ftsz{$\lambda_3$}}
\Text(18,20)[]{\ftsz{$(\lambda_B,\lambda_8)$}}
\Text(8,-40)[]{$Z$}\Text(23,-40)[]{$(Z')$}
\Text(0,-11)[]{\small{$\tilde{\ell}_{L_\alpha}$}}
\Text(42,33)[]{(7)}
\end{picture}
\begin{picture}(180,80)(80,-40)
\ArrowLine(-35,0)(-49,0) \ArrowLine(-35,0)(-20,0)
\ArrowLine(0,0)(-20,0) \ArrowLine(0,0)(20,0)
\ArrowLine(35,0)(20,0) \ArrowLine(35,0)(49,0)
\DashLine(0,0)(0,15){1} \DashArrowLine(0,30)(0,15){1}
\Photon(-20,0)(-20,30){2}{3} \GCirc(-20,0){2}{0.5}
\DashArrowArc(0,20)(40,210,330){3} \Text(-42,-8)[]{$\tau^c$}
\Text(42,-7)[]{$\mu$} \Text(-9,25)[]{\small{$\rho^0$}}
\Text(-30,8)[]{\ftsz{$\tilde{\rho}^{\prime0}$}}
\Text(-8,8)[]{\ftsz{$\tilde{\rho}^0$}}
\Text(22,9)[]{\ftsz{$\lambda_3$}}
\Text(22,20)[]{\ftsz{$(\lambda_B,\lambda_8)$}}
\Text(-28,25)[]{$Z$}\Text(-28,38)[]{$(Z')$}
\Text(0,-11)[]{\small{$\tilde{\ell}_{L_\alpha}$}}
\Text(42,33)[]{(8)}
\end{picture}
%
\begin{picture}(0,80)(150,-40)
\ArrowLine(-49,0)(-35,0) \ArrowLine(-18,0)(-35,0)
\ArrowLine(-18,0)(0,0) \ArrowLine(18,0)(0,0)
\ArrowLine(18,0)(35,0) \ArrowLine(49,0)(35,0)
\DashLine(0,0)(0,15){1} \DashArrowLine(0,15)(0,30){1}
\Photon(0,-20)(0,-40){2}{3} \DashArrowArc(0,20)(40,210,270){3}
\DashArrowArc(0,20)(40,270,330){3} \Text(-42,-7)[]{$\tau$}
\Text(44,-8)[]{$\mu^c$} \Text(-10,25)[]{\small{$\rho^0$}}
\Text(-26,8)[]{\ftsz{$\tilde{\rho}^{\prime0}$}}
\Text(-9,8)[]{\ftsz{$\tilde{\rho}^0$}}
\Text(18,9)[]{\ftsz{$\lambda_B$}}
\Text(8,-40)[]{$Z$}\Text(23,-40)[]{$(Z')$}
\Text(0,-11)[]{\small{$\tilde{\ell}_{R_\alpha}$}}
\Text(42,33)[]{(9)}
\end{picture}
%
\begin{picture}(-50,80)(40,-40)
\ArrowLine(-49,0)(-35,0) \ArrowLine(-20,0)(-35,0)
\ArrowLine(-20,0)(0,0) \ArrowLine(20,0)(0,0)
\ArrowLine(20,0)(35,0) \ArrowLine(49,0)(35,0)
\DashLine(0,0)(0,15){1} \DashArrowLine(0,15)(0,30){1}
\Photon(-20,0)(-20,30){2}{3} \GCirc(-20,0){2}{0.5}
\DashArrowArc(0,20)(40,210,330){3} \Text(-42,-7)[]{$\tau$}
\Text(44,-8)[]{$\mu^c$} \Text(-9,25)[]{\small{$\rho^0$}}
\Text(-32,8)[]{\ftsz{$\tilde{\rho}^{\prime0}$}}
\Text(-9,8)[]{\ftsz{$\tilde{\rho}^0$}}
\Text(22,9)[]{\ftsz{$\lambda_B$}}
\Text(-28,25)[]{$Z$}\Text(-28,37)[]{$(Z')$}
\Text(0,-11)[]{\small{$\tilde{\ell}_{R_\alpha}$}}
\Text(42,33)[]{(10)}
\end{picture}
\ec \caption{\ftsz Diagrams contributing to $D^{Z (b)}_{L}$
($D^{Z' (b)}_{L}$) (two first lines) and $D^{Z (b)}_{R}$ ($D^{Z'
(b)}_{R}$) (the last line). Noting that the first diagram only
contributes to $D^{Z (b)}_{L}$ while the sixth  only contributes
to $D^{Z' (b)}_{L}$.}
\label{dzlrb}
\end{figure}

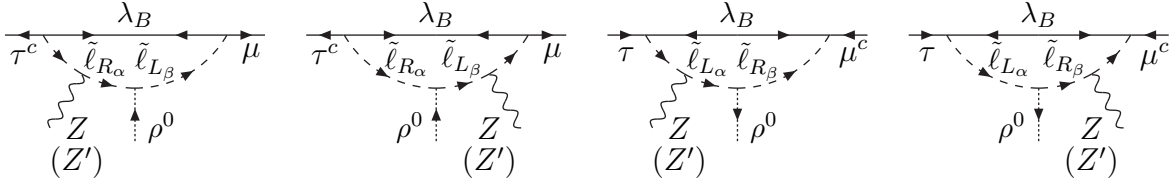
\begin{figure}[h]
%
\begin{center}
\begin{picture}(200,80)(-10,-40)
\ArrowLine(-35,0)(-49,0) \ArrowLine(-35,0)(0,0)
\ArrowLine(35,0)(0,0) \ArrowLine(35,0)(49,0)
\Photon(-32,-35)(-20,-15){2}{3} \DashArrowLine(0,-40)(0,-20){1}
\DashArrowArc(0,20)(40,210,240){3}
\DashArrowArc(0,20)(40,240,270){3}
\DashArrowArc(0,20)(40,270,330){3} \Text(-42,-7)[]{$\tau^c$}
\Text(44,-7)[]{$\mu$} \Text(0,8)[]{\ftsz{$\lambda_B$}}
\Text(-22,-35)[]{$Z$}\Text(-22,-48)[]{$(Z')$}
\Text(10,-35)[]{\small{$\rho^{0}$}}
\Text(-11,-9)[]{\small{$\tilde{\ell}_{R_\alpha}$}}
\Text(8,-10)[]{\small{$\tilde{\ell}_{L_\beta}$}}
\end{picture}
%
\begin{picture}(180,80)(80,-40)
\ArrowLine(-35,0)(-49,0) \ArrowLine(-35,0)(0,0)
\ArrowLine(35,0)(0,0) \ArrowLine(35,0)(49,0)
\Photon(32,-35)(20,-15){2}{3} \DashArrowLine(0,-40)(0,-20){1}
\DashArrowArc(0,20)(40,210,270){3}
\DashArrowArc(0,20)(40,270,300){3}
\DashArrowArc(0,20)(40,300,330){3} \Text(-42,-7)[]{$\tau^c$}
\Text(44,-7)[]{$\mu$} \Text(0,8)[]{\ftsz{$\lambda_B$}}
\Text(20,-35)[]{$Z$}\Text(20,-48)[]{$(Z')$}
\Text(-10,-35)[]{\small{$\rho^0$}}
\Text(-11,-9)[]{\small{$\tilde{\ell}_{R_\alpha}$}}
\Text(10,-9)[]{\small{$\tilde{\ell}_{L_\beta}$}}
\end{picture}
%
%
\begin{picture}(0,80)(150,-40)
\ArrowLine(-49,0)(-35,0) \ArrowLine(0,0)(-35,0)
\ArrowLine(0,0)(35,0) \ArrowLine(49,0)(35,0)
\Photon(-32,-35)(-20,-15){2}{3} \DashArrowLine(0,-20)(0,-40){1}
\DashArrowArc(0,20)(40,210,240){3}
\DashArrowArc(0,20)(40,240,270){3}
\DashArrowArc(0,20)(40,270,330){3} \Text(-42,-7)[]{$\tau$}
\Text(44,-7)[]{$\mu^c$} \Text(0,8)[]{\ftsz{$\lambda_B$}}
\Text(-22,-35)[]{$Z$}\Text(-22,-48)[]{$(Z')$}
\Text(10,-35)[]{\small{$\rho^{0}$}}
\Text(-11,-9)[]{\small{$\tilde{\ell}_{L_\alpha}$}}
\Text(8,-10)[]{\small{$\tilde{\ell}_{R_\beta}$}}
\end{picture}
\begin{picture}(-50,80)(40,-40)
\ArrowLine(-49,0)(-35,0) \ArrowLine(0,0)(-35,0)
\ArrowLine(0,0)(35,0) \ArrowLine(49,0)(35,0)
\Photon(32,-35)(20,-15){2}{3} \DashArrowLine(0,-20)(0,-40){1}
\DashArrowArc(0,20)(40,210,270){3}
\DashArrowArc(0,20)(40,270,300){3}
\DashArrowArc(0,20)(40,300,330){3} \Text(-42,-7)[]{$\tau$}
\Text(44,-8)[]{$\mu^c$} \Text(0,8)[]{\ftsz{$\lambda_B$}}
\Text(20,-35)[]{$Z$}\Text(20,-48)[]{$(Z')$}
\Text(-10,-35)[]{\small{$\rho^{0}$}}
\Text(-11,-9)[]{\small{$\tilde{\ell}_{L_\alpha}$}}
\Text(10,-9)[]{\small{$\tilde{\ell}_{R_\beta}$}}
\end{picture}
\end{center}
\hspace{0.3cm}
 \caption{\ftsz Diagrams that contribute to $D^{Z
(c)}_{L,R}$ ($D^{Z' (c)}_{L,R}$).}
\label{dzlrc}
\end{figure}

For $D^Z_{L,R}$, we have $D^Z_{L,R} = D^{Z(b)}_{L,R}
+D^{Z(c)}_{L,R}$. They are presented by diagrams in
figs.~\ref{dzlrb} and ~\ref{dzlrc}.

 Formulas for $D^{Z(b)}$:
\bea  D^{Z(b)}_L &=&
(s_Lc_L)\frac{g^{2}}{16\pi^2}\times \frac{1}{6}\times \mu_{\rho}
m_\lambda \tan\gamma
  \crn&\times&\left[ 2J_5(m_{\lambda}^2,\mu^2_{\rho},
 \mu^2_{\rho},\mu^2_{\rho},m^2_{\tilde{l}_{L2}})
+J_5(m_{\lambda}^2,m_{\lambda}^2,\mu^2_{\rho},\mu^2_{\rho}
,m^2_{\tilde{l}_{L2}}) \right] \crn
&+&(s_Lc_L) \frac{ g^{2}}{16\pi^2}\times \frac{1}{3}\mu_{\rho}
m_\lambda\tan\gamma  \crn&\times&
  c_{2W}\left[m^2_{\tilde{l}_{L2}} I_5(m_{\lambda}^2,\mu_{\rho}^2,
 m^2_{\tilde{l}_{L2}},m^2_{\tilde{l}_{L2}},m^2_{\tilde{l}_{L2}})\right]
\crn&+&(s_{\nu_L}c_{\nu_L})\frac{
g^{2}}{16\pi^2}\times\frac{1}{2}\mu_{\rho}
m_\lambda\tan\gamma\crn&\times&
  m^2_{\tilde{\nu}_{L2}}I_5(m_{\lambda}^2,\mu_{\rho}^2,
 m^2_{\tilde{\nu}_{L2}},m^2_{\tilde{\nu}_{L2}},m^2_{\tilde{\nu}_{L2}})
\crn&-&(s_{\nu_L}c_{\nu_L})\frac{g^2 }{16\pi^2}\times
\frac{1}{2}\left(\mu_{\rho} m_{\lambda} \tan\gamma \right) \crn
 &\times&c^2_W \left[ 2J_5(m_{\lambda}^2,m_{\lambda}^2,m_{\lambda}^2,\mu^2_{\rho},m^2_{\tilde{\nu}_{L2}})
+J_5(m_{\lambda}^2,m_{\lambda}^2,\mu^2_{\rho},\mu^2_{\rho}
,m^2_{\tilde{\nu}_{L2}}) \right]
\crn&-& (s_{\nu_R}c_{\nu_R}) \frac{g^2 }{16\pi^2}\times\frac{1}{4}
\left(\mu_{\rho} m_{\lambda} \tan\gamma \right) \times\crn
 &\times& c_{2W}\left[ 2J_5(m_{\lambda}^2,m_{\lambda}^2,m_{\lambda}^2,\mu^2_{\rho},m^2_{\tilde{\nu}_{R2}})
+J_5(m_{\lambda}^2,m_{\lambda}^2,\mu^2_{\rho},\mu^2_{\rho}
,m^2_{\tilde{\nu}_{R2}}) \right]
\crn&+& (s_{\nu_L}c_{\nu_L})\frac{g^2 }{16\pi^2}\times
\left(\mu_{\rho} m_{\lambda} \tan\gamma \right) \times
\frac{1}{4}\crn
 &\times& \left(-1+2s^2_W\right) \left[ 2J_5(m_{\lambda}^2,
 \mu^2_{\rho},\mu^2_{\rho},\mu^2_{\rho},m^2_{\tilde{\nu}_{L2}})
+J_5(m_{\lambda}^2,m_{\lambda}^2,\mu^2_{\rho},\mu^2_{\rho}
,m^2_{\tilde{\nu}_{L2}}) \right]
\crn&+& (s_{\nu_R}c_{\nu_R})\frac{g^2 }{16\pi^2}\times
\left(\mu_{\rho} m_{\lambda} \tan\gamma \right) \times
\frac{1}{4}\crn
 & \times& s^2_W\left[ 2J_5(m_{\lambda}^2,\mu^2_{\rho},
 \mu^2_{\rho},\mu^2_{\rho},m^2_{\tilde{\nu}_{R2}})
+J_5(m_{\lambda}^2,m_{\lambda}^2,\mu^2_{\rho},\mu^2_{\rho}
,m^2_{\tilde{\nu}_{R2}}) \right]
\crn&+&(s_Lc_L)\frac{g^{\prime2}}{16\pi^2}\times
\frac{1}{27}\times \mu_{\rho} m_B \tan\gamma
  \crn&\times& c_{2W} m^2_{\tilde{l}_{L2}} I_5(m_B^2,\mu_{\rho}^2,
 m^2_{\tilde{l}_{L2}},m^2_{\tilde{l}_{L2}},m^2_{\tilde{l}_{L2}})
\crn&+&(s_Lc_L)\frac{g^{\prime2}}{16\pi^2}\times
\frac{1}{54}\times \mu_{\rho} m_B \tan\gamma
  \crn&\times&\left[ 2J_5(m_{B}^2,\mu^2_{\rho},
 \mu^2_{\rho},\mu^2_{\rho},m^2_{\tilde{l}_{L2}})
+J_5(m_{B}^2,m_{B}^2,\mu^2_{\rho},\mu^2_{\rho}
,m^2_{\tilde{l}_{L2}}) \right] \crn
 &-& (L_2\rightarrow L_3, ~R_2\rightarrow R_3),
 \eea
\bea D^{Z(b)}_R&=&
-(s_Rc_R)\frac{ g^{\prime2}}{16\pi^2}\times \frac{1}{18} m_B
\mu_{\rho}\tan\gamma\crn&\times&\left[(-4s^2_W)m^2_{\tilde{l}_{R2}}
  I_5(m_B^2,\mu_{\rho}^2,
 m^2_{\tilde{l}_{R2}},m^2_{\tilde{l}_{R2}},m^2_{\tilde{l}_{R2}})\right.
 \crn&+&\left. 2J_5(m_B^2,\mu_{\rho}^2,\mu_{\rho}^2,\mu_{\rho}^2
,m^2_{\tilde{l}_{R2}})+
  J_5(m_B^2,m_B^2,\mu_{\rho}^2,\mu_{\rho}^2,m^2_{\tilde{l}_{R2}}) \right]
  \crn  &-& (R_2\rightarrow R_3). \label{dzlrbf}\eea

Formulas for $D^{Z(c)}_{L,R} $:
 \bea  D^{Z(c)}_L&=& -
(s_Lc_L)\frac{g^{\prime2}}{16\pi^2}
 m_B\mu_{\rho}\tan\gamma \times\frac{1}{72}\crn
 &\times&\left[
 (1-2s^2_W) \left(s^2_RJ_5(m_B^2,m_B^2,m^2_{\tilde{l}_{L2}},m^2_{\tilde{l}_{L2}},
 m^2_{\tilde{l}_{R2}})\right.\right.\crn
 &+&\left.c^2_RJ_5(m_B^2,m_B^2,m^2_{\tilde{l}_{L2}},m^2_{\tilde{l}_{L2}},
 m^2_{\tilde{l}_{R3}})\left.\right)\right.\crn
 &+&\left. 2s^2_W \left(s^2_RJ_5(m_B^2,m_B^2,m^2_{\tilde{l}_{L2}},m^2_{\tilde{l}_{R2}},
 m^2_{\tilde{l}_{R2}})\right.\right.\crn
 &+&\left.c^2_RJ_5(m_B^2,m_B^2,m^2_{\tilde{l}_{L2}},m^2_{\tilde{l}_{R3}},
 m^2_{\tilde{l}_{R3}})\left.\right)\right] -(L_2\rightarrow L_3),\eea
 \bea
 D^{Z(c)}_R&=& -(s_Rc_R) \times \frac{g^{\prime2}}{16\pi^2}
 m_B\mu_{\rho}\tan\gamma \times\frac{1}{72}\crn
 &\times&\left[
 (1-2s^2_W) \left(s^2_LJ_5(m_B^2,m_B^2,m^2_{\tilde{l}_{L2}},m^2_{\tilde{l}_{L2}},
 m^2_{\tilde{l}_{R2}})\right.\right.\crn
 &+&\left.c^2_LJ_5(m_B^2,m_B^2,m^2_{\tilde{l}_{L3}},m^2_{\tilde{l}_{L3}},
 m^2_{\tilde{l}_{R2}})\left.\right)\right.\crn
 &+&\left. 2s^2_W \left(s^2_LJ_5(m_B^2,m_B^2,m^2_{\tilde{l}_{L2}},m^2_{\tilde{l}_{R2}},
 m^2_{\tilde{l}_{R2}})\right.\right.\crn
 &+&\left.c^2_LJ_5(m_B^2,m_B^2,m^2_{\tilde{l}_{L3}},m^2_{\tilde{l}_{R2}},
 m^2_{\tilde{l}_{R2}})\left.\right)\right] -(R_2\rightarrow R_3).
 \label{dzclr1}  \eea
\section{\label{apzpmutau}Contributions to $Z'\rightarrow \mu\tau$}
\subsection{\label{apa1zpmutau}Contributions to  $A^{1Z'}_{L,R}$}

To determine the values of $A^{Z}_{L,R}$, $A^{1Z'}_{L,R}$ and
$A^{2Z'}_{L,R}$, we use techniques mentioned in \cite{Anna2}. From
formulas of covariant derivatives of neutral Higgs in Appendix
\ref{gaugeinteration1}, it is easy to  see that two terms relating
with $Z$ and $Z'$ bosons appearing in these covariant derivatives
are different from each others one factor $(-1)$. For
$A^{1Z'}_{L,R}$ which relates with $\rho^0$ and $\rho^{\prime0}$
we have  $ A^{1Z'}_{L,R} =A^{1Z'(a)}_{L,R}+A^{Z'(b)}_{L,R}
+A^{Z'(c)}_{L,R}$ and $A^{Z'}_{L(R)}= (m^2_Z/m^2_{Z'})
A^{1Z'}_{L(R)}+A^{2Z'}_{L(R)}$. This leads to the results:
 \bea
   A^{1Z'(a)}_{L,R}= -A^{Z(a)}_{L,R},\crn
 A^{Z'(b)}_{L,R}= -A^{Z(b)}_{L,R},\crn
 A^{Z'(c)}_{L,R}= -A^{Z(c)}_{L,R}.
  \eea
  \subsection{\label{apa2zpmutau}Contributions to  $A^{2Z'}_{L,R}$}
  Diagrams  contributing to $A^{2Z'}_{L,R}$ are quite similar to those shown in
Fig.\ref{alra}.
\begin{figure}[h]
\begin{center}
\begin{picture}(200,70)(-10,-30)
\ArrowLine(-49,0)(-35,0) \ArrowLine(-21,0)(-35,0)
\ArrowLine(-21,0)(-7,0) \ArrowLine(7,0)(-7,0)
\ArrowLine(7,0)(21,0) \ArrowLine(35,0)(21,0)
\ArrowLine(35,0)(49,0) \DashArrowLine(-21,30)(-21,15){1}
\DashLine(-21,15)(-21,0){1} \DashArrowLine(7,30)(7,15){1}
\DashLine(7,15)(7,0){1} \DashArrowArcn(0,20)(40,330,210){3}
\Text(-42,-7)[]{$\tau$} \Text(42,-7)[]{$\mu$}
\Text(-29,10)[]{\ftsz{$\tilde{Y}^+$}}
\Text(-13,9)[]{\footnotesize{$\tilde{\chi}^{-}$}}
\Text(0,9)[]{\ftsz{$\tilde{\chi}^{\prime+}$}}
\Text(16,10)[]{\ftsz{$\tilde{Y}^-$}}
\Text(32,10)[]{\ftsz{$\tilde{Y}^+$}}
\Text(-21,32)[b]{\small{$\chi^{0}_2$}}
\Text(7,32)[b]{\small{$\chi^{\prime0}_2$}}
\Text(0,-12)[]{\small{$\tilde{\nu}_{R_\alpha}$}}\Text(35,40)[]{(1)}
\end{picture}
\begin{picture}(180,70)(80,-30)
\ArrowLine(-49,0)(-35,0) \ArrowLine(-21,0)(-35,0)
\ArrowLine(-21,0)(-7,0) \ArrowLine(7,0)(-7,0)
\ArrowLine(7,0)(21,0) \ArrowLine(35,0)(21,0)
\ArrowLine(35,0)(49,0) \DashArrowLine(-7,15)(-7,30){1}
\DashLine(-7,15)(-7,0){1} \DashArrowLine(21,15)(21,30){1}
\DashLine(21,15)(21,0){1} \DashArrowArcn(0,20)(40,330,210){3}
\Text(-42,-7)[]{$\tau$} \Text(42,-7)[]{$\mu$}
\Text(-30,10)[]{\ftsz{$\tilde{Y}^+$}}
\Text(-14,10)[]{\ftsz{$\tilde{Y}^-$}}
\Text(1,9)[]{\ftsz{$\tilde{\chi}^{\prime+}$}}
\Text(15,9)[]{\ftsz{$\tilde{\chi}^-$}}
\Text(32,10)[]{\ftsz{$\tilde{Y}^+$}}
\Text(-7,32)[b]{\small{$\chi^{\prime0}_2$}}
\Text(21,32)[b]{\small{$\chi^{0}_2$}}
\Text(0,-12)[]{\small{$\tilde{\nu}_{R_\alpha}$}}\Text(44,40)[]{(2)}
\end{picture}
\begin{picture}(0,70)(150,-30)
\ArrowLine(-49,0)(-35,0) \ArrowLine(-9,0)(-35,0)
\ArrowLine(-9,0)(9,0) \ArrowLine(35,0)(9,0) \ArrowLine(35,0)(49,0)
\DashArrowLine(-9,30)(-9,15){1} \DashLine(-9,15)(-9,0){1}
\DashArrowLine(9,15)(9,30){1} \DashLine(9,15)(9,0){1}
\DashArrowArcn(0,20)(40,330,210){3} \Text(-42,-7)[]{$\tau$}
\Text(42,-7)[]{$\mu$} \Text(-21,10)[]{\ftsz{$\tilde{Y}^+$}}
\Text(0,9)[]{\ftsz{$\tilde{\chi}^{-}$}}
\Text(24,10)[]{\ftsz{$\tilde{Y}^+$}}
\Text(-9,32)[b]{\small{$\chi^{0}_2$}}
\Text(9,32)[b]{\small{$\chi^{0}_2$}}
\Text(0,-12)[]{\small{$\tilde{\nu}_{R_\alpha}$}}\Text(44,40)[]{(3)}
\end{picture}
\begin{picture}(-50,70)(40,-30)
\ArrowLine(-49,0)(-35,0) \ArrowLine(-22,0)(-35,0)
\ArrowLine(-22,0)(-9,0) \ArrowLine(9,0)(-9,0)
\ArrowLine(9,0)(22,0) \ArrowLine(35,0)(22,0)
\ArrowLine(35,0)(49,0) \DashArrowLine(-9,15)(-9,30){1}
\DashLine(-9,15)(-9,0){1} \DashArrowLine(9,30)(9,15){1}
\DashLine(9,15)(9,0){1} \DashArrowArcn(0,20)(40,330,210){3}
\Text(-42,-7)[]{$\tau$} \Text(42,-7)[]{$\mu$}
\Text(-32,10)[]{\ftsz{$\tilde{Y}^+$}}
\Text(-16,10)[]{\ftsz{$\tilde{Y}^-$}}
\Text(0,9)[]{\ftsz{$\tilde{\chi}^{\prime+}$}}
\Text(18,10)[]{\ftsz{$\tilde{Y}^-$}}
\Text(34,10)[]{\ftsz{$\tilde{Y}^+$}}
\Text(-9,32)[b]{\small{$\chi^{\prime0}_2$}}
\Text(9,32)[b]{\small{$\chi^{\prime0}_2$}}
\Text(0,-12)[]{\small{$\tilde{\nu}_{R_\alpha}$}}\Text(44,40)[]{(4)}
\end{picture}
\end{center}

\begin{center}
\begin{picture}(200,70)(0,-30)
\ArrowLine(-49,0)(-35,0) \ArrowLine(-9,0)(-35,0)
\ArrowLine(-9,0)(9,0) \ArrowLine(35,0)(9,0) \ArrowLine(35,0)(49,0)
\DashArrowLine(-9,30)(-9,15){1} \DashLine(-9,15)(-9,0){1}
\DashArrowLine(9,15)(9,30){1} \DashLine(9,15)(9,0){1}
\DashArrowArc(0,20)(40,210,330){3} \Text(-42,-7)[]{$\tau$}
\Text(42,-7)[]{$\mu$} \Text(-21,10)[]{\ftsz{$\lambda_i$}}
\Text(0,9)[]{\ftsz{$\tilde{H}^0_k$}}
\Text(24,10)[]{\ftsz{$\lambda_i$}}
\Text(-9,32)[b]{\small{$H^0_k$}} \Text(9,32)[b]{\small{$H^0_k$}}
\Text(0,-11)[]{\small{$\tilde{\ell}_{L_\alpha}$}}\Text(44,30)[]{(5)}
\end{picture}
\begin{picture}(180,70)(80,-30)
\ArrowLine(-49,0)(-35,0) \ArrowLine(-22,0)(-35,0)
\ArrowLine(-22,0)(-9,0) \ArrowLine(9,0)(-9,0)
\ArrowLine(9,0)(22,0) \ArrowLine(35,0)(22,0)
\ArrowLine(35,0)(49,0) \DashArrowLine(-9,15)(-9,30){1}
\DashLine(-9,15)(-9,0){1} \DashArrowLine(9,30)(9,15){1}
\DashLine(9,15)(9,0){1} \DashArrowArc(0,20)(40,210,330){3}
\Text(-42,-7)[]{$\tau$} \Text(42,-7)[]{$\mu$}
\Text(-21,10)[]{\ftsz{$\lambda_i$}}
\Text(0,9)[]{\ftsz{$\tilde{H}^0_k$}}
\Text(24,10)[]{\ftsz{$\lambda_i$}}
\Text(-9,32)[b]{\small{$H^0_k$}} \Text(9,32)[b]{\small{$H^0_k$}}
\Text(0,-11)[]{\small{$\tilde{\ell}_{L_\alpha}$}}\Text(44,30)[]{(6)}
\end{picture}
\begin{picture}(0,70)(150,-30)
\ArrowLine(-49,0)(-35,0) \ArrowLine(-9,0)(-35,0)
\ArrowLine(-9,0)(9,0) \ArrowLine(35,0)(9,0) \ArrowLine(35,0)(49,0)
\DashArrowLine(-9,30)(-9,15){1} \DashLine(-9,15)(-9,0){1}
\DashArrowLine(9,15)(9,30){1} \DashLine(9,15)(9,0){1}
\DashArrowArc(0,20)(40,210,330){3} \Text(-42,-7)[]{$\tau$}
\Text(42,-7)[]{$\mu$} \Text(-21,10)[]{\ftsz{$\lambda_i$}}
\Text(0,9)[]{\ftsz{$\tilde{H}^0_k$}}
\Text(24,10)[]{\ftsz{$\lambda_j$}}
\Text(-9,32)[b]{\small{$H^0_k$}} \Text(9,32)[b]{\small{$H^0_k$}}
\Text(0,-11)[]{\small{$\tilde{\ell}_{L_\alpha}$}}\Text(44,30)[]{(7)}
\end{picture}
\begin{picture}(-50,70)(40,-30)
\ArrowLine(-49,0)(-35,0) \ArrowLine(-22,0)(-35,0)
\ArrowLine(-22,0)(-9,0) \ArrowLine(9,0)(-9,0)
\ArrowLine(9,0)(22,0) \ArrowLine(35,0)(22,0)
\ArrowLine(35,0)(49,0) \DashArrowLine(-9,15)(-9,30){1}
\DashLine(-9,15)(-9,0){1} \DashArrowLine(9,30)(9,15){1}
\DashLine(9,15)(9,0){1} \DashArrowArc(0,20)(40,210,330){3}
\Text(-42,-7)[]{$\tau$} \Text(42,-7)[]{$\mu$}
\Text(-21,10)[]{\ftsz{$\lambda_i$}}
\Text(0,9)[]{\ftsz{$\tilde{H}^0_k$}}
\Text(24,10)[]{\ftsz{$\lambda_j$}}
\Text(-9,32)[b]{\small{$H_k^0$}} \Text(9,32)[b]{\small{$H_k^0$}}
\Text(0,-11)[]{\small{$\tilde{\ell}_{L_\alpha}$}}\Text(44,30)[]{(8)}
\end{picture}
\end{center}
%
%
\begin{center}
\begin{picture}(110,70)(-55,-30)
\ArrowLine(-35,0)(-49,0) \ArrowLine(-35,0)(-9,0)
\ArrowLine(9,0)(-9,0) \ArrowLine(9,0)(35,0) \ArrowLine(49,0)(35,0)
\DashArrowLine(-9,15)(-9,30){1} \DashLine(-9,15)(-9,0){1}
\DashArrowLine(9,30)(9,15){1} \DashLine(9,15)(9,0){1}
\DashArrowArc(0,20)(40,210,330){3} \Text(-42,-7)[]{$\tau^c$}
\Text(44,-8)[]{$\mu^c$} \Text(-21,10)[]{\ftsz{$\lambda_B$}}
\Text(0,9)[]{\ftsz{$\tilde{H}^0_k$}}
\Text(24,10)[]{\ftsz{$\lambda_B$}}
\Text(-9,32)[b]{\small{$H^0_k$}} \Text(9,32)[b]{\small{$H^0_k$}}
\Text(0,-11)[]{\small{$\tilde{\ell}_{R_\alpha}$}}\Text(44,30)[]{(9)}
\end{picture}
\begin{picture}(110,70)(-55,-30)
\ArrowLine(-35,0)(-49,0) \ArrowLine(-35,0)(-22,0)
\ArrowLine(-9,0)(-22,0) \ArrowLine(-9,0)(9,0)
\ArrowLine(22,0)(9,0) \ArrowLine(22,0)(35,0)
\ArrowLine(49,0)(35,0) \DashArrowLine(-9,30)(-9,15){1}
\DashLine(-9,15)(-9,0){1} \DashArrowLine(9,15)(9,30){1}
\DashLine(9,15)(9,0){1} \DashArrowArc(0,20)(40,210,330){3}
\Text(-42,-7)[]{$\tau^c$} \Text(44,-8)[]{$\mu^c$}
\Text(-21,10)[]{\ftsz{$\lambda_B$}}
\Text(0,9)[]{\ftsz{$\tilde{H}^0_k$}}
\Text(24,10)[]{\ftsz{$\lambda_B$}}
\Text(-9,32)[b]{\small{$H^0_k$}} \Text(9,32)[b]{\small{$H^0_k$}}
\Text(0,-11)[]{\small{$\tilde{\ell}_{R_\alpha}$}}\Text(44,30)[]{(10)}
\end{picture}
\end{center}
\caption{\ftsz Diagrams contributing to $A^{(2Z')}_L$ (first and
second  rows) and $A^{(2Z')}_R$ (third row). Here we denote
$H^0_k\in\{\chi^0_2,~\chi^{\prime0}_2\}$ while $\lambda_{i,j}$
implies $i,j= \{B,8\}$ and $i\neq j$.}
\label{alrap2}
\end{figure}
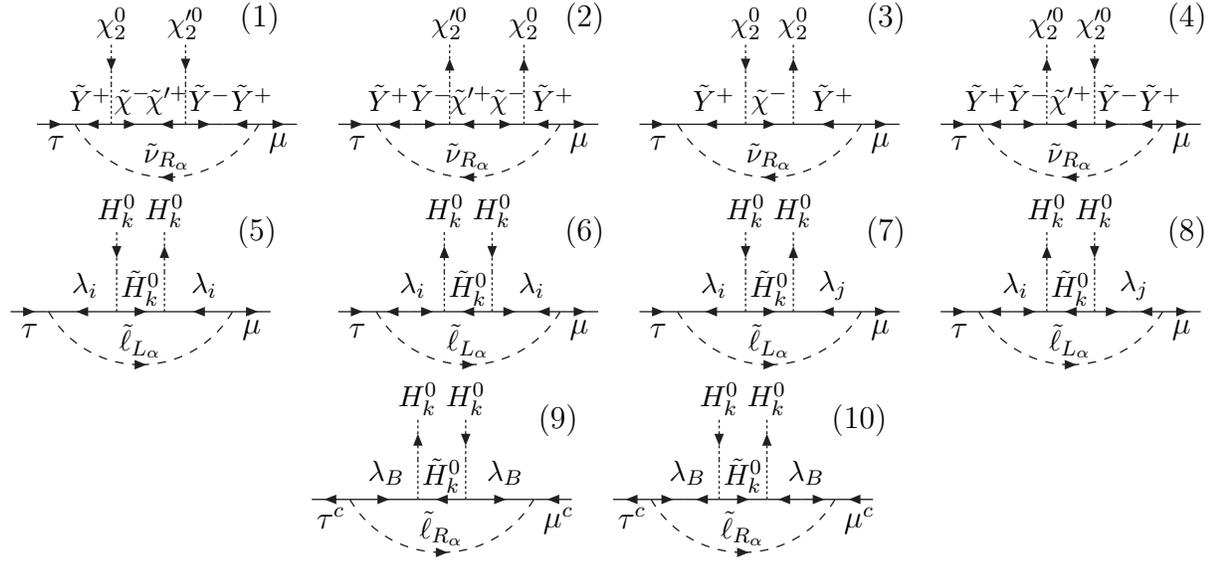
There is a interesting point in the SUSYE331 model that  both
Higgs $\chi$ and $\chi'$ do not couple to leptons and sleptons. As
a consequence, $A^{(2Z')}_{L,R}$ give contributions from only the
class of diagrams depicted in Fig.\ref{alrap2} where $\rho^0$,
$\rho^{\prime0}$ and $Z$ boson in Fig.\ref{alra} are
correspondingly replaced with $\chi^0_2$, $\chi^0_2$ and $Z'$
boson.  We use the equivalence role between $\rho^0\leftrightarrow
\chi^0_2$ and $\rho^{\prime0}\leftrightarrow \chi^{\prime0}_2$ for
the calculation ( see App.\ref{gaugeinteration1}). The results
are: \bea A^{(2Z')}_{L}&=&
(s_{\nu_R}c_{\nu_R})\frac{g^2\kappa_1^2}{16\pi^2}\times
\frac{1}{4} m_{\lambda}\mu_{\chi}s_{2\beta}~
J_5(m^2_{\lambda},m^2_{\lambda},\mu^2_{\chi},\mu^2_{\chi},
\tilde{m}^2_{\tilde{\nu}_{R2}})
\crn&-&(s_{\nu_R}c_{\nu_R})\frac{g^2\kappa_1^2}{16\pi^2}\times
\frac{1}{4}s^2_{\beta}\left[~~2
J_4(m^2_{\lambda},m^2_{\lambda},\mu^2_{\chi},
\tilde{m}^2_{\tilde{\nu}_{R2}})
\right.\crn&&\left.~~~~~~~~~~~~~~~~~~~~~~~~~~~
+\mu^2_{\chi}J_5(m^2_{\lambda},m^2_{\lambda},\mu^2_{\chi},\mu^2_{\chi},
\tilde{m}^2_{\tilde{\nu}_{R2}})\right]
\crn&-&(s_{\nu_R}c_{\nu_R})\frac{g^2\kappa_1^2}{16\pi^2}\times
\frac{1}{4}m^2_{\lambda}c^2_{\beta}
\left[~~\mu^2_{\chi}I_5(m^2_{\lambda},m^2_{\lambda},\mu^2_{\chi},\mu^2_{\chi},
\tilde{m}^2_{\tilde{\nu}_{R2}})
\right.\crn&&\left.~~~~~~~~~~~~~~~~~~~~~~~~~~~~~~~ -
I_4(m^2_{\lambda},m^2_{\lambda},\mu^2_{\chi},
\tilde{m}^2_{\tilde{\nu}_{R2}})\right]
\crn&+&(s_Lc_L)\frac{g^{\prime2}t^2\kappa_1^2}{16\pi^2}\times
\frac{1}{2916}c_{2\beta}\left[~2
J_4(m^2_{B},m^2_{B},\mu^2_{\chi}, \tilde{m}^2_{\tilde{\ell}_{L2}})
\right.\crn&&\left.~~~~~~~~~~~~~~~~~~~~~~~~~~~~~~~
+\mu^2_{\chi}J_5(m^2_{B},m^2_{B},\mu^2_{\chi},\mu^2_{\chi},
\tilde{m}^2_{\tilde{\ell}_{L2}})\right]
\crn&+&(s_Lc_L)\frac{g^{2}\kappa_1^2}{16\pi^2}\times
\frac{1}{9}c_{2\beta}\left[~2
J_4(m^2_{\lambda},m^2_{\lambda},\mu^2_{\chi},
\tilde{m}^2_{\tilde{\ell}_{L2}})
\right.\crn&&\left.~~~~~~~~~~~~~~~~~~~~~~~~~~
+\mu^2_{\chi}J_5(m^2_{\lambda},m^2_{\lambda},\mu^2_{\chi},\mu^2_{\chi},
\tilde{m}^2_{\tilde{\ell}_{L2}})\right]
\crn&-&(s_Lc_L)\frac{g^{\prime2} t^2\kappa_1^2}{16\pi^2}\times
\frac{1}{2916}m^2_{B}c_{2\beta}
\left[~~\mu^2_{\chi}I_5(m^2_{B},m^2_{B},\mu^2_{\chi},\mu^2_{\chi},
\tilde{m}^2_{\tilde{\ell}_{L2}})
\right.\crn&&\left.~~~~~~~~~~~~~~~~~~~~~~~~~~~~~~~ -
I_4(m^2_{B},m^2_{B},\mu^2_{\chi},
\tilde{m}^2_{\tilde{\ell}_{L2}})\right]
\crn&-&(s_Lc_L)\frac{g^2\kappa_1^2}{16\pi^2}\times
\frac{1}{9}m^2_{\lambda}c_{2\beta}
\left[~~\mu^2_{\chi}I_5(m^2_{\lambda},m^2_{\lambda},\mu^2_{\chi},\mu^2_{\chi},
\tilde{m}^2_{\tilde{\ell}_{L2}})
\right.\crn&&\left.~~~~~~~~~~~~~~~~~~~~~~~~~~~~~~~ -
I_4(m^2_{\lambda},m^2_{\lambda},\mu^2_{\chi},
\tilde{m}^2_{\tilde{\ell}_{L2}})\right]
\crn&-&(s_Lc_L)\frac{g^{\prime2}\kappa_1^2}{16\pi^2}\times
\frac{1}{162}c_{2\beta}\left[~2
J_4(m^2_{B},m^2_{\lambda},\mu^2_{\chi},
\tilde{m}^2_{\tilde{\ell}_{L2}})
\right.\crn&&\left.~~~~~~~~~~~~~~~~~~~~~~~~~~
+\mu^2_{\chi}J_5(m^2_{B},m^2_{\lambda},\mu^2_{\chi},\mu^2_{\chi},
\tilde{m}^2_{\tilde{\ell}_{L2}})\right]
\crn&+&(s_Lc_L)\frac{g^{\prime2}\kappa_1^2}{16\pi^2}\times
\frac{1}{162}m_{B}m_{\lambda}c_{2\beta}
\left[~~\mu^2_{\chi}I_5(m^2_{B},m^2_{\lambda},\mu^2_{\chi},\mu^2_{\chi},
\tilde{m}^2_{\tilde{\ell}_{L2}})
\right.\crn&&\left.~~~~~~~~~~~~~~~~~~~~~~~~~~~~~~~ -
I_4(m^2_{B},m^2_{\lambda},\mu^2_{\chi},
\tilde{m}^2_{\tilde{\ell}_{L2}})\right]
\crn&-&\left(L_2\rightarrow L_3, R_2\rightarrow R_3\right),
 \label{azp2l}\eea
\bea A^{(2Z')}_{R}&=&
-(s_Rc_R)\frac{g^{\prime2} t^2\kappa_1^2}{16\pi^2}\times
\frac{1}{324}c_{2\beta}\left[~2
J_4(m^2_{B},m^2_{B},\mu^2_{\chi}, \tilde{m}^2_{\tilde{\ell}_{R2}})
\right.\crn&&\left.~~~~~~~~~~~~~~~~~~~~~~~~~~
+\mu^2_{\chi}J_5(m^2_{B},m^2_{B},\mu^2_{\chi},\mu^2_{\chi},
\tilde{m}^2_{\tilde{\ell}_{R2}})\right]
\crn&+&(s_Rc_R)\frac{g^{\prime2}\kappa_1^2}{16\pi^2}\times
\frac{1}{324}m^2_{B}c_{2\beta}
\left[~~\mu^2_{\chi}I_5(m^2_{B},m^2_{B},\mu^2_{\chi},\mu^2_{\chi},
\tilde{m}^2_{\tilde{\ell}_{R2}})
\right.\crn&&\left.~~~~~~~~~~~~~~~~~~~~~~~~~~~~~~~ -
I_4(m^2_{B},m^2_{B},\mu^2_{\chi},
\tilde{m}^2_{\tilde{\ell}_{R2}})\right]\crn
&-&\left(R_2\rightarrow R_3\right).
 \label{azp2r}\eea

\subsection{\label{apczpmutau}Contributions to  $C^{Z'}_{L,R}$}
Diagrams contributing to $C^{Z'}_{L,R}$  are those from  2-6 in
the Fig. \ref{cz}. Comparing  with the case of the $Z$ boson we
easily deduce the formulas as:
\bea
 C^{Z'}_{L} &=&
-(c_{\nu_L}s_{\nu_L})\times
\frac{g^{2}}{16\pi^2}\times\frac{1}{12}c_{2W}\left[K_5
    (m^{2}_\lambda,m^2_{\tilde{\nu}_{L2}},m^2_{\tilde{\nu}_{L2}},
    m^2_{\tilde{\nu}_{L2}},m^2_{\tilde{\nu}_{L2}})\right]
    \crn&-& (c_Ls_L)\times\frac{g^2 }{16\pi^2}\times \frac{1}{18}c_{2W}
\left[-K_5
(m^2_{\lambda},m^2_{\tilde{l}_{L2}},m^2_{\tilde{l}_{L2}},m^2_{\tilde{l}_{L2}},m^2_{\tilde{l}_{L2}})
\right]
\crn&-&(c_{\nu_R}s_{\nu_R})\times\frac{g^2 }{16\pi^2}\times
\frac{4c^2_W-1}{12}
\left[-2K_{5}(m^2_{\lambda},m^2_{\lambda},m^2_{\lambda},m^2_{\lambda}
,\tilde{m}^2_{\nu_{R2}})\right.\crn&+&\left.3 m^2_{\lambda}
J_5(m^2_{\lambda},m^2_{\lambda},m^2_{\lambda},m^2_{\lambda}
,\tilde{m}^2_{\nu_{R2}})\right]
 \crn&+&(c_Ls_L )\times\frac{g^{\prime2}}{16
\pi^2}\times\frac{1}{324}(1-2s^2_W) \left[K_5
    (m^2_B,m^2_{\tilde{l}_{L2}},m^2_{\tilde{l}_{L2}},
    m^2_{\tilde{l}_{L2}},m^2_{\tilde{l}_{L2}})\right]
\crn&+&(c_{\nu_R}s_{\nu_R})\times
\frac{g^{2}}{16\pi^2}\times\frac{1}{12}c^2_{W}\left[K_5
    (m^{2}_\lambda,m^2_{\tilde{\nu}_{R2}},m^2_{\tilde{\nu}_{R2}},
    m^2_{\tilde{\nu}_{R2}},m^2_{\tilde{\nu}_{R2}})\right]
 \crn&-&[L_2\rightarrow L_3],
\label{czpl}\eea
\be C^{Z'}_R=- C^{Z}_R.\label{czpr}\ee

\subsection{\label{apdzpmutau}Contributions to  $D^{Z'}_{L,R}$}
Contribution to $D^{Z'}_{L,R}$ can be deduced from diagrams shown
for $D^{Z}_{L,R}$ in
figs. \ref{dzlrb} and \ref{dzlrc}. We also write
$D^{Z'}_{L,R}= D^{Z'(b)}_{L,R}+D^{Z'(c)}_{L,R}$.
From Fig. \ref{dzlrb} we can deduce formulas to determine
$D^{Z'(b)}$:
\bea  D^{Z'(b)}_L &=&
 -(s_{\nu_L}c_{\nu_L})\frac{g^2 }{16\pi^2}\times
\left(\mu_{\rho} m_{\lambda} \tan\gamma \right) \times
\frac{1}{4}\crn
 &\times& \left[ 2J_5(m_{\lambda}^2,
\mu^2_{\rho},\mu^2_{\rho},\mu^2_{\rho},m^2_{\tilde{\nu}_{L2}})
+J_5(m_{\lambda}^2,m_{\lambda}^2,\mu^2_{\rho},\mu^2_{\rho}
,m^2_{\tilde{\nu}_{L2}}) \right]
\crn&+&(s_{\nu_L}c_{\nu_L})\frac{
g^{2}}{16\pi^2}\times\frac{1}{2}\mu_{\rho}
m_\lambda\tan\gamma\crn&\times&
 c_{2W} \left[m^2_{\tilde{\nu}_{L2}}I_5(m_{\lambda}^2,\mu_{\rho}^2,
 m^2_{\tilde{\nu}_{L2}},m^2_{\tilde{\nu}_{L2}},m^2_{\tilde{\nu}_{L2}})\right]
\crn&+& (s_{\nu_R}c_{\nu_R}) \frac{g^2 }{16\pi^2}\times\frac{1}{4}
\left(\mu_{\rho} m_{\lambda} \tan\gamma \right)\crn
 &\times& \kappa_1^2\left[ 2J_5(m_{\lambda}^2,
 m_{\lambda}^2,m_{\lambda}^2,\mu^2_{\rho},m^2_{\tilde{\nu}_{R2}})
+J_5(m_{\lambda}^2,m_{\lambda}^2,\mu^2_{\rho},\mu^2_{\rho}
,m^2_{\tilde{\nu}_{R2}}) \right]
\crn&+& (s_{\nu_R}c_{\nu_R})\frac{g^2 }{16\pi^2}\times
\left(\mu_{\rho} m_{\lambda} \tan\gamma \right) \times
\frac{1}{4}\crn
 & \times& c_{2W}\left[ 2J_5(m_{\lambda}^2,\mu^2_{\rho},
\mu^2_{\rho},\mu^2_{\rho},m^2_{\tilde{\nu}_{R2}})
+J_5(m_{\lambda}^2,m_{\lambda}^2,\mu^2_{\rho},\mu^2_{\rho}
,m^2_{\tilde{\nu}_{R2}}) \right]
\crn&+&(s_{\nu_R}c_{\nu_R})\frac{ g^{2}}{16\pi^2}\times\mu_{\rho}
m_\lambda\tan\gamma\crn&\times&
  c^2_W\left[m^2_{\tilde{\nu}_{R2}}I_5(m_{\lambda}^2,\mu_{\rho}^2,
 m^2_{\tilde{\nu}_{R2}},m^2_{\tilde{\nu}_{R2}},m^2_{\tilde{\nu}_{R2}})\right]
\crn&-&(s_Lc_L)\frac{g^{\prime2}}{16\pi^2}\times
\frac{1}{27}\times \mu_{\rho} m_B \tan\gamma
  \crn&\times& c_{2W}\left[ m^2_{\tilde{l}_{L2}} I_5(m_B^2,\mu_{\rho}^2,
 m^2_{\tilde{l}_{L2}},m^2_{\tilde{l}_{L2}},m^2_{\tilde{l}_{L2}})\right]
\crn&-&(s_Lc_L) \frac{ g^{2}}{16\pi^2}\times \frac{1}{3}\mu_{\rho}
m_\lambda\tan\gamma  \crn&\times&
  c_{2W}\left[m^2_{\tilde{l}_{L2}} I_5(m_{\lambda}^2,\mu_{\rho}^2,
 m^2_{\tilde{l}_{L2}},m^2_{\tilde{l}_{L2}},m^2_{\tilde{l}_{L2}})\right]
\crn&-&(s_Lc_L)\frac{g^{\prime2}}{16\pi^2}\times
\frac{1}{54}\times \mu_{\rho} m_B \tan\gamma
  \crn&\times&\left[ 2J_5(m_{B}^2,\mu^2_{\rho},
 \mu^2_{\rho},\mu^2_{\rho},m^2_{\tilde{l}_{L2}})
+J_5(m_{B}^2,m_{B}^2,\mu^2_{\rho},\mu^2_{\rho}
,m^2_{\tilde{l}_{L2}}) \right]
\crn&-&(s_Lc_L)\frac{g^{2}}{16\pi^2}\times \frac{1}{6}\times
\mu_{\rho} m_\lambda \tan\gamma
  \crn&\times&\left[ 2J_5(m_{\lambda}^2,\mu^2_{\rho},
 \mu^2_{\rho},\mu^2_{\rho},m^2_{\tilde{l}_{L2}})
+J_5(m_{\lambda}^2,m_{\lambda}^2,\mu^2_{\rho},\mu^2_{\rho}
,m^2_{\tilde{l}_{L2}}) \right] \crn &-& \left( L_2\rightarrow
L_3,~R_2\rightarrow R_3\right),
  \label{dzpb1}\eea

\be D^{Z'(b)}_R= -D^{Z(b)}_R\label{dzplrbf}\ee
 \be  D^{Z'(c)}_L= -D^{Z(c)}_L\label{dzcpl1}\ee
 \be
 D^{Z'(c)}_R=- D^{Z(c)}_R\label{dzpcr1}  \ee

\section{\label{apbtaumu}Contribution from  $B^{\mu_{L,R}}_{L,R}$ to $\tau\rightarrow 3\mu$}
\begin{figure}[h]
\begin{center}
\begin{picture}(200,60)(-10,-30)
\ArrowLine(-45,15)(-25,15) \ArrowLine(-25,-15)(-25,15)
\ArrowLine(-45,-15)(-25,-15) \ArrowLine(25,15)(45,15)
\ArrowLine(25,15)(25,-15)
 \ArrowLine(25,-15)(45,-15)\DashArrowLine(-25,15)(25,15){3}
\DashArrowLine(25,-15)(-25,-15){3} \Text(-42,23)[]{$\tau$}
\Text(42,23)[]{$\mu$} \Text(-42,-24)[]{$\mu$}
\Text(42,-24)[]{$\mu$} \Text(-14,0)[]{\ftsz{$\tilde{W}^+$}}
\Text(14,0)[]{\ftsz{$\tilde{W}^+$}}
\Text(0,25)[]{\small{$\tilde{\nu}_{L_\alpha}$}}
\Text(0,-25)[]{\small{$\tilde{\nu}_{L_\alpha}$}}
\end{picture}
\begin{picture}(180,60)(80,-30)
\ArrowLine(-45,15)(-25,15) \ArrowLine(-25,-15)(-25,15)
\ArrowLine(-45,-15)(-25,-15) \ArrowLine(25,15)(45,15)
\ArrowLine(25,15)(25,-15)
 \ArrowLine(25,-15)(45,-15)\DashArrowLine(-25,15)(25,15){3}
\DashArrowLine(25,-15)(-25,-15){3} \Text(-42,23)[]{$\tau$}
\Text(42,23)[]{$\mu$} \Text(-42,-24)[]{$\mu$}
\Text(42,-24)[]{$\mu$} \Text(-14,0)[]{\ftsz{$\tilde{Y}^+$}}
\Text(14,0)[]{\ftsz{$\tilde{Y}^+$}}
\Text(0,25)[]{\small{$\tilde{\nu}_{R_\alpha}$}}
\Text(0,-25)[]{\small{$\tilde{\nu}_{R_\alpha}$}}
\end{picture}
\begin{picture}(0,60)(150,-30)
\ArrowLine(-45,15)(-25,15) \ArrowLine(-25,-15)(-25,15)
\ArrowLine(-25,-15)(-45,-15) \ArrowLine(25,15)(45,15)
\ArrowLine(25,15)(25,-15) \ArrowLine(45,-15)(25,-15)
\DashArrowLine(-25,15)(25,15){3}
\DashArrowLine(25,-15)(-25,-15){3} \Text(-42,23)[]{$\tau$}
\Text(42,23)[]{$\mu$} \Text(-42,-24)[]{$\mu$}
\Text(42,-24)[]{$\mu$}
\Text(-15,0)[]{\ftsz{$\lambda_i$}}
\Text(15,0)[]{\ftsz{$\lambda_i$}}
\Text(0,25)[]{\small{$\tilde{\ell}_{L_\alpha}$}}
\Text(0,-25)[]{\small{$\tilde{\ell}_{L_\beta}$}}
\end{picture}
\begin{picture}(-50,60)(40,-30)
\ArrowLine(-45,15)(-25,15) \ArrowLine(-25,0)(-25,15)
\ArrowLine(-25,0)(-25,-15) \ArrowLine(-45,-15)(-25,-15)
\ArrowLine(25,15)(45,15) \ArrowLine(25,15)(25,0)
\ArrowLine(25,-15)(25,0) \ArrowLine(25,-15)(45,-15)
\DashArrowLine(-25,15)(25,15){3}
\DashArrowLine(-25,-15)(25,-15){3} \Text(-42,23)[]{$\tau$}
\Text(42,23)[]{$\mu$} \Text(-42,-24)[]{$\mu$}
\Text(42,-24)[]{$\mu$} \Text(-15,0)[]{\ftsz{$\lambda_i$}}
\Text(15,0)[]{\ftsz{$\lambda_i$}}
\Text(0,25)[]{\small{$\tilde{\ell}_{L_\alpha}$}}
\Text(0,-25)[]{\small{$\tilde{\ell}_{L_\beta}$}}
\end{picture}
\end{center}
 \hspace{0.3 cm}
\begin{center}
\begin{picture}(200,60)(-10,-30)
\ArrowLine(-45,15)(-25,15) \ArrowLine(-25,-15)(-25,15)
\ArrowLine(-25,-15)(-45,-15) \ArrowLine(25,15)(45,15)
\ArrowLine(25,15)(25,-15) \ArrowLine(45,-15)(25,-15)
\DashArrowLine(-25,15)(25,15){3}
\DashArrowLine(25,-15)(-25,-15){3} \Text(-42,23)[]{$\tau$}
\Text(42,23)[]{$\mu$} \Text(-42,-24)[]{$\mu$}
\Text(42,-24)[]{$\mu$} \Text(-15,0)[]{\ftsz{$\lambda_i$}}
\Text(15,0)[]{\ftsz{$\lambda_j$}}
\Text(0,25)[]{\small{$\tilde{\ell}_{L_\alpha}$}}
\Text(0,-25)[]{\small{$\tilde{\ell}_{L_\beta}$}}
\end{picture}
%
\begin{picture}(180,60)(80,-30)
\ArrowLine(-45,15)(-25,15) \ArrowLine(-25,0)(-25,15)
\ArrowLine(-25,0)(-25,-15) \ArrowLine(-45,-15)(-25,-15)
\ArrowLine(25,15)(45,15) \ArrowLine(25,15)(25,0)
\ArrowLine(25,-15)(25,0) \ArrowLine(25,-15)(45,-15)
\DashArrowLine(-25,15)(25,15){3}
\DashArrowLine(-25,-15)(25,-15){3} \Text(-42,23)[]{$\tau$}
\Text(42,23)[]{$\mu$} \Text(-42,-23)[]{$\mu$}
\Text(42,-23)[]{$\mu$} \Text(-15,0)[]{\ftsz{$\lambda_i$}}
\Text(15,0)[]{\ftsz{$\lambda_j$}}
\Text(0,25)[]{\small{$\tilde{\ell}_{L_\alpha}$}}
\Text(0,-25)[]{\small{$\tilde{\ell}_{L_\beta}$}}
\end{picture}
%
\begin{picture}(0,60)(150,-30)
\ArrowLine(-45,15)(-25,15) \ArrowLine(-25,-15)(-25,15)
\ArrowLine(-25,-15)(-45,-15) \ArrowLine(25,15)(45,15)
\ArrowLine(25,15)(25,-15) \ArrowLine(45,-15)(25,-15)
\DashArrowLine(-25,15)(25,15){3}
\DashArrowLine(-25,-15)(25,-15){3} \Text(-42,23)[]{$\tau$}
\Text(42,23)[]{$\mu$} \Text(-42,-24)[]{$\mu^c$}
\Text(42,-24)[]{$\mu^c$} \Text(-15,0)[]{\ftsz{$\lambda_B$}}
\Text(15,0)[]{\ftsz{$\lambda_B$}}
\Text(0,25)[]{\small{$\tilde{\ell}_{L_\alpha}$}}
\Text(0,-25)[]{\small{$\tilde{\ell}_{R_\beta}$}}
\end{picture}
\begin{picture}(-50,60)(40,-30)
\ArrowLine(-45,15)(-25,15) \ArrowLine(-25,0)(-25,15)
\ArrowLine(-25,0)(-25,-15) \ArrowLine(-45,-15)(-25,-15)
\ArrowLine(25,15)(45,15) \ArrowLine(25,15)(25,0)
\ArrowLine(25,-15)(25,0) \ArrowLine(25,-15)(45,-15)
\DashArrowLine(-25,15)(25,15){3}
\DashArrowLine(25,-15)(-25,-15){3} \Text(-42,23)[]{$\tau$}
\Text(42,23)[]{$\mu$} \Text(-42,-24)[]{$\mu^c$}
\Text(42,-24)[]{$\mu^c$} \Text(-15,0)[]{\ftsz{$\lambda_B$}}
\Text(15,0)[]{\ftsz{$\lambda_B$}}
\Text(0,25)[]{\small{$\tilde{\ell}_{L_\alpha}$}}
\Text(0,-25)[]{\small{$\tilde{\ell}_{R_\beta}$}}
\end{picture}
\end{center}
\vspace{0.3 cm} \hspace{0.3 cm}
\begin{center}
\begin{picture}(200,60)(-10,-40)
\ArrowLine(-25,15)(-45,15) \ArrowLine(-25,15)(-25,-15)
\ArrowLine(-45,-15)(-25,-15) \ArrowLine(45,15)(25,15)
\ArrowLine(25,-15)(25,15) \ArrowLine(25,-15)(45,-15)
\DashArrowLine(-25,15)(25,15){3}
\DashArrowLine(-25,-15)(25,-15){3} \Text(-42,23)[]{$\tau^c$}
\Text(44,23)[]{$\mu^c$} \Text(-42,-24)[]{$\mu$}
\Text(42,-24)[]{$\mu$} \Text(-15,0)[]{\ftsz{$\lambda_B$}}
\Text(15,0)[]{\ftsz{$\lambda_B$}}
\Text(0,25)[]{\small{$\tilde{\ell}_{R_\alpha}$}}
\Text(0,-25)[]{\small{$\tilde{\ell}_{L_\beta}$}}
\end{picture}
\begin{picture}(180,60)(80,-40)
\ArrowLine(-25,15)(-45,15) \ArrowLine(-25,15)(-25,0)
\ArrowLine(-25,-15)(-25,0) \ArrowLine(-25,-15)(-45,-15)
\ArrowLine(45,15)(25,15) \ArrowLine(25,0)(25,15)
\ArrowLine(25,0)(25,-15) \ArrowLine(45,-15)(25,-15)
\DashArrowLine(-25,15)(25,15){3}
\DashArrowLine(25,-15)(-25,-15){3} \Text(-42,23)[]{$\tau^c$}
\Text(44,23)[]{$\mu^c$} \Text(-42,-24)[]{$\mu$}
\Text(42,-24)[]{$\mu$} \Text(-15,0)[]{\ftsz{$\lambda_B$}}
\Text(15,0)[]{\ftsz{$\lambda_B$}}
\Text(0,25)[]{\small{$\tilde{\ell}_{R_\alpha}$}}
\Text(0,-25)[]{\small{$\tilde{\ell}_{L_\beta}$}}
\end{picture}
\begin{picture}(0,60)(150,-40)
\ArrowLine(-25,15)(-45,15) \ArrowLine(-25,15)(-25,-15)
\ArrowLine(-45,-15)(-25,-15) \ArrowLine(45,15)(25,15)
\ArrowLine(25,-15)(25,15) \ArrowLine(25,-15)(45,-15)
\DashArrowLine(-25,15)(25,15){3}
\DashArrowLine(25,-15)(-25,-15){3} \Text(-42,23)[]{$\tau^c$}
\Text(44,23)[]{$\mu^c$} \Text(-42,-24)[]{$\mu^c$}
\Text(42,-24)[]{$\mu^c$} \Text(-15,0)[]{\ftsz{$\lambda_B$}}
\Text(15,0)[]{\ftsz{$\lambda_B$}}
\Text(0,25)[]{\small{$\tilde{\ell}_{R_\alpha}$}}
\Text(0,-25)[]{\small{$\tilde{\ell}_{R_\beta}$}}
\end{picture}
\begin{picture}(-50,60)(40,-40)
\ArrowLine(-25,15)(-45,15) \ArrowLine(-25,15)(-25,0)
\ArrowLine(-25,-15)(-25,0) \ArrowLine(-25,-15)(-45,-15)
\ArrowLine(45,15)(25,15) \ArrowLine(25,0)(25,15)
\ArrowLine(25,0)(25,-15) \ArrowLine(45,-15)(25,-15)
\DashArrowLine(-25,15)(25,15){3}
\DashArrowLine(-25,-15)(25,-15){3} \Text(-42,23)[]{$\tau^c$}
\Text(44,23)[]{$\mu^c$} \Text(-42,-24)[]{$\mu^c$}
\Text(42,-24)[]{$\mu^c$} \Text(-15,0)[]{\ftsz{$\lambda_B$}}
\Text(15,0)[]{\ftsz{$\lambda_B$}}
\Text(0,25)[]{\small{$\tilde{\ell}_{R_\alpha}$}}
\Text(0,-25)[]{\small{$\tilde{\ell}_{R_\beta}$}}
\end{picture}
\end{center}
\caption{\ftsz Diagrams contributing to $B^{\mu_{L,R}}_L$ (first
and second rows) and $B^{\mu_{L,R}}_R$ (third row). $\lambda_i$
and $\lambda_j$  ($i\neq j$ in each above diagram) are gauginos in
which $\lambda_i$ and $\lambda_j \in \{ \lambda_B,
\lambda_3,\lambda_8\}$.} \label{blrf}
\end{figure}
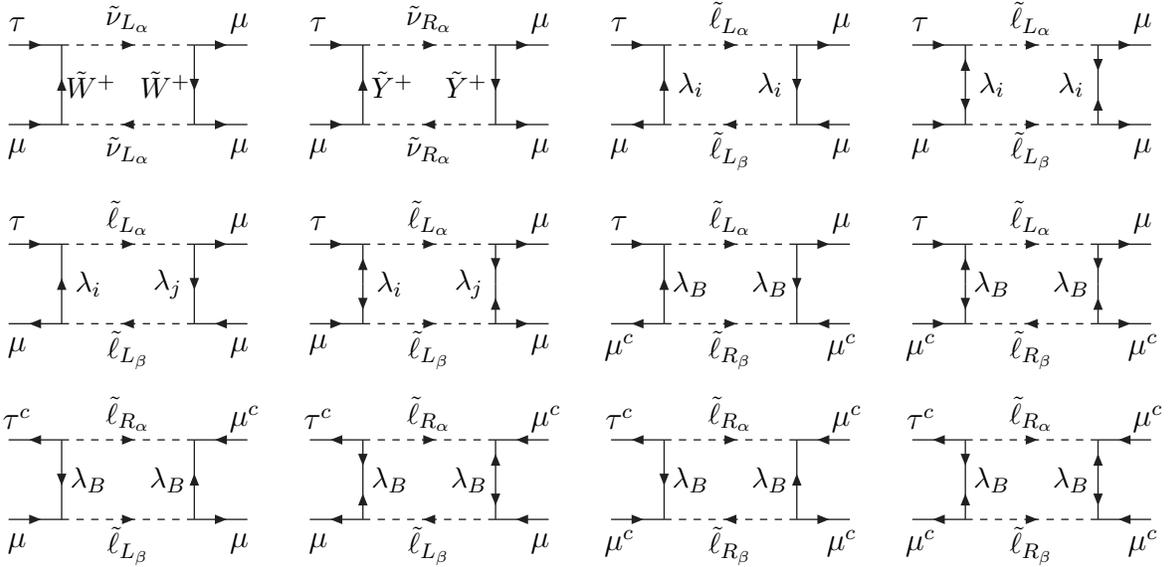

 Contributions to
$B^{\mu_{L,R}}_{L,R}$ arise from the diagrams in
Fig.~\ref{blrf}.
%
The formulas are:
\bea  B^{\mu_L}_L&=&
(s_{\nu_L}c_{\nu_L})\times\frac{g^4}{16\pi^2}\times\frac{1}{8}
\left[-c^2_{\nu_L}
J_4(m^2_\lambda,m^2_\lambda,m^2_{\tilde{\nu}_{L2}}
,m^2_{\tilde{\nu}_{L2}})\right.\crn&+& \left.s^2_{\nu_L}
J_4(m^2_\lambda,m^2_\lambda,m^2_{\tilde{\nu}_{L3}}
,m^2_{\tilde{\nu}_{L3}})\right.\crn &+&\left.
(c^2_{\nu_L}-s^2_{\nu_L})J_4(m^2_\lambda,m^2_\lambda,m^2_{\tilde{\nu}_{L2}}
,m^2_{\tilde{\nu}_{L3}}) \right]
\crn&+&(s_{\nu_R}c_{\nu_R})\frac{g^4}{16\pi^2}\times\frac{1}{8}
\left[-c^2_{\nu_R}
J_4(m^2_\lambda,m^2_\lambda,m^2_{\tilde{\nu}_{R2}}
,m^2_{\tilde{\nu}_{R2}})\right.\crn&+&\left. s^2_{\nu_R}
J_4(m^2_\lambda,m^2_\lambda,m^2_{\tilde{\nu}_{R3}}
,m^2_{\tilde{\nu}_{R3}})\right.\crn &+&\left.
(c^2_{\nu_R}-s^2_{\nu_R})J_4(m^2_\lambda,m^2_\lambda,m^2_{\tilde{\nu}_{R2}}
,m^2_{\tilde{\nu}_{R3}}) \right]
\crn &+&(s_Lc_L)\times\frac{g^4}{16\pi^2}\times \frac{1}{18}\crn
&\times& \left[-c^2_L\left(J_4(m_\lambda^2,m_\lambda^2,
 m^2_{\tilde{l}_{L2}},m^2_{\tilde{l}_{L2}})+2 m_{\lambda}^2I_4(m_\lambda^2,m_\lambda^2,
 m^2_{\tilde{l}_{L2}},m^2_{\tilde{l}_{L2}}) \right)\right.\crn
 &+&\left.s^2_L\left(J_4(m_\lambda^2,m_\lambda^2,
 m^2_{\tilde{l}_{L3}},m^2_{\tilde{l}_{L3}})+2m_{\lambda}^2I_4(m_\lambda^2,m_\lambda^2,
 m^2_{\tilde{l}_{L3}},m^2_{\tilde{l}_{L3}}) \right)
 \right.\crn
 &+&\left.(c^2_L-s^2_L)\left(J_4(m_\lambda^2,m_\lambda^2,
 m^2_{\tilde{l}_{L2}},m^2_{\tilde{l}_{L3}})+2m_{\lambda}^2I_4(m_\lambda^2,m_\lambda^2,
 m^2_{\tilde{l}_{L2}},m^2_{\tilde{l}_{L3}}) \right) \right]
\crn&+&(s_Lc_L)\times\frac{g^2g^{\prime2}}{16\pi^2}\times
\frac{1}{162}\crn&\times& \left[-c^2_L\left(J_4(m_B^2,m_\lambda^2,
 m^2_{\tilde{l}_{L2}},m^2_{\tilde{l}_{L2}})+2m_B~m_{\lambda}I_4(m_B^2,m_\lambda^2,
 m^2_{\tilde{l}_{L2}},m^2_{\tilde{l}_{L2}}) \right)\right.\crn
 &+&\left.s^2_L\left(J_4(m_B^2,m_\lambda^2,
 m^2_{\tilde{l}_{L3}},m^2_{\tilde{l}_{L3}})+2m_B~m_{\lambda}I_4(m_B^2,m_\lambda^2,
 m^2_{\tilde{l}_{L3}},m^2_{\tilde{l}_{L3}}) \right)
 \right.\crn
 &+&\left.(c^2_L-s^2_L)\left(J_4(m_B^2,m_\lambda^2,
 m^2_{\tilde{l}_{L2}},m^2_{\tilde{l}_{L3}})+2m_B~m_{\lambda}I_4(m_B^2,m_\lambda^2,
 m^2_{\tilde{l}_{L2}},m^2_{\tilde{l}_{L3}}) \right) \right]\crn
&+&(s_Lc_L)\times \frac{g^{\prime4}}{16\pi^2}\times
\frac{1}{216}\crn&\times& \left[-c^2_L\left(J_4(m_B^2,m_B^2,
 m^2_{\tilde{l}_{L2}},m^2_{\tilde{l}_{L2}})+2m_B^2I_4(m_B^2,m_B^2,
 m^2_{\tilde{l}_{L2}},m^2_{\tilde{l}_{L2}}) \right)\right.\crn
 &+&\left.s^2_L\left(J_4(m_B^2,m_B^2,
 m^2_{\tilde{l}_{L3}},m^2_{\tilde{l}_{L3}})+2m_B^2I_4(m_B^2,m_B^2,
 m^2_{\tilde{l}_{L3}},m^2_{\tilde{l}_{L3}}) \right)
 \right.\crn
 &+&\left.(c^2_L-s^2_L)\left(J_4(m_B^2,m_B^2,
 m^2_{\tilde{l}_{L2}},m^2_{\tilde{l}_{L3}})+2m_B^2I_4(m_B^2,m_B^2,
 m^2_{\tilde{l}_{L2}},m^2_{\tilde{l}_{L3}}) \right) \right],\crn\eea
 \bea \frac{B^{\mu_R}_L}{s_Lc_L}&=&\frac{g^{\prime4}}{16\pi^2}\times\frac{1}{648}
 \left[c^2_R \left(
 J_4(m_B^2,m_B^2,
 m^2_{\tilde{l}_{R2}},m^2_{\tilde{l}_{L2}})+2m_B^2I_4(m_B^2,m_B^2,
 m^2_{\tilde{l}_{R2}},m^2_{\tilde{l}_{L2}})\right)\right.
 \crn&+&\left.s^2_R \left(
 J_4(m_B^2,m_B^2,
 m^2_{\tilde{l}_{R3}},m^2_{\tilde{l}_{L2}})+2m_B^2I_4(m_B^2,m_B^2,
 m^2_{\tilde{l}_{R3}},m^2_{\tilde{l}_{L2}})\right)\right]\crn
 &-& (L_2\rightarrow L_3),\eea
 \bea
  \frac{B^{\mu_L}_R}{s_Rc_R}&=&\frac{g^{\prime4}}{16\pi^2}\times \frac{1}{648}\left[c^2_L \left(
 J_4(m_B^2,m_B^2,
 m^2_{\tilde{l}_{L2}},m^2_{\tilde{l}_{R2}})+2m_B^2I_4(m_B^2,m_B^2,
 m^2_{\tilde{l}_{L2}},m^2_{\tilde{l}_{R2}})\right)\right.\crn
 &+&\left. s^2_L \left(
 J_4(m_B^2,m_B^2,
 m^2_{\tilde{l}_{L3}},m^2_{\tilde{l}_{R2}})+2m_B^2I_4(m_B^2,m_B^2,
 m^2_{\tilde{l}_{L3}},m^2_{\tilde{l}_{R2}})\right)\right]\crn
 &-& (R_2\rightarrow R_3) ,\eea
 \bea
  \frac{ B^{\mu_R}_R}{s_Rc_R}&=&\frac{g^{\prime4}}{16\pi^2} \times \frac{1}{18}\left[-c^2_R \left(
 J_4(m_B^2,m_B^2,
 m^2_{\tilde{l}_{R2}},m^2_{\tilde{l}_{R2}})+2m_B^2I_4(m_B^2,m_B^2,
 m^2_{\tilde{l}_{R2}},m^2_{\tilde{l}_{R2}})\right)\right.\crn
 &+&\left.s^2_R \left(
 J_4(m_B^2,m_B^2,
 m^2_{\tilde{l}_{R3}},m^2_{\tilde{l}_{R3}})+2m_B^2I_4(m_B^2,m_B^2,
 m^2_{\tilde{l}_{R3}},m^2_{\tilde{l}_{R3}})\right) \right.\crn
 &+&\left.\left(c^2_R-s^2_R \right)\left(
 J_4(m_B^2,m_B^2,
 m^2_{\tilde{l}_{R2}},m^2_{\tilde{l}_{R3}})+2m_B^2I_4(m_B^2,m_B^2,
 m^2_{\tilde{l}_{R2}},m^2_{\tilde{l}_{R3}})\right) \right].\crn
 \label{Bmulr1}\eea

\end{document}